\documentclass[notitlepage, nofootinbib, no reprint, amsmath, amssymb, aps, twocolumn, pra]{revtex4-2}

\usepackage{amssymb}
\usepackage{hyperref}
\usepackage{bm}
\usepackage{soul}
\usepackage{dsfont}
\usepackage{amsthm}
\usepackage{amsmath}
\usepackage{amssymb}
\usepackage{verbatim} 
\usepackage{qcircuit}
\usepackage{circuitikz}
\usepackage{tabularx}
\usepackage{color,soul}
\usepackage{esint}
\usepackage{physics}
\usepackage{braket}
\usepackage{mathtools}
\usepackage{bbold}
\usepackage{float}
\usepackage[normalem]{ulem}

\usepackage{todonotes}

\def\R{\mathbb{R}}

\def\Id{\mathbb{1}}

\newcommand{\rrem}[1]{\ifmmode\text{\textcolor{red}{\sout{\ensuremath{#1}}}}\else\textcolor{red}{\sout{#1}}\fi} 

\theoremstyle{remark}

\def\f{f_{\alpha}}
\def\fa{\partial_{\alpha}\f}

\def\ver{\vec r_\alpha}
\def\da{\partial_{\alpha}}

\def\wl{w_\mathrm{ls}}
\def\wv{w_\mathrm{var}}

\def\hil{\mathcal H}

\newcommand\myangle[2]{
  \setbox0=\hbox{$\!\vec{#1}\,\vec{#2}\!$}
  \ht0=\dimexpr\ht0-1pt\relax
  \,\widehat{\copy0}\,}

\begin{document}

\author{Andrey Kardashin}
\email{andrey.kardashin@skoltech.ru}
\affiliation{Skolkovo Institute of Science and Technology, Moscow 121205, Russia}
\author{Yerassyl Balkybek}
\affiliation{Skolkovo Institute of Science and Technology, Moscow 121205, Russia}
\author{Vladimir V. Palyulin}
\affiliation{Skolkovo Institute of Science and Technology, Moscow 121205, Russia}
\author{Konstantin Antipin}
\affiliation{Faculty of Physics, M.V. Lomonosov Moscow State University,\\
Leninskie gory, GSP-1, Moscow 119991,  Russia}
\affiliation{Skolkovo Institute of Science and Technology, Moscow 121205, Russia}

\title{Predicting properties of quantum systems by regression on a quantum computer}

\begin{abstract}
    Quantum computers can be considered as a natural means for performing machine learning tasks for inherently quantum labeled data.
    Many quantum machine learning techniques have been developed for solving classification problems, such as distinguishing between phases of matter or quantum processes.
    Similarly, one can consider a more general problem of regression, when the aim is to predict continuous labels quantifying properties of quantum states, such as purity or entanglement.
    In this work, we propose a method for predicting such properties.
    The method is based on the notion of parametrized quantum circuits, and it seeks to find an observable the expectation of which gives the prediction of the property of interest with a low variance.
    We numerically test our approach in learning to predict (i) the parameter of a parametrized channel given its output state, (ii) entanglement of two-qubit states, and (iii) the parameter of a parametrized Hamiltonian given its ground state. 
    The results show that the proposed method is able to find observables such that they provide highly accurate predictions of the considered properties, and in some cases even saturate the Cramer-Rao bound, which characterizes the prediction error.
    We also compare our method with the Bayesian approach, and find that the latter prefers to minimize the prediction variance, having therefore a larger bias.
\end{abstract}

\maketitle

\section{Introduction}
    
    Quantum machine learning (QML) is a promising application for quantum computers \cite{schuld2015introduction, biamonte2017quantum, lloyd2018quantum, schuld2019quantum, mitarai2018quantum}.
    In QML one commonly considers datasets of labeled classical data points mapped to quantum states via various encoding strategies \cite{perez2020data, perez2021one, schuld2021effect, goto2021universal}.
    Often data labeled with discrete labels need to be classified \cite{rebentrost2014quantum, park2020theory, banchi2021generalization}, or a regression problem has to be solved for predicting continuous labels \cite{schuld2016prediction, suzuki2020predicting, reddy2021hybrid, nagano2023quantum}. 
    
    One approach for solving these problems uses kernel-based methods centered around computing a similarity measure (inner products) between data points \cite{havlivcek2019supervised, blank2020quantum, jerbi2023quantum}.
    The other approach of processing quantum states in QML is built upon the notion of parametrized (sometimes called variational) quantum circuits \cite{benedetti2019parameterized, schuld2020circuit}.
    Essentially these circuits are parametrized unitary operators used for transforming quantum states \cite{cerezo2021variational, biamonte2021universal}.
    For instance, one can find the ground state of a given Hamiltonian \cite{peruzzo2014variational, kandala_hardware-efficient_2017, Kardashin_2021}, solve linear algebra problems \cite{xu2021variational, bravo2023variational}, perform state \cite{larose2019variational} and Hamiltonian diagonalization \cite{commeau2020variational, cirstoiu2020variational, zeng2021variational}, etc.
    Additionally, variational quantum computing methods have been applied for quantum metrology and parameter estimation \cite{koczor2020variational, meyer2021variational, kaubruegger2021quantum}, i.e. to find optimal input states and measurements that provide an estimation of parameters of a quantum channel as precise as possible \cite{dorner2009optimal, pirandola2018advances, chen2022quantum}. 

    QML techniques can also be used for processing data which are initially quantum. For instance, one can classify phases of matter \cite{carrasquilla2017machine, van2017learning, uvarov2020machine}, detect anomalies in sets of quantum states \cite{liu2018quantum, liang2019quantum}, and discern between quantum states \cite{patterson2021quantum, chen2021universal} and quantum channels \cite{qiu2019solving, kardashin2022quantum, kundu2022variational}.
    Additionally, QML helps with simulating the dynamics of quantum systems \cite{gibbs2024dynamical}, detecting and classifying entanglement in quantum states \cite{qiu2019detecting, sanavio2023entanglement}, and even with discovery of quantum algorithms \cite{cincio2018learning} as well as error correction codes \cite{johnson2017qvector}.  
    
    In this work, we propose a method for solving regression problems for labeled data represented by quantum states $\rho_\alpha$ with labels $\alpha$.
    The method is based on variational quantum circuits, and assumes no knowledge about the connection between the labeled states and their labels.
    Among such connections, we consider: (i) the labeled state $\rho_\alpha$ is output state of a parametrized channel $\Phi_\alpha[\rho]$ for some fixed input $\rho$, (ii) $\alpha$ is an entanglement measure of $\rho_\alpha$, and (iii) $\rho_\alpha = \ketbra{\psi_\alpha}$ is such that $\ket{\psi_\alpha}$ is the ground state of a parametrized Hamiltonian $H_\alpha$.
    Case (i) is often considered in quantum metrology and channel estimation, (ii) is similar to the problem of entanglement quantification \cite{roik2022entanglement, wang2022detecting}, and (iii) can be viewed as a generalization of the classification of phases of matter \cite{nagano2023quantum}.
    
    The performance of our method is assessed with quantum Fisher information, a key concept of quantum metrology \cite{liu2020quantum, meyer2021fisher}.
    Using that notion, one can lower-bound the error of the prediction of the label. We compare the results given by our method with this bound and with the Bayesian approach \cite{sidhu2020geometric}.

    This work is structured as follows.
    In Section~\ref{sec:methods} we formulate the problem setting, describe the method of predicting the labels of labeled states, and introduce the Fisher information and Cramer-Rao bounds, as well as the Bayesian approach.
    Section~\ref{sec:literature} reviews the literature and states the contribution of our work.
    Section~\ref{sec:theory} contains analytical considerations of the method we propose.
    Section~\ref{sec:results} shows the results of the numerical application of the method for solving the regression problems (i)-(iii) mentioned above.
    The work is concluded with Section~\ref{sec:discussion} where we discuss the results and outline directions for further studies.

\section{Methods}
\label{sec:methods}

    We start with establishing the notation and the formulation of the regression problem. For this problem, we present a method for predicting the label of a given datum. 
    Then we discuss the importance of the variance of this  prediction, introduce both classical and quantum Fisher information, and use the latter to bound the prediction error.

    \subsection{Problem statement}

        Suppose we are given the following training set:
        \begin{equation}
            \label{eq:training_set}
            \mathcal{T} = \left\{ \rho_{\alpha_j}, \alpha_j \right\}_{j=1}^{T},
        \end{equation}
        where $\rho_{\alpha_j}$ are labeled data points and $\alpha_j \in \R$ are their corresponding labels.
        We consider the data points to be quantum states described by density operators, i.e. $\rho_{\alpha} \geqslant 0$ and $\Tr \rho_\alpha = 1$.
        Hereinafter, we assume that $\rho_{\alpha}$ describes a state of $n$ qubits, so that $\rho_\alpha$ acts in a Hilbert space of dimensionality $2^n$.
        Given the training set $\mathcal{T}$, the goal is to learn the prediction of parameters $\alpha$ for unseen data $\rho_\alpha$ (the data not present in $\mathcal{T}$).
        
        Essentially, we formulate a regression problem with data points being represented by quantum states. It should be emphasized that we assume no knowledge about the connection of the states $\rho_\alpha$ with their labels $\alpha$.
        One instance of such connection is $\alpha$ quantifying the entanglement of the state $\rho_\alpha$. 
        Thus, our aim is to devise a method for predicting the labels having as little information about the data as possible.
        
    \subsection{The estimator and the cost function}

        Since the data points $\rho_\alpha$ are quantum states, it is natural to obtain the label $\alpha$ from the expected value of an observable measured in $\rho_\alpha$.
        Generally, such expectation will be a function $\mathsf{a}$ of $\alpha$, which we represent as 
        \begin{equation}
            \label{eq:estimator}
            \mathsf{a}(\boldsymbol{x}, \Pi, \rho_\alpha) = \sum_i x_i \Tr\left(\Pi_i \rho_\alpha \right),
        \end{equation}
        where $\boldsymbol{x}=\{x_i\}_i \subset \R$ and the operators $\Pi=\{\Pi_i\}_i$ form a positive operator-valued measure (POVM),
        \begin{equation*}
            \Pi_i \geqslant 0, \quad \sum_i \Pi_i = \Id,
        \end{equation*}
        where $\Id$ is the identity operator.
        For a given state $\rho_\alpha$, an estimate of $\alpha$ is formally $\hat{\alpha} =\mathsf{a}^{-1}(\hat{\mathsf{a}})$, where $\hat{\mathsf{a}}$ is an estimate of the expectation $\mathsf{a}$ obtained in a measurement experiment.
        Essentially, $\mathsf{a}$ is the expected value of an observable represented by a POVM $\Pi$ with the corresponding measurement outcomes $\boldsymbol{x}$ measured with respect to $\rho_\alpha$.

        To avoid confusion, we list the notations and terminology we use in this work:
        \begin{itemize}
            \item $\alpha$ is the true label of a labeled state $\rho_\alpha$.
            \item $\mathsf{a}$ we call a \textit{prediction} of the label $\alpha$; it is obtained as the expected value \eqref{eq:estimator} and generally is a function of $\alpha$.
            \item $\hat{\mathsf{a}}$ is an \textit{estimate of the expectation} $\mathsf{a}$ obtained in a measurement experiment; essentially, it is generally a \textit{biased} estimate of the true label $\alpha$.
            \item $\hat{\alpha}$ is an \textit{estimate of the true label} $\alpha$, which can be obtained by inverting the function $\mathsf{a}$, i.e., $\hat{\alpha} = \mathsf{a}^{-1}(\hat{\mathsf{a}})$.
        \end{itemize}
        
        To find optimal $\Pi$ and $\boldsymbol{x}$ in \eqref{eq:estimator}, we minimize the least squares between the labels $\alpha_j$ given in $\mathcal{T}$ and our predictions $\mathsf{a}_j (\boldsymbol{x}, \Pi) \equiv \mathsf{a} (\boldsymbol{x}, \Pi, \rho_{\alpha_j})$,
        \begin{equation}
            \label{eq:ls_min}
            (\boldsymbol{x}^*, \Pi^*) = \arg\min_{\boldsymbol{x}, \Pi} \sum_{j=1}^{T} \Big(\alpha_j - \mathsf{a}_j(\boldsymbol{x}, \Pi) \Big)^2.
        \end{equation}
        In other words, the operators $\Pi^*$ with the corresponding outcomes $\boldsymbol{x}^*$ form an engineered observable.
        If this observable is measured with respect to a state $\rho_{\alpha_j}$, it gives a prediction $\mathsf{a}_j$ presumably as close to a given $\alpha_j$ as possible.
        
        It is not difficult to optimize over a set of real numbers $\boldsymbol{x}$. However, the problem is more challenging for $\Pi$.
        Nonetheless, the POVM elements $\Pi_i \in \Pi$ can be found with the variational quantum computing approach.
        Let us assume that $\Pi_i \Pi_j = \delta_{ij}\Pi_i$ with the operators $\Pi_i$ being rank-one projectors.
        With this assumption, we can obtain POVM elements as
        \begin{equation}
            \label{eq:orth_POVM_variational}
            \Pi_i \equiv \Pi_i(\boldsymbol{\theta}) = U^\dagger(\boldsymbol{\theta}) \ketbra{i} U(\boldsymbol{\theta}),
        \end{equation}
        where $\ketbra{i}$ is the projector onto the $i$th state of the computational basis, and $U(\boldsymbol{\theta})$ is a variational ansatz, i.e., a quantum circuit parametrized by $\boldsymbol{\theta} = \{\theta_k\}_k \subset \R$.
        That is, instead of optimizing over the operators $\Pi$, we optimize over real numbers $\boldsymbol{\theta}$.
        Denoting $\Pi(\boldsymbol{\theta}) = \{\Pi_i(\boldsymbol{\theta})\}_i$, we rewrite \eqref{eq:ls_min} as
        \begin{equation}
            \label{eq:ls_min_variational}
            (\boldsymbol{x}^*, \boldsymbol{\theta}^*) = \arg\min_{\boldsymbol{x}, \boldsymbol{\theta}} \sum_{j=1}^{T} \Big(\alpha_j - \mathsf{a}_j \big(\boldsymbol{x}, \Pi(\boldsymbol{\theta})\big) \Big)^2.
        \end{equation}
    
        Note that in the case of an orthogonal POVM \eqref{eq:orth_POVM_variational}, one can associate the measurement operators $\Pi$ and the corresponding measurement outcomes $\boldsymbol{x}$ with an $n$-qubit Hermitian operator $H$, which has $\boldsymbol{x}$ and $\Pi$ as its eigenvalues and eigenprojectors, respectively,
        \begin{equation}
            \label{eq:povm_ham}
            H(\boldsymbol{x}, \boldsymbol{\theta}) = \sum_{i=1}^{2^n} x_i \Pi_i(\boldsymbol{\theta}).
        \end{equation}
        Omitting the dependence on $\boldsymbol{x}$ and $\boldsymbol{\theta}$, we can rewrite \eqref{eq:estimator} simply as $\mathsf{a} = \Tr H \rho_\alpha$, i.e., the expected value of $H$ in the state $\rho_\alpha$.
        Equivalently, we can say that we transform the state $\rho_\alpha$ with a unitary $U$, measure the resulting state $U \rho_\alpha U^\dagger$ in the computational basis $\{\ketbra{i}\}_i$, and associate the outcome $i$ with the eigenvalue $x_i$. 
        Note also that if the ansatz $U(\boldsymbol{\theta})$ is expressive enough, there is always an assignment for $\boldsymbol{x}$ and $\boldsymbol{\theta}$ such that the decomposition \eqref{eq:povm_ham} gives any Hermitian matrix of the size $2^n \times 2^n$.
        
        In order to obtain a non-orthogonal POVM, one can utilize the Naimark's extension theorem.
        Namely, one can attach an auxiliary quantum system to a pure state, say, $\ketbra{0}$, and find a variational ansatz $U(\boldsymbol{\theta})$ acting on the joint state $\rho_\alpha\otimes\ketbra{0}$ such that \cite{rethinasamy2023estimating, banchi2021generalization}
        \begin{equation}
            \label{eq:naimark-variational}
            \Tr \Pi_i \rho_\alpha = \Tr\Big[U^\dagger(\boldsymbol{\theta}) \big(\Id \otimes \ketbra{i}\big) U(\boldsymbol{\theta}) \, \big(\rho_\alpha\otimes\ketbra{0}\big) \Big].
        \end{equation}
        Thus, one can measure a non-orthogonal POVM by measuring an orthogonal POVM in an extended Hilbert space.
        The dimensionality of this extension should coincide with the number of the POVM elements.
        As we show further in Sections~\ref{sec:iso} and \ref{sec:ent-pure}, Naimark's extension may be advantageous in solving certain regression problems.

    \subsection{Number of measured qubits}
    \label{sec:m-qubits}

        It was assumed above that we measure all qubits of the $n$-qubit state $\rho_\alpha$ making the number of varied parameters $|\boldsymbol{x}| = 2^n$ exponentially large.
        In some cases, however, it may be sufficient to measure only $m$ out of $n$ qubits.
        Hence, we can modify the observable \eqref{eq:povm_ham} to
        \begin{equation}
            \label{eq:povm_ham_m}
            H(\boldsymbol{x}, \boldsymbol{\theta}) = \sum_{i=1}^{2^m} x_i \Pi_i(\boldsymbol{\theta}),
        \end{equation}
        for which one can take the POVM elements to be, e.g.,
        \begin{equation}
            \label{eq:orth_POVM_variational_m}
            \Pi_i(\boldsymbol{\theta}) = U^\dagger(\boldsymbol{\theta}) \left( \Id^{\otimes (n - m)} \otimes \ketbra{i} \right) U(\boldsymbol{\theta}),
        \end{equation}
        being now orthogonal projectors of the rank $2^{(n-m)}$.
        Schematically, the process of measuring such $H(\boldsymbol{x}, \boldsymbol{\theta})$ with respect to $\rho_\alpha$ can be depicted as follows:
        \begin{equation*}
            \Qcircuit @C=1em @R=1em{
                & \qw & \qw/_{n-m} & \qw & \multigate{1}{U(\boldsymbol{\theta})} &        & \\
                & \qw & \qw/_m     & \qw & \ghost{U(\boldsymbol{\theta})}        & \meter & \;\;\qquad \overset{\,p_i}{\underset{}{\longrightarrow}} i \mapsto x_i 
                \inputgroupv{1}{2}{.75em}{1.em}{\rho_\alpha}
            }
        \end{equation*}        
        That is, we measure the last $m \leqslant n$ qubits of the state $\rho_\alpha(\boldsymbol{\theta}) \equiv U(\boldsymbol{\theta})\rho_\alpha U^\dagger(\boldsymbol{\theta})$ in the computational basis, and with probability $p_i = \Tr\! \left[\left( \Id^{\otimes (n - m)} \otimes \ketbra{i} \right) \rho_\alpha(\boldsymbol{\theta})\right]$ 
        get the outcome $i$ associated with $x_i$, evaluating the expectation therefore to $\Tr H \rho_\alpha = \sum_{i=1}^{2^m} x_i p_i$.

        In the case of measuring all $n$ qubits, the parametrized unitary $U(\boldsymbol{\theta})$ can be, for example, a hardware efficient ansatz \cite{kandala_hardware-efficient_2017}.
        If we measure only $m=1$ or 2 qubits, the circuit may take the form of a quantum convolutional neural network (QCNN) \cite{cong2019quantum}.
        In Appendix~\ref{app:hea}, we show examples of both such ans\"atze.
        In addition to having fixed depth and being resistant to the phenomenon of vanishing gradients \cite{pesah2021absence}, QCNNs in our setting also allow to decrease the number of varied parameters to $|\boldsymbol{x}|=2^m$.
        However, as we show in this work, to achieve better results in some cases, one needs to measure all qubits of the $n$-qubit states;
        we discuss one of such cases in Section~\ref{sec:iso}.

    \subsection{Estimation variance}
    \label{sec:estimation_variance}

        One can estimate the label $\alpha$ of a given $\rho_\alpha$ from the expected value of an observable $H$ measured with respect to the state $\rho_\alpha$, i.e. from an estimate $\hat{\mathsf{a}}$ of $\mathsf{a} = \Tr H \rho_\alpha$ the desired estimation of the label is obtained as $\hat{\alpha} = \mathsf{a}^{-1}(\hat{\mathsf{a}})$ \cite{shettell2022cryptographic}.
        Let us now consider the variance of this estimation.
        
        As a simple example, assume that $\rho_\alpha$ is an output state of a single-qubit depolarizing channel acting on an input state $\rho$:
        \begin{equation}
            \label{eq:depol_chan}
            \rho_\alpha
            = (1 - \alpha) \rho + \frac{\alpha}{2} \Id,
        \end{equation}
        where the parameter $\alpha$ can take values from $0$ to $4/3$ \cite{king2003capacity}.
        Let us also assume that the input state is $\rho = (\Id + \sigma_x)/2= \ketbra{+}$, where $\sigma_x$ is a Pauli operator and $\ket{+} = (\ket{0} + \ket{1})/\sqrt{2}$.
        Consider a single-qubit Hermitian observable $H$ decomposed in the Pauli basis:
        \begin{equation*}
            H = \sum_{i=0}^3 h_i \sigma_i, \quad h_i \in \R,
        \end{equation*}
        where $\sigma_0 \equiv \Id$ is the single-qubit identity. 
        One can show that $h_0 = 1$, $ h_1 = -1$, and arbitrary $h_2$ and $h_3$ give the expectation $\langle H \rangle_{\rho_\alpha} \equiv \Tr H \rho_\alpha = \alpha$.
        At the same time, the variance of $H$ in the state $\rho_\alpha$ is
        \begin{align}
            \Delta_{\rho_\alpha}^2 H 
            \equiv& \langle H^2 \rangle_{\rho_\alpha} - \langle H \rangle_{\rho_\alpha}^2 \label{eq:observable_variance} \\
            =& -\alpha^2 + 2\alpha + h_2^2 + h_3^2. \label{eq:depol_chan_obs_var}
        \end{align}
        Thus, even if an observable produces a target expectation, the variance could be substantial.
        This is crucial since the variance $\Delta_{\rho_\alpha}^2 H$ is connected to the mean squared error (MSE) of the estimation $\Delta^2 \hat{\alpha} \equiv \left\langle (\alpha - \hat{\alpha})^2 \right\rangle_{\hat{\alpha}}$ via the error propagation formula \cite{shettell2022quantum, sidhu2020geometric, pezze2018quantum, toth2014quantum}
        \begin{equation}
            \label{eq:estimation_variance}
            \Delta^2 \hat{\alpha} 
            = \frac{\Delta_{\rho_\alpha}^2 H} {\mu \big| \partial_\alpha \langle H \rangle_{\rho_\alpha} \big|^2}
            = \frac{\Delta^2 \hat{\mathsf{a}}}{\big| \partial_\alpha \langle H \rangle_{\rho_\alpha} \big|^2},
        \end{equation}
        where we have used the shorthand $\partial_\alpha \equiv \partial/\partial\alpha$ and $\mu$ is the number of measurements. 
        For the derivation of the second equality, and also for a similar expression for a biased estimator, see Appendix \ref{app:estimation_bias}.

        Hence, the variance should be included in our label prediction approach. We first rewrite the expressions \eqref{eq:estimator} and \eqref{eq:observable_variance}.
        Let us denote the probability of obtaining the outcome $x_i$ after measuring $\Pi$ in the state $\rho_\alpha$ as $p_i = \Tr \Pi_i \rho_\alpha$.
        Since $H$ has the form \eqref{eq:povm_ham}, we get
        \begin{equation}
            \label{eq:expectation_probs}
            \mathsf{a} = \sum_i x_i p_i.
        \end{equation}
        Similarly, for the variance we obtain
        \begin{equation}
            \label{eq:observable_variance_probs}
            \Delta_{\rho_\alpha}^2 H = \sum_i x_i^2 p_i - \left(\sum_i x_i p_i\right)^2.
        \end{equation}
        It is important that both the expectation \eqref{eq:expectation_probs} and the variance \eqref{eq:observable_variance_probs} can be calculated from the same measurement results.
        Indeed, experimentally, one obtains only the probabilities as frequencies $p_i \approx \mu_i/\mu$ with $\mu_i$ being the number of times the outcome $x_i$ was observed, and $\mu$ the total number of measurements (or shots).
        Therefore, with little post-processing overhead we modify the optimization procedure \eqref{eq:ls_min_variational} such that it takes into account the variance of our observable,
        \begin{equation}
        \label{eq:ls_min_variational_mod}
            (\boldsymbol{x}^*, \boldsymbol{\theta}^*) 
            = \arg\min_{\boldsymbol{x}, \boldsymbol{\theta}} \Big( w_\mathrm{ls} f_\mathrm{ls}(\boldsymbol{x}, \boldsymbol{\theta}) + w_\mathrm{var} f_\mathrm{var}(\boldsymbol{x}, \boldsymbol{\theta}) \Big),
        \end{equation}
        where
        \begin{equation}
            f_\mathrm{ls}(\boldsymbol{x}, \boldsymbol{\theta}) = \sum_{j=1}^T \Big(\alpha_j - \mathsf{a}_j\big(\boldsymbol{x}, \Pi(\boldsymbol{\theta})\big) \Big)^2, \label{eq:f_ls}
        \end{equation}
        \begin{equation}
            f_\mathrm{var}(\boldsymbol{x}, \boldsymbol{\theta}) = \sum_{j=1}^T \Delta_{\rho_{\alpha_j}}^2 H(\boldsymbol{x}, \boldsymbol{\theta}) \label{eq:f_var}, 
        \end{equation}
        with $w_\mathrm{ls}, w_\mathrm{var} > 0$ being the weights. We seek to simultaneously minimize the least squares between the given labels and expectations in \eqref{eq:f_ls}, and the variance of the observable in \eqref{eq:f_var}.
        Therefore, a weighted sum of these two expressions forms a cost function which is to be minimized in \eqref{eq:ls_min_variational_mod}.
        In Sections~\ref{sec:ad_chan-res} and~\ref{sec:ising}, we show how by adjusting the weights $w_\mathrm{ls}$ and $w_\mathrm{var}$ one can trade off between a higher estimation accuracy and a lower variance.

    \subsection{Summary and remarks on the method}
        Let us outline the procedure for solving a regression problem with the proposed method, which may essentially be considered a variational quantum algorithm \cite{cerezo2021variational}.
        Having access to a training set $\mathcal{T} = \left\{ \rho_{\alpha_j}, \alpha_j \right\}_{j=1}^{T}$, for each quantum state we compute the expectations $\mathsf{a}_j(\boldsymbol{x}, \boldsymbol{\theta}) \equiv \Tr H(\boldsymbol{x}, \boldsymbol{\theta}) \rho_{\alpha_j}$ of the observable \eqref{eq:povm_ham} or \eqref{eq:povm_ham_m}.
        Similarly, for this observable, we compute the variances $\Delta_{\rho_{\alpha_j}}^2 H(\boldsymbol{x}, \boldsymbol{\theta})$ defined in \eqref{eq:observable_variance_probs}.
        These two quantities and the true labels $\alpha_j$ are then inserted into the cost function in \eqref{eq:ls_min_variational_mod}.
        Having access to this function, a classical minimization algorithm evaluates it in order to iteratively update the parameters $(\boldsymbol{x}, \boldsymbol{\theta})$ towards optimal $(\boldsymbol{x}^*, \boldsymbol{\theta}^*)$, e.g., using a gradient-based optimization algorithm.

        We emphasize again that by measuring an observable $H$ in a state $\rho_\alpha$, one obtains not the exact expectation $\mathsf{a} = \Tr H \rho_\alpha$ itself, but a $\mu$-shot estimation $\hat{\mathsf{a}}$ of it. $\mathsf{a}$ is, generally, a function of $\alpha$ due to a possible bias. 
        After the measurement, the state reduces to one of the eigenstates of the observable.
        In variational quantum computing, such post-measurement state is normally discarded (not reused for further computations). Thus, one needs to be able to prepare the labeled states $\rho_\alpha$, or have a number of their replicas\footnote{To avoid confusion, here we did not use a more common term ``copies'' as it appears further in a slightly different context.}.
        However, in most of numerical experiments presented in this work, we assume that we can perform an arbitrary number of measurement shots $\mu$ on the labeled states $\rho_\alpha$, which results in $\hat{\mathsf{a}} = \mathsf{a}$ in the limit $\mu \to \infty$.
        In other words, it is supposed that one can compute the expectations $\mathsf{a} = \Tr H \rho_\alpha$ exactly.
        Nonetheless, in Appendix~\ref{sec:shots}, we numerically study the performance of the method assuming finite $\mu$.

    It also noteworthy that once an optimal observable $H(\boldsymbol{x}^*, \boldsymbol{\theta}^*)$ is obtained, one can estimate the number of measurements $\mu$ to achieve a given MSE $\Delta^2 \hat{\alpha}$ or $\Delta^2 \hat{\mathsf{a}}$ using formula \eqref{eq:estimation_variance}.
    As common for variational quantum algorithms, the post-measurement labeled states $\rho_\alpha$ are again assumed to be discarded.
    Therefore, we need $\mu$ identical replicas of the states to achieve a desired MSE with a given observable.

    \subsection{Fisher information}
    \label{sec:fisher_info}

        The MSE \eqref{eq:estimation_variance} is known to be lower-bounded by the classical Cramer-Rao bound (cCRB) \cite{sidhu2020geometric}:
        \begin{equation}
            \label{eq:cramer-rao_classical}
            \Delta^2 \hat{\alpha} \geqslant \frac{1}{\mu I_c(\Pi, \rho_\alpha)},
        \end{equation}
        where $I_c(\Pi, \rho_\alpha)$ is the classical Fisher information (cFI) which in our setup can be written as \cite{meyer2021variational} 
        \begin{equation}
            \label{eq:fisher_classical}
            I_c(\Pi, \rho_\alpha) = \sum_i \frac{1}{p_i} \left( \frac{\partial p_i}{\partial \alpha} \right)^2
        \end{equation}
        with $p_i$ depending on $\alpha$ through $\rho_\alpha$.

        Classical Fisher information $I_c(\Pi, \rho_\alpha)$
        is known to be upper-bounded by its quantum counterpart $I_q(\rho_\alpha)$.
        To obtain the latter one introduces the symmetric logarithmic derivative (SLD) operator $L$,
        \begin{equation}
            \label{eq:sld_def}
            \partial_\alpha \rho_\alpha = \frac{1}{2} \big( \rho_\alpha L  + L \rho_\alpha\big),
        \end{equation}
        and calculates the quantum Fisher information (qFI) as
        \begin{equation}
            \label{eq:qfi}
            I_q(\rho_\alpha) = \Tr L^2 \rho_\alpha.
        \end{equation}
        In Appendix~\ref{app:qfi_formulas} we list various ways of finding $L$ and computing $I_q(\rho_\alpha)$.
        
        With qFI, one can obtain a tighter version of the bound \eqref{eq:cramer-rao_classical} called the quantum Cramer-Rao bound (qCRB), namely,
        \begin{equation}
            \label{eq:cramer-rao_quantum}
            \Delta^2 \hat{\alpha} \geqslant \frac{1}{\mu I_c(\Pi, \rho_\alpha)} \geqslant \frac{1}{\mu I_q(\rho_\alpha)}.
        \end{equation}
        Thus, being dependent on both the measurements $\Pi$ and state $\rho_\alpha$ cFI $I_c(\Pi, \rho_\alpha)$ is upper bounded by qFI $I_q(\rho_\alpha)$, which depends solely on the state $\rho_\alpha$.
        One can find the measurement operators such that cFI attains its maximum \cite{liu2020quantum},
        \begin{equation*}
            \max_{\Pi}\, I_c(\Pi, \rho_\alpha) = I_q(\rho_\alpha).
        \end{equation*}
        In order to find such optimal measurements $\Pi$, one could compute the eigenprojectors of the SLD operator $L$ introduced in \eqref{eq:sld_decomp} \cite{sidhu2020geometric}.
        Additionally, one can use $L$ itself for constructing an optimal observable
        \begin{equation}
            \label{eq:H_opt}
            H = \alpha \Id + \frac{1}{I_q(\rho_\alpha)} L
        \end{equation}
        resulting in
        \begin{equation*}
            \langle H \rangle_{\rho_\alpha} = \alpha, \quad \partial_\alpha \langle H \rangle_{\rho_\alpha} = 1, \quad \Delta^2_{\rho_\alpha} H = \frac{1}{I_q(\rho_\alpha)}.
        \end{equation*}
        However, the eigenprojectors of $L$ (i.e. also the optimal measurements $\Pi$) generally depend on the parameter $\alpha$ to be estimated.
        Therefore, it is often impossible to find a POVM such that it saturates qCRB over the entire range of $\alpha$.

        One peculiarity of the theory presented in this Section is that it is devised on the assumption that the parameter to be estimated is unknown but fixed \cite{demkowicz2020multi}.
        Such a setting is often considered together with the maximum likelihood estimator (MLE).
        This estimator is unbiased in the limit of large $\mu$. In this limit any other estimator is no better than MLE \cite{pezze2018quantum}.
        Additionally, in the limit $\mu \to \infty$ MLE saturates the Cramer-Rao bound, which therefore is said to be asymptotically attainable \cite{pezze2014quantum, ChapB15}.
        However, the expressions for CRB \eqref{eq:cramer-rao_quantum} themselves are valid for unbiased estimators (the biased case is discussed later in Section~\ref{sec:func_dep}) and any $\mu$.
        Combined with \eqref{eq:estimation_variance}, it allows to characterize the observable as
        \begin{equation}
            \label{eq:var-qcrb}
            \frac{\Delta_{\rho_\alpha}^2 H} {\big| \partial_\alpha \langle H \rangle_{\rho_\alpha} \big|^2} \geqslant \frac{1}{I_q(\rho_\alpha)},
        \end{equation}
        As we show in Appendix~\ref{app:estimation_bias}, this relation is preserved in the presence of bias as well.
        Therefore, we will treat the Cramer-Rao bounds \eqref{eq:cramer-rao_quantum} as a figure of merit for our method for predicting the label $\alpha$ for a given $\rho_\alpha$.

   \subsection{Bayesian approach}
   \label{sec:theory-bayes}
   
        As we noted above, the estimation theory based on the Fisher information  
        is ``local''.       Therefore it generally produces  a ``local'' optimal observable when using \eqref{eq:H_opt}.
        In practice, the parameter itself can be a random variable with a specific prior probability distribution.
        Then one can apply the Bayesian approach for finding a ``global'' optimal observable.
        In this approach, one minimizes the (Bayesian) MSE for a parameter $\theta \in [a, b]$ obtained by the measurement of an observable $M$~\cite{Personick71, sidhu2020geometric}:
        \begin{equation}\label{eq:bmse}
           \Delta^2_B\theta= \int_a^b \mathrm{Pr}(\theta)\Tr \rho(\theta)\left(M -\theta \Id\right)^2 \,d \theta,
        \end{equation}
        where $\mathrm{Pr}(\theta)$ is the prior probability distribution for $\theta$.
        Note that here we denote the parameter to be estimated and the corresponding observable as $\theta$ and $M$, respectively, for a better distinguishability with the proposed method \eqref{eq:ls_min_variational_mod}. 
        The optimal observable $M_0$ minimizing the MSE~(\ref{eq:bmse}) satisfies
        \begin{equation}\label{eq:opt_ba}
            \frac12\left(\overline{\rho}M_0 + M_0\overline{\rho}\right) = \overline{\theta\rho},
        \end{equation}
        where 
        \begin{equation}
            \overline{\rho} = \int_a^b\mathrm{Pr}(\theta)\rho(\theta)\,d \theta
        \end{equation}
        is a prior-weighted density operator and
        \begin{equation}
            \overline{\theta\rho} = \int_a^b\mathrm{Pr}(\theta)\theta\rho(\theta)\,d \theta
        \end{equation}
        is the prior mean operator~\cite{sidhu2020geometric}. If $\overline{\rho}$ is strictly positive, the solution $M_0$ of~(\ref{eq:opt_ba}) is unique~\cite{Personick71}.

        The optimal observable $M_0$ gives rise to the Bayesian quantum Cramer-Rao bound~\cite{sidhu2020geometric}
        \begin{equation}
            \label{eq:bqcrb}
            \Delta^2_B\theta\geqslant \Delta^2_p\theta - I_B,
        \end{equation}
        where 
        \begin{equation}
            \Delta^2_p\theta = \int_a^b\mathrm{Pr}(\theta)\theta^2\,d\theta
        \end{equation}
        and
        \begin{equation}
            \label{eq:bayes_info}
            I_B = \Tr \overline{\rho}M_0^2,
        \end{equation}
        the \emph{quantum Bayes information}.
        Therefore, one can use \eqref{eq:bqcrb} to assess the quality of the solution obtained, but the optimal observable $M_0$ has to be known.
        Alternatively for this purpose, the van Trees bound can be used \cite{li2018frequentist, demkowicz2020multi, kaubruegger2021quantum}.
        In addition to these bounds, one can characterize the performance of the Bayesian approach by entropic quantities, such as the mutual information between the parameter and the measurement \cite{lecamwasam2024relative}.

        In this work, we also compare the observables found by minimizing the Bayesian MSE \eqref{eq:bmse} to the ones obtained with our method \eqref{eq:ls_min_variational_mod}.
        Importantly, in Section~\ref{sec:bayes-comp} we show that under specific assumptions the method \eqref{eq:ls_min_variational_mod} reduces to the Bayesian MSE \eqref{eq:bmse} if one puts equal weights $w_\mathrm{ls} = w_\mathrm{var}$ and the flat prior $\mathrm{Pr}(\theta) = 1/(b - a)$.

    \subsection{Biased estimator}
    \label{sec:func_dep}

        The expected value of an observable $H$ measured in the state $\rho_\alpha$ allows an estimation of $\alpha$ for a given $\rho_\alpha$. This estimation could also contain a bias $b(\alpha)$, i.e. \cite{sidhu2020geometric}
        \begin{equation*}
            \mathsf{a} \equiv \Tr \rho_\alpha H = \alpha + b(\alpha),
        \end{equation*}
        with the estimation $\hat{\mathsf{a}}$ obtained from a measurement with the MSE $\Delta^2 \hat{\mathsf{a}} = \Delta^2 H / \mu$ \cite{shettell2022quantum, shettell2022delegated, shettell2022cryptographic}. 
        In this biased case, the earlier expressions have to be adjusted.
        Indeed, if $b\neq 0$, instead of the MSE $\Delta^2 \hat{\alpha}$, one would rather observe $\langle (\hat{\mathsf{a}} - \alpha)^2 \rangle$ in an experiment. We can transform the corresponding formulas \eqref{eq:estimation_variance} and \eqref{eq:cramer-rao_classical} to \cite{ChapB15}
        \begin{gather}
            \langle ( \hat{\mathsf{a}} - \alpha)^2 \rangle = \frac{\Delta^2 H}{\mu} + b^2, \label{eq:error_prop-biased} \\
            \langle ( \hat{\mathsf{a}} - \alpha)^2 \rangle \geqslant \frac{|\partial_\alpha \langle H \rangle|^2}{\mu I_q} + b^2. \label{eq:ccrb-biased}
        \end{gather}
        See Appendix~\ref{app:estimation_bias} for a detailed derivation. 
        In Appendix~\ref{sec:shots}, these expressions are used to verify our numerical experiments with finite $\mu$.
        
        We note again that similarly to \eqref{eq:estimation_variance} and \eqref{eq:cramer-rao_quantum}, for an optimized observable $H$ from the expression \eqref{eq:error_prop-biased} one can estimate the number of measurements $\mu$ needed for achieving a desired error of estimating the label $\alpha$ \cite{zhang2024inferring}.
        Indeed, since we engineer the observable $H$ ourselves, we know its variance on the states of the training set, as well as the bias $b$, from which one can also approximate  $\partial_\alpha \langle H \rangle_\alpha  = \partial_\alpha b + 1$ (see Appendix~\ref{app:estimation_bias}).
        
        The bias of estimation could be worked around by inversion of the function $\mathsf{a}$ \cite{alderete2022inference}.
        Indeed, if $\mathsf{a}$ is bijective within a range of $\alpha$, then one gets the desired estimation as $\hat{\alpha} = \mathsf{a}^{-1}(\hat{\mathsf{a}})$.
        Alternatively, one could try to compensate the bias $b(\alpha)$, as one extracts it from the difference between the prediction $\mathsf{a}$ and the true labels $\alpha$ given in the training set $\mathcal{T}$ (see Appendix~\ref{app:bias_comp}).

        Another strategy of sorting out an unwanted bias is to  modify the prediction itself,
        \begin{equation}
            \label{eq:expectation-c}
            \mathsf{a}(\boldsymbol{x}, \boldsymbol{\theta}, c, \rho_\alpha) = \Tr H(\boldsymbol{x}, \boldsymbol{\theta}) \rho_\alpha^{\otimes c},
        \end{equation}
        i.e. we process $c$ copies of the $n$-qubit state $\rho_\alpha$ \textit{simultaneously} as one joint state $\rho_\alpha^{\otimes c}$, and so that $H$ is now a $cn$-qubit observable.
        In line with Sec.~\ref{sec:m-qubits}, calculating the quantity \eqref{eq:expectation-c} can be schematically depicted as follows:
        \begin{equation*}
            \qquad\quad
            \begin{rcases}       
                \Qcircuit @C=1em @R=0.25em{
                    & \rho_\alpha & & \qw/_n & \multigate{3}{U(\boldsymbol{\theta})} & \qw & \meter \\
                    & \rho_\alpha & & \qw/_n &        \ghost{U(\boldsymbol{\theta})} & \qw & \meter \\
                    &      \vdots & &        &       \nghost{U(\boldsymbol{\theta})} &     &        \\
                    & \rho_\alpha & & \qw/_n &        \ghost{U(\boldsymbol{\theta})} & \qw & \meter 
                    \inputgroupv{1}{4}{.25em}{1.75em}{\rho_\alpha^{\otimes c}}
                } 
            \end{rcases}
            \overset{\,p_i}{\underset{}{\longrightarrow}} i \mapsto x_i         
        \end{equation*}
        That is, here we seek to find an observable $H$ acting on the the composite state $\rho_\alpha^{\otimes c}$, which is now treated as a \textit{one} labeled state with the label $\alpha$.
        This introduces a non-linearity to the prediction \cite{wu2021expressivity, holmes2023nonlinear}, 
        and similar techniques are utilized in QML to improve its performance \cite{schuld2020circuit, goto2021universal, banchi2021generalization, jerbi2023quantum, banchi2023statistical}.
        In Appendix \ref{app:unitary_chan} we apply this method to a problem of predicting $\alpha$ for $\rho_\alpha = e^{-i \alpha \sigma_z /2} \ketbra{+} e^{i \alpha \sigma_z /2}$, in which case one cannot find $H$ giving the expectation $\Tr H \rho_\alpha = \alpha$.
        However, we show that for any $c \geqslant 1$ there is an observable $H_c$ such that the expectation $\Tr H_c \rho_\alpha^{\otimes c}$ gives the first $c$ terms of the Fourier series of the linear function $f(\alpha) = \alpha$.

\section{Relation to the literature}
\label{sec:literature}

    There is a vast amount of literature dedicated to QML and its various aspects.
    One of such aspects is the capability of QML-models of generalization, i.e., the performance of trained models on unseen data.
    The ability of a model to generalize has been studied with respect to, e.g., the size of the training set \cite{huang2021power, banchi2023statistical} or number of accesses to the process preparing a data point \cite{huang2021information}, number of parametrized gates in the variational circuit \cite{caro2022generalization}, data encoding strategy \cite{caro2021encoding}, 
    and classical and quantum information-theoretic quantities measuring correlation between training data and learner’s hypothesis \cite{banchi2021generalization, caro2024information}.
    In our work, we do not study this aspect of our method analytically.
    However, we observe that the number of states for training depends on the connection between the labeled state and its label.
    Among the considered problems only entanglement quantification requires a substantial training set size which we study numerically.

    Another aspect of QML considered in the literature is the construction of models.
    QML models can be based on shadow tomography \cite{aaronson2018shadow, buadescu2021improved}.
    This approach can be used for predicting the expectation of an exponential number of arbitrary two-outcome observables in state $\rho$ by \textit{simultaneously} measuring a polynomial number of copies of $\rho$.
    Based on shadow tomography and other clever measurement and post-processing techniques, the work \cite{huang2021information} proposes an algorithm such that given $n$ copies of an $n$-qubit state $\rho$, it efficiently predicts expectations $\Tr P \rho$ for $\textit{all}$ Pauli strings $P$.
    Moreover, as shown in \cite{huang2022quantum}, conducting collective measurements on only two copies of state allows classification of observables and quantum processes.
    Regarding the latter, the work \cite{fanizza2024learning} describes, inter alia, an algorithm which learns a quantum process from collective measurements on not necessarily identical state copies.

    Shadow tomography and related techniques require to measure many copies of the state as a whole, which demands for a quantum memory to store the copies.
    To circumvent this obstacle, the work \cite{huang2020predicting} introduced the notion of classical shadows.
    A classical shadow is essentially a collection of results of randomized measurements conducted for a given state.
    Similarly to the shadow tomography, this method also allows to predict the expectations of observables, but the copies of a state can be measured \textit{independently}.
    However, the number of copies needed for accurate prediction depends on the so-called shadow norm of the observables in question as well as on the way random measurements are performed.
    For QML, classical shadows were utilized in \cite{zhao2024learning} for learning states and unitaries.
    Closer to our setting is one of the algorithms described in \cite{huang2023learning}, where an unknown observable $O$ is learned assuming a training set of the form $\{\rho_{\alpha_j}, \Tr O \rho_{\alpha_j}\}_j$.
    This may not hold for some cases, for instance, if the labels $\alpha_j$ enter the states $\rho_{\alpha_j}$ only non-linearly. Nonetheless, this algorithm should still be applicable for labels $\alpha$ with arbitrary connection to $\rho_\alpha$.

    The notion of classical shadows gives rise to the concept \textit{``measure first, ask questions later''} \cite{elben2023randomized}.
    Indeed, a classical shadow is essentially a classical representation of a quantum state.
    Once obtained, it can be utilized for various purposes, including solving machine learning tasks.
    For example, classical shadows can be fed into a classical neural network which learns to classify phases of matter \cite{huang2022provably}, even in the unsupervised setting \cite{huang2022quantum, elben2023randomized}.
    More generally, for classical data, one can train a so-called shadow model \cite{jerbi2024shadows} by using the classical shadows technique for training a quantum device to produce bit-string ``advices'' for a classical algorithm. The latter is then used for evaluating the labels without the need for the quantum device.
    Although such methods are appealing as they minimize the use of quantum resources, they also have their limits of applicability \cite{gyurik2023limitations, jerbi2024shadows}.

    As in the present work, QML models can also be designed in a different setting, the setting of variational quantum computing \cite{cerezo2021variational, cerezo2022challenges}.
    This approach is believed to be suitable for noisy intermediate scale quantum (NISQ) devices.
    Although we also consider processing several copies of a labeled state, doing so for many copies (as required for shadow tomography and related methods) may be prohibitive for such devices \cite{gebhart2023learning}.
    As for classical shadows, one could try to perform random Clifford measurements for estimating the expectations of \eqref{eq:povm_ham_m}.
    In our setting, however, estimating this expectation in a given labeled state with a desired precision $\epsilon$ would require about $m 2^{n-m} \max_i (x_i^2/\epsilon^2)$ measurements (see Theorem 1 in \cite{huang2020predicting}).
    Therefore, the larger $m$ is, the fewer measurements we need to conduct, but the more parameters we need to vary.
    Here we do not apply this technique and leave it for future considerations.

    Within the variational quantum computing approach the work \cite{nagano2023quantum} considers quantum simulation of multiparticle states governed by a specific Lagrangian aiming at prediction of one of its parameters. The authors apply a QCNN to transform the states and measure a single-qubit observable $\sigma_z$, the expectation of which is used as a prediction for the parameter.
    We consider a similar problem of predicting the strength of the transverse field $h$ of the Ising Hamiltonian $H_h$ given its ground state $\ket{\psi_h}$.
    One difference between our work and \cite{nagano2023quantum} is that we accompany our numerical experiments with the simulation of finite number of measurement shots (see Appendix~\ref{sec:shots}), and compare the variance of the optimized observable against qCRB.

    A QCNN was also utilized in \cite{alderete2022inference} to approximate a $\mu$-shot estimate $\hat{\mathsf{a}}$ of the expectation $\mathsf{a} = \Tr \rho_\alpha O$ by a trigonometric polynomial $f(\alpha)$ and obtain the desired parameter as its inverse.
    The authors prove that the approach works well (in terms of required $\mu$) for any non-trivial observable $O$ and certain dependencies of $\rho_\alpha$ and $\mathsf{a}$ on $\alpha$.
    In one of the numerical experiments, they considered finding an optimal observable for predicting the angle $\alpha$ of a unitary rotation of a state of $n$ qubits.
    Similarly to \cite{nagano2023quantum}, the observable had the form  of a single-qubit operator $\sigma_z$ transformed by a QCNN.
    The MSE given by the optimized observable scales as $1/n^2$, i.e. it follows the ultimate Heisenberg limit \cite{pirandola2018advances}.
    To optimize the observable, the authors minimized the MSE between $f/n$ and $\alpha$, where the factor $1/n$ was introduced for achieving the Heisenberg limit. 

    Our approach of optimizing the observable is different, as we account for its variance (linked to the MSE of estimation via \eqref{eq:estimation_variance} or \eqref{eq:error_prop-biased}) through explicit addition of it to the cost function. 
    Additionally, we test our method on a variety of tasks, including entanglement quantification.
    Another difference with \cite{nagano2023quantum} and \cite{alderete2022inference} is in our observable being many-qubit. This may be important as we notice that measuring one qubit out of many may result in higher prediction variance. We emphasize that what matters here is apparently not the observable's locality itself, but the dimensionality of its eigenprojectors, see Sections~\ref{sec:iso} and~\ref{sec:ent-pure}.
    Additionally, we can vary the spectrum of our observable which brings more flexibility.

    The idea of parametrizing the observable's spectrum was recently proposed in \cite{quantumKrepRoth2024}. 
    In the reference, classical data $\alpha$ are encoded into quantum states via a parametrized unitary $V(\alpha)$. 
    This allows to utilize the data re-uploading techniques \cite{perez2020data, perez2021one, schuld2021effect}, i.e. one can alternate between the encoding unitary $V(\alpha)$ and variational circuits $U(\boldsymbol{\theta}_i)$ for achieving higher prediction accuracy.
    In contrast to \cite{quantumKrepRoth2024}, in our regression tasks we consider data which are initially quantum.
    For such data, the re-uploading techniques are not applicable, unless we have access to the process $\Phi_\alpha$ taking as input an initial state $\rho$ and producing the labeled state $\Phi_\alpha[\rho]$ for a given label $\alpha$. 
    However, we consider processing $c$ copies of the labeled state as one joint state (i.e., $\rho_\alpha^{\otimes c}$) which may introduce a desired non-linearity \cite{wu2021expressivity, banchi2023statistical}.

    Also in \cite{quantumKrepRoth2024} it is proposed to parametrize the observable in the Pauli basis with variable coefficients. 
    In our approach, we represent the observable as a linear combination of eigenprojectors with parameters being the eigenvalues.
    Although the two techniques are similar, ours is
    more compatible with the notion of SLD operators \eqref{eq:sld_def}, as well as
    more transparent in application to Naimark's extension \eqref{eq:naimark-variational} which readily translates into working with general POVM elements.
    Naimark's extension may play an important role if one wants to measure fewer qubits with no harm to the prediction variance (see Section~\ref{sec:iso}). 
    Additionally, in contrast to the observable parametrization proposed in \cite{quantumKrepRoth2024}, our parametrization \textit{always} allows to compute $\Tr H^2 \rho_\alpha$ by only a classical post-processing of the measurement results obtained for computing $\Tr H \rho_\alpha$.

    We also provide analytical results for the proposed method.
    Assuming a large size of the training set, we write an equation for finding an optimal observable and derive upper bounds on its variance.
    Interestingly, the obtained optimal observable cannot give unbiased predictions on the whole range of the labels.
    Importantly, we also show that with equal weights in \eqref{eq:ls_min_variational_mod}, our method is reducible to minimizing the Bayesian MSE, an approach for finding a global observable.
    For this approach, variational quantum algorithms have also been developed \cite{kaubruegger2021quantum, kaubruegger2023optimal, thurtell2024optimizing}, as well as executed experimentally \cite{marciniak2022optimal}.

    In both our work and \cite{quantumKrepRoth2024}, the variance of the observable is taken into account.
    We propose the assessment of the quality of the solution by comparing the variance against the Cramer-Rao bounds.
    The idea comes from quantum metrology \cite{toth2014quantum, pirandola2018advances, pezze2018quantum}. 
    One is given a usually \textit{known}, but possibly noised parametrized channel $\Phi_\alpha$ with an aim to find an optimal input state $\rho$ and measurements for the output state $\Phi_\alpha[\rho]$ allowing a more accurate prediction of the parameter $\alpha$.
    For this setting, there exist variational algorithms as well \cite{meyer2021variational, koczor2020variational}.
    One typically varies the input states and measurements to maximize the Fisher information about the \textit{fixed} parameter $\alpha$.
    The Fisher information under maximization may be classical or quantum, or it may be an easier-to-compute surrogate of the latter \cite{cerezo2021sub}.
    Our setting is slightly different, as the parameter $\alpha$ may take values from a fairly wide interval.
    Additionally, we assume no knowledge about the channel $\Phi_\alpha$, nor do we control the input states $\rho$ for it.

\section{Theoretical arguments}
\label{sec:theory}

    In this Section, we analyze the method introduced in \eqref{eq:ls_min_variational_mod}.
    Assuming certain approximations, we first obtain an equation for an optimal observable $H$.
    Then, we derive upper bounds on the total variance of the parameter prediction.
    Finally, we show that for the condition $w_\mathrm{ls} = w_\mathrm{var}$ and uniform distribution of the labels, our method is equivalent to the minimization of the Bayesian MSE.

    \subsection{Approximation for the optimal observable}
    
        When a training set comprises some fixed family of states $\{\rho_\alpha\}_\alpha$, the solution of the optimization procedure~(\ref{eq:ls_min_variational_mod}) can be approximated by the solution of the minimization problem
        \begin{multline}
            \label{eq:min_cost_appx}
                \min_H\:  w_\mathrm{ls} \int_{a}^{b} \left( \Tr \rho_\alpha H - \alpha\right)^2 \,d \alpha  
                \\ 
                +
                w_\mathrm{var} \int_{a}^{b}\big(\Tr \rho_\alpha H^2-(\Tr\rho_\alpha H)^2\big) \,d \alpha  
        \end{multline}
        with fixed non-negative weights $w_\mathrm{ls}, w_\mathrm{var}$. Here it is supposed that finite sums in Eqs.~(\ref{eq:f_ls}) and~(\ref{eq:f_var}) can be approximated by integrals over the range $[a, b]$ of possible values of the predicted parameter $\alpha$ provided that the distribution of the training sample's labels is dense enough.

        The optimization problem~(\ref{eq:min_cost_appx}) can be treated analytically with the use of the method which was  demonstrated in the context of  quantum state discrimination in Chapter 2 of Ref.~\cite{Holevo_2012}. Considering a small perturbation $\epsilon$ around the optimal observable $H_0$ by an arbitrary Hermitian operator $Y$, $H=H_0+ \epsilon Y$, we obtain that $H^2=H_0^2+ \epsilon (H_0 Y+Y H_0) + \epsilon^2 Y^2$.  Substituting expressions for $H$ and $H^2$ into (\ref{eq:min_cost_appx}) and setting the contribution in front of  $\epsilon$ to zero yields
        \begin{multline}
        \label{eq:epsilon_zero}    
            \wv\int_a^b\Tr\rho_\alpha (H_0Y+Y H_0)\,d\alpha \\ + 2(\wl-\wv) \int_a^b \Tr(\rho_\alpha Y)\Tr (H_0\rho_\alpha) \,d\alpha \\
            - 2 \wl \int_a^b \alpha \Tr(\rho_\alpha Y) \,d\alpha = 0.
        \end{multline}

        The contribution in front of $\epsilon^2$ reads
        \begin{multline}\label{eq:eps2}
            \wl\int_a^b (\Tr \rho_\alpha Y)^2\,d\alpha \\
            + \wv\int_a^b\left(\Tr \rho_\alpha Y^2-(\Tr\rho_\alpha Y)^2\right) \,d \alpha.
        \end{multline}
        This expression is non-negative since 
        $(\Tr\rho_\alpha Y)^2 \leqslant \Tr \rho_\alpha Y^2$ as a particular case of the inequality~\cite{HayQI17}
        \begin{equation}
        \label{eq:trconv}
            f(\Tr\rho X)\leqslant\Tr\rho f(X),
        \end{equation}
        which holds for a convex function $f$, a Hermitian operator $X$, and a state $\rho$.
        Non-negativity of the expression (\ref{eq:eps2}) implies that the operator $H_0$, which satisfies (\ref{eq:epsilon_zero}) for any $Y$, is the solution  to the minimization problem~(\ref{eq:min_cost_appx}). 
        Making use of the fact that (\ref{eq:epsilon_zero}) holds for any Hermitian $Y$, inside the trace we can gather all the terms that are  multiplied by $Y$ and set their sum to zero.
        We arrive at an operator equation for the optimal observable $H_0$:
        \begin{multline}\label{eq:opt_obs}
            \frac12\left(\Tilde{\rho} H_0+H_0\Tilde{\rho}\right) - \frac{k}{L}\int_a^b \alpha \rho_\alpha \,d\alpha \\  + \frac{(k-1)}{L} \int_a^b \Tr (H_0\rho_\alpha) \rho_\alpha\,d\alpha 
             = 0,
        \end{multline}
        where
        \begin{equation}
        \label{eq:rhotil}
            \Tilde{\rho} = \frac1{L}\int_a^b \rho_\alpha \,d\alpha, \quad 
            L = b - a, \quad 
            k=\frac{\wl}{\wv},
        \end{equation}
        with dependence on weights being isolated in the factor $k$, and hence only the ratio of weights matters.

        The structure of (\ref{eq:opt_obs}) can be used to get the following general property of the optimal observable $H_0$. Applying the trace to both sides and simplifying the result one gets
        \begin{equation}\label{eq:estsquare}
            \int_a^b \Tr (H_0\rho_\alpha) \,d\alpha = \int_a^b \alpha  \,d\alpha = \frac{b^2 - a^2}{2},
        \end{equation}
        i.e. irrespective of weights the area under the prediction curve equals to the area under the straight line corresponding to the true values of the label. 
        The manifestation of this property   can be  most clearly seen in Fig.~\ref{fig:ising-w_var}~(left) and Fig.~\ref{fig:z-rot}~(left).

    \subsection{Upper bound on the total variance} 
    \label{sec:theor-var}
        
        When the ratio of weights $k\leqslant 1$, Eq.~(\ref{eq:opt_obs}) also implies an upper estimate on the total variance of the optimal observable $H_0$.
        In Appredix~\ref{app:theor-var}, we derive the following bound:
        \begin{equation}
            \label{eq:totvar_bound}
            \int_a^b\Delta^2H_0\,d\alpha\,\leqslant\,k\left(\frac{b^3 - a^3}{3} - \frac{(b^2 - a^2)^2}{4L}\right).
        \end{equation}
        This expression shows that the total variance can be made arbitrarily small by choosing a sufficiently small weight ratio $k$. Assuming $a=0$ and $b = L$, for the \emph{average} total variance one obtains
        \begin{equation}\label{eq:totvarprac}
           \frac{1}{L}\int_0^L\Delta^2H_0\,d\alpha\,\leqslant\,\frac{kL^2}{12},
        \end{equation}
        i.e. the average total variance for this particular case can be made less than $\epsilon$ by selecting the weight ratio as small as $k\leqslant \min(1,\,12\epsilon/L^2)$.

        Eqs.~(\ref{eq:totvar_bound}) and~(\ref{eq:totvarprac}) point to the fact that the procedure~(\ref{eq:ls_min_variational_mod}) can be regarded as the \emph{variance regularization}, the term used in Ref.~\cite{quantumKrepRoth2024}. 

        We note that the total variance of the optimal observable \eqref{eq:totvar_bound} for \emph{all finite} $k$ can be expressed as (see Appendix~\ref{app:theor-var})
        \begin{multline} \label{eq:totvar_bound-2}
        \int_a^b\Delta^2H_0\,d\alpha\\
        = k\int_a^b\left(\alpha \Tr\rho_\alpha H_0 -(\Tr\rho_\alpha H_0)^2\right)\,d\alpha.
        \end{multline}
        This expression shows that at finite $k$ the prediction $\Tr\rho_\alpha H_0$ will never coincide exactly with the true values $\alpha$ at each point of the domain $[a, b]$. 
        Indeed, let us suppose that there is no bias at each point $\alpha$.
        Then, according to \eqref{eq:totvar_bound-2}, the total variance and hence the variance for all $\alpha$ is equal to zero, which implies an observable $H_0$ giving a constant prediction $\langle H_0 \rangle_{\rho_\alpha} = c$.
        This, in turn, leads to a bias $b = c - \alpha $, which contradicts the initial assumption that $\Tr\rho_\alpha H_0 = \alpha$ for any $\alpha \in [a, b]$.
        Therefore, in the prediction of the parameter according to the the procedure~(\ref{eq:ls_min_variational_mod}), there will always be a bias with finite $k$.

   \subsection{Comparison with the Bayesian approach}
   \label{sec:bayes-comp}
   
        At equal weights $\wl = \wv = 1$, the sum of the two terms in (\ref{eq:min_cost_appx}) simplifies, and the optimization problem reduces to
        \begin{equation}\label{eq:min_eq_w}
           \min_H\:   \int_{a}^{b}  \Tr \rho_\alpha \left(H -\alpha \Id\right)^2 \,d \alpha. 
        \end{equation}
        This is a minimization of MSE of the observable $H$ over the interval $[a, b],$ which can be compared with the optimization task from the Bayesian approach introduced in Section~\ref{sec:theory-bayes}.
        For more instructive comparison, for the Bayesian approach we denote the parameter under prediction as $\theta$ and the corresponding observable as $M$. 
        Recall that in this approach, the parameter $\theta$ is a random variable which follows a prior probability distribution $\mathrm{Pr}(\theta)$.
        The Bayesian~MSE for $\theta$, which is obtained by the measurement of an observable $M$, reads
        \begin{equation}\label{eq:bmse-comp}
           \Delta^2_B\theta= \int_a^b \mathrm{Pr}(\theta)\Tr \rho(\theta)\left(M -\theta \Id\right)^2 \,d \theta.
        \end{equation}
        From the last two expressions, it is seen that the minimization task~(\ref{eq:min_eq_w}) is equivalent to the minimization of~(\ref{eq:bmse-comp}) if the prior is uniform~(or \emph{flat}), i.e., $\mathrm{Pr}(\theta)=1/(b-a)$. 
        Consequently, the optimization problem~(\ref{eq:min_cost_appx}) with equal weights~($w_\mathrm{ls} = \wv$) is equivalent to minimization of the Bayesian MSE with the flat prior.

        The  correspondence stated above can also be seen at the level of the solutions to the optimization problems. 
        Setting $k = 1$ (equal weights) in (\ref{eq:opt_obs}), one obtains
        \begin{equation}\label{eq:opt_fi}
            \frac12\left(\Tilde{\rho} H_0+H_0\Tilde{\rho}\right) = \widetilde{\alpha\rho},
        \end{equation}
        where 
        \begin{equation}
           \widetilde{\alpha\rho} =\frac{1}{L}\int_a^b \alpha \rho_\alpha \,d\alpha.
        \end{equation}
        The equation (\ref{eq:opt_fi}) for the optimal observable $H_0$ resembles the equation \eqref{eq:opt_ba} for the optimal observable $M_0$ minimizing Bayesian MSE~(\ref{eq:bmse}).
        When the prior $\mathrm{Pr}(\theta)$ is uniform, (\ref{eq:opt_ba}) exactly coincides with (\ref{eq:opt_fi}).
        At the same time, the solution of the problem~(\ref{eq:min_cost_appx}) with non-equal weights can differ substantially from the Bayesian solution for flat prior in terms of approximation error and variance (see  Section~\ref{sec:ad_chan-res}).

\section{Results}
\label{sec:results}

    In this Section, we present the results of numerical experiments of the regression problem solution for the data of the form \eqref{eq:training_set} with our method, i.e., we solve the optimization problem \eqref{eq:ls_min_variational_mod}, minimizing the least squares \eqref{eq:f_ls} between the true parameters $\alpha$ and our predictions $\mathsf{a}$ as well as the variance of the observable \eqref{eq:f_var} simultaneously.
    That is, we show the results of applying our method to predict the parameter of a channel $\Phi_\alpha$ given its output state $\rho_\alpha$, the entanglement of various families of states, and the parameter of a Hamiltonian $H_\alpha$ given its ground state $\ket{\psi_\alpha}$. 
    In our numerical experiments, we assume that we can calculate the expectations $\mathsf{a} = \Tr H \rho_\alpha$ exactly. However, in Appendix~\ref{sec:shots}, we also perform simulations with $\mathsf{a}$ obtained as $\mu$-shot estimates $\hat{\mathsf{a}}$.
    
    Where applicable, the classical and quantum Fisher informations were calculated using formulas \eqref{eq:fisher_classical} and \eqref{eq:qfi-fidelity}, respectively.
    If not stated otherwise, we use a hardware-efficient ansatz described in Appendix~\ref{app:hea} to represent the parametrized unitary $U(\boldsymbol{\theta})$, and set $w_\mathrm{ls} = 1$ and $w_\mathrm{var} = 10^{-4}$.
    For optimization, the BFGS algorithm from the SciPy library \cite{2020SciPy-NMeth} is applied.
    Additionally, we use tools from the QuTiP package \cite{johansson2012qutip, JOHANSSON20131234}.

    \subsection{Predicting the parameter of the amplitude-damping channel}
    \label{sec:ad_chan-res}

        \begin{figure*}
            \centering
            \includegraphics[width=.495\textwidth]{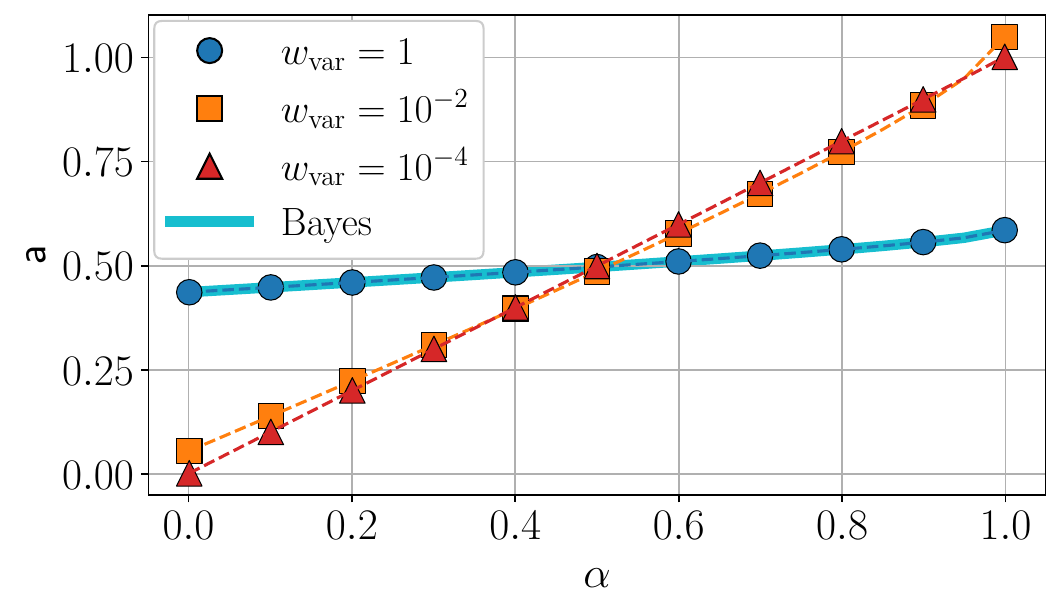}
            \includegraphics[width=.495\textwidth]{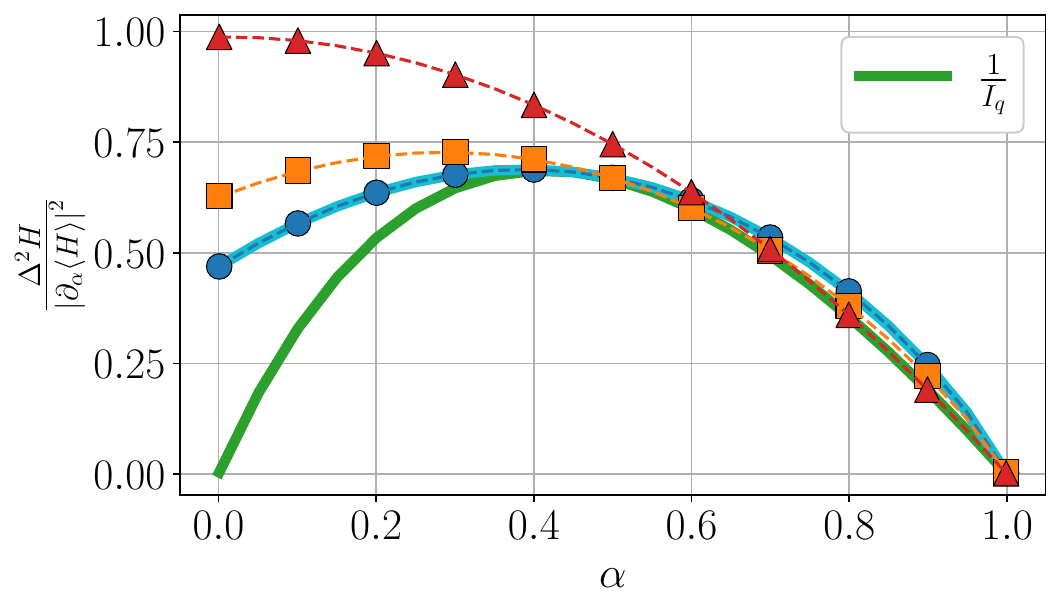}
            \caption{
                Left: Predicted $\mathsf{a}$ vs. true $\alpha$ parameter of the AD channel \eqref{eq:ad_chan} for fixed $w_\mathrm{ls}=1$ different weights $w_\mathrm{var}$.
                Right: Variance \eqref{eq:var-qcrb} of the optimized observable $H$ vs. true $\alpha$.
                In both panels, the solid blue line shows the data produced by the observable obtained by minimizing the Bayesian MSE \eqref{eq:bayes_min}; the dashed lines indicate the data given by the observable obtained from the solution of \eqref{eq:sol_ad}.
            }
            \label{fig:ad_chan-bayes}
        \end{figure*}
    
        As the first test case, let us consider the amplitude damping (AD) channel
        \begin{equation}
            \label{eq:ad_chan}
            \Phi_\alpha[\rho] = \sum_{k=1}^2 V_k(\alpha) \rho V_k^\dagger(\alpha)
        \end{equation}
        with Kraus operators
        \begin{equation*}
            V_1(\alpha) = \sqrt{\alpha} \ketbra{0}{1}, \quad V_2(\alpha) = \ketbra{0} + \sqrt{1 - \alpha} \ketbra{1}.
        \end{equation*}
            The input state $\rho=\ketbra{+}$ is used to obtain the output
        \begin{equation}
            \rho_\alpha \equiv \Phi_\alpha[\rho]=\frac{1}{2} (\Id + \sqrt{1-\alpha} \sigma_x +\alpha\sigma_z)
        \end{equation}
        As we show in Appendix~\ref{app:ad-chan}, the SLD operator for this state has eigenvectors dependent on the parameter $\alpha$.
        Hence, there may be no option to find an optimal observable from the construction~(\ref{eq:H_opt}), as such an observable will depend on $\alpha$ and hence will not be global.

        In order to find the global optimal observable, we need to solve the equation~(\ref{eq:opt_obs}) directly. After several transformations, one gets
        \begin{multline}
            \label{eq:ad_main_eq}
            \frac{1}{2}(\Tilde{\rho}H_0+H_0\Tilde{\rho})+\frac{(k-1)}{L}\int_a^b \Tr(\rho_\alpha H_0)\rho_\alpha \,d \alpha \\
            -\frac{k}{2L} \big(c_2 \Id +(c_1+c_3)\sigma_x+c_4\sigma_z\big)=0,
        \end{multline}
        where
        \begin{equation}
            \Tilde{\rho} = \frac{1}{2L} ( L + c_1 \sigma_x +c_2 \sigma_z)
        \end{equation}
        and
        \begin{align*}
                &c_1=\frac23\big((1-a)^{3/2}-(1-b)^{3/2}\big),\\
                &c_3=\frac{2}{5}\big((1-b)^{5/2}-(1-a)^{5/2}\big),  \\
                &c_2 = \frac{b^2 - a^2}{2},\quad c_4=\frac{b^3-a^3}{3}.
        \end{align*}
        We  look for the observable in the form 
        \begin{equation}
            \label{eq:H0_ad}
            H_0 = \frac{1}{2}(h_0 \Id + h_1 \sigma_x +h_2 \sigma_y+h_3\sigma_z).
        \end{equation}
        Substituting it into~(\ref{eq:ad_main_eq}), after simple but tedious calculations with Pauli matrices we obtain $h_2 = 0$ and the following linear system of equations  
        \begin{equation} \label{eq:sol_ad}
            \begin{cases}
                L h_0 + c_1 h_1 +c_2 h_3=2 c_2, \\
                k c_1 h_0 + \big(L+(k-1)(L-c_2)\big)h_1 \\
                \qquad\qquad\quad + (k-1)(c_1+c_3)h_3 = 2 k (c_1+c_3), \\
                k c_2 h_0 + (k-1)(c_1+c_3)h_1 \\
                \qquad\qquad\quad + \big(L + (k-1)c_4\big)h_3 = 2 k c_4. 
            \end{cases} 
        \end{equation}
        Solving~(\ref{eq:sol_ad}) for $h_0, h_1, h_3$, and for given values of $a$ and $b$, we can construct our global optimal observable $H_0$ from (\ref{eq:H0_ad}).

        In Fig.~\ref{fig:ad_chan-bayes}, we compare the predictions $\mathsf{a} = \Tr H(\boldsymbol{x}^*, \boldsymbol{\theta}^*) \rho_\alpha$ of the parameter $\alpha$ of the AD channel for the observables $H(\boldsymbol{x}^*, \boldsymbol{\theta}^*)$ obtained with the two optimization procedures: 
        Via our method (\ref{eq:ls_min_variational_mod}) with $w_\mathrm{ls} = 1$ and three different weights $w_\mathrm{var}$, and via the Bayesian approach \eqref{eq:bmse} with uniform prior. 
        Namely, in the former, we trained our model on a set $\mathcal{T} = \{\rho_{\alpha_j}, \alpha_j\}_{j=1}^{500}$ of (deliberately) excessively large size $T=500$ with equidistant labels $\alpha_j$.
        In the latter, we minimize the Bayesian MSE:
       \begin{equation}
        \label{eq:bayes_min}
            (\boldsymbol{x}^*, \boldsymbol{\theta}^*) = \arg\min_{\boldsymbol{x}, \boldsymbol{\theta}} \int_0^1 \Tr \rho_\alpha\big(H(\boldsymbol{x}, \boldsymbol{\theta}) -\alpha \Id\big)^2 \,d \alpha.
        \end{equation}
        In order to execute this procedure, one needs an access to the process which prepares $\rho_\alpha$ for a given $\alpha$. This may be impossible if, e.g. $\alpha$ is an entanglement measure of a random state $\rho_\alpha$.
        However, the integral in \eqref{eq:bayes_min} can be approximated by a sum for discrete values of $\alpha$ present in the training set \cite{marciniak2022optimal}.

        Fig.~\ref{fig:ad_chan-bayes} shows that our procedure with equal weights $w_\mathrm{ls} = w_\mathrm{bar}$ coincides with the Bayesian one with the flat prior. 
        With no preliminary knowledge about the parameter, the Bayesian method produces a rather large prediction error in exchange for smaller variance. 
        The former can be decreased by replacing the Bayesian method with the procedure~(\ref{eq:ls_min_variational_mod}) and by choosing smaller weights $w_{\mathrm{var}}$, again, in exchange for larger variance.

        In Fig.~\ref{fig:ad_chan-bayes} the obtained results are also compared with theoretical predictions according to (\ref{eq:opt_obs}), (\ref{eq:H0_ad}), and~(\ref{eq:sol_ad}).
        As we see, the solutions to (\ref{eq:sol_ad}) match well the numerical results of application of our method \eqref{eq:ls_min_variational_mod} and minimizing the Bayesian MSE \eqref{eq:bayes_min}.
        Furthermore, with the solution to (\ref{eq:sol_ad}) for $k=1$, one can calculate the Bayesian qCRB \eqref{eq:bqcrb} to obtain 
        \begin{equation*}
            \Delta^2_B\alpha = \Delta^2_p\alpha - I_B = \frac{178}{2475} \approx 0.0719
        \end{equation*}
        This value of Bayesian MSE was achieved by the variational procedure \eqref{eq:bayes_min} in our numerical results, signifying the saturation of the bound.
    
        In order to numerically test our analytical expressions obtained in Section~\ref{sec:theory}, here we trained our model \eqref{eq:ls_min_variational_mod} on a set of a large size $T=500$.
        However, as we show in Appendix~\ref{app:ad-chan}, one can get good results having a training set of a much smaller size of $T=5$ data points.

        Additionally for this regression problem, in Appendix~\ref{sec:ad_shots} we train the model assuming that the expectations $\mathsf{a} = \Tr H\rho_\alpha$ are obtained as $\mu$-shot estimates $\hat{\mathsf{a}}$.
        As this problem is relatively simple for our method, even $\mu=2^6$ is enough to achieve good prediction accuracy.

    \subsection{Predicting the entanglement of two-qubit states}
    \label{sec:ent}

        Here we test our method for entanglement quantification~\cite{QuantifyEnt97,  HorEntRev09, GUHNE20091, ZiewEltRevEnt14, roik2022entanglement, wang2022detecting} for various families of two-qubit states.
        Namely, we train our model to predict the negativity $N$ of a given two-qubit state $\rho$,\footnote{We dropped an additional factor $1/2$ which often appears in the literature. All related expressions have been modified accordingly.}
        \begin{equation}
            \label{eq:neg}
            N(\rho) = \norm{\rho^{T_2}}_1 - 1,
        \end{equation}
        where $\norm{\,\cdot\,}_1$ is the trace norm, and $\rho^{T_2} \equiv (\Id \otimes T)[\rho]$ is the partial transpose of $\rho$ with respect to the second qubit.
        For more details on the negativity and bipartite entanglement overall, see Appendix~\ref{app:negativity}.

        \subsubsection{Bell-type states}
        \label{sec:bell_ent}        
    
            \begin{figure*}
                \centering
                \includegraphics[width=.48\textwidth]{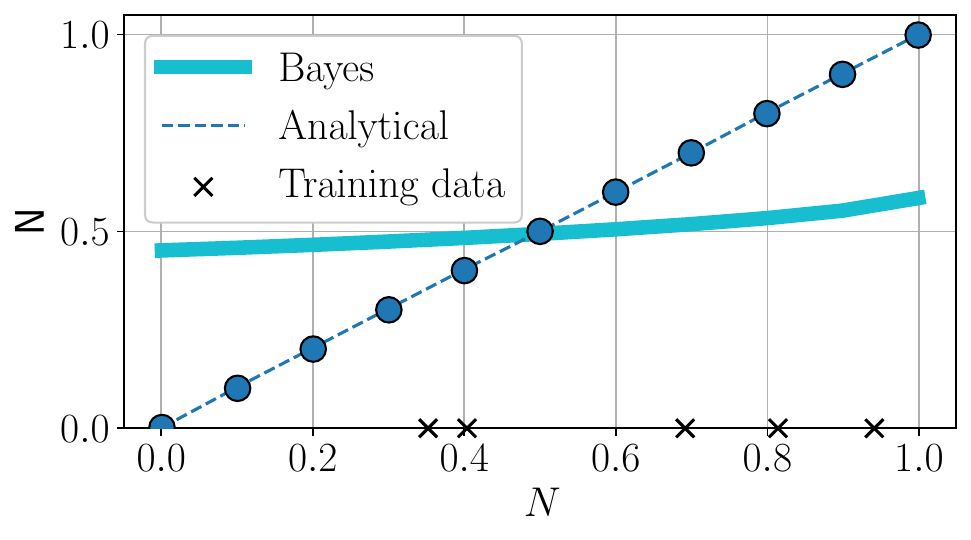}
                \includegraphics[width=.48\textwidth]{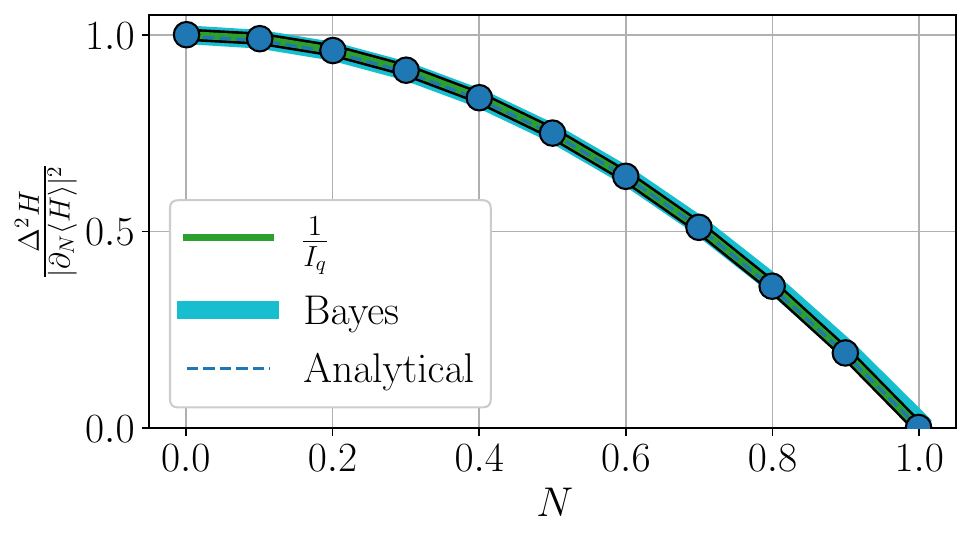}
                \caption{
                    Left: Predicted $\mathsf{N}$ vs. the true $N$ negativity of the Bell-type state \eqref{eq:bell}; the black crosses indicate the states from the training set. Right: Variance \eqref{eq:var-qcrb} together with the quantum Cramer-Rao bound \eqref{eq:cramer-rao_quantum}.
                    In both panels, the solid blue line stands for the results obtained via the minimization of the Bayesian MSE \eqref{eq:bmse}, and the dashed line indicates the results for the optimal observable \eqref{eq:opt_ham_bell0}; in the right panel, the both overlap with qCRB.
                }
                \label{fig:bell_ent}
            \end{figure*}
            
            First, we train the model \eqref{eq:ls_min_variational_mod} to predict the negativity \eqref{eq:neg} of Bell-type states of the form
            \begin{equation}
                \label{eq:bell}
                \ket{\Phi^+_p} = \sqrt{p}\ket{00} + \sqrt{1 - p} \ket{11}.
            \end{equation}
            The training set is $\mathcal{T} = \big\{|\Phi^+_{p_j}\rangle, N_j \big\}_{j=1}^{5}$, where the coefficients $p_j$ are picked randomly from a uniform distribution.
            The observable $H$ is composed in the form \eqref{eq:povm_ham_m} and only $m=1$ qubit of a two-qubit state \eqref{eq:bell} is measured.        
    
            Fig.~\ref{fig:bell_ent} shows the performance of the trained model on the testing set $\mathcal{V}= \big\{|\Phi^+_{p_j}\rangle, N_j \big\}_{j=1}^{10}$ with equidistant negativities $N_j$.
            As can be seen, not only the accurate predictions $\mathsf{N} = \langle \Phi^+_p | H(\boldsymbol{x}^*, \boldsymbol{\theta}^*) | \Phi^+_p\rangle$ of the negativity $N(\Phi^+_p)$ are obtained, but also the optimized observable $H$ saturates qCRB \eqref{eq:cramer-rao_quantum}.
    
            The explanation for such good performance could be the following. 
            The negativity of the state \eqref{eq:bell} reads $N(\Phi^+_p) = 2\sqrt{p(1 - p)}$. 
            Then one gets $p = \big(1 \pm \sqrt{1 - N^2}\big)/2$.
            Therefore, \eqref{eq:bell} can be rewritten as
            \begin{equation}
                \label{eq:bell_neg}
                \ket{\Phi^+_N} = \frac{1}{\sqrt{2}}\big(c_1\ket{00} + c_2 \ket{11}\big),
            \end{equation}
            where $c_{1,2} = \sqrt{1 \pm \sqrt{1 - N^2}}$.
            Using \eqref{eq:sld_pure} one finds an SLD operator $L$ with a spectral decomposition
            \begin{gather*}
                L = \frac{1}{\sqrt{1-N^2}} \ketbra{l_1} - \frac{1}{\sqrt{1-N^2}} \ketbra{l_2}\\
                \ket{l_{1,2}}= \sqrt{\frac{1 \mp N}{2}} \ket{00} \pm \sqrt{\frac{1 \pm N}{2}} \ket{11}
            \end{gather*}
            with the eigenvectors $\ket{l_i}$ being dependent on $N$, giving therefore the observable \eqref{eq:H_opt} which cannot saturate the qCRB in the whole range of $N$.
    
            However, as we see in Fig.~\ref{fig:bell_ent}, the bound is saturated, which suggests that our model finds a better observable.
            Indeed,
            solving (\ref{eq:opt_obs}) for the state (\ref{eq:bell_neg}) yields the following optimal observable:
            \begin{equation}
                \label{eq:opt_ham_bell}
                    H_0 =
                \begin{pmatrix}
                    \frac{t_3}{t_1} & 0 & 0 & \frac{t_2}{t_1} \\
                    0 &  2h_{00} - 2h_{03} + \frac{t_4}{t_1} & 2h_{11} + 2i h_{12} - \frac{t_2}{t_1} & 0 \\
                    0 & 2h_{11} - 2i h_{12} - \frac{t_2}{t_1} & 2h_{00} + 2h_{03} - \frac{t_3}{t_1} & 0 \\
                    \frac{t_2}{t_1} & 0 & 0 & -\frac{t_4}{t_1}
                \end{pmatrix}
            \end{equation}
            where $h_{00}$, $h_{03}$, $h_{11}$ and $h_{02}$ are free variables and
            \begin{align*}
                t_1 &= 3k^2\pi^2 + 6k\pi^2 - 12k^2\pi + 12k\pi + 8k^2 - 100k - 16, \\
                t_2 &= 3k^2\pi^2 - 12k^2\pi + 6k\pi + 8k^2 - 20k, \\
                t_3 &= 24k\pi - 76k - 8, \\
                t_4 &= 4k + 8.
            \end{align*}
            If we put a sufficiently small variance weight $w_\mathrm{var}$ in (\ref{eq:ls_min_variational_mod}) and (\ref{eq:opt_obs}), which implies $k\rightarrow \infty$, then \eqref{eq:opt_ham_bell} becomes
            \begin{equation}
                \label{eq:opt_ham_bell0}
                 H_0 =
                \begin{pmatrix}
                    0 & 0 & 0 & 1 \\
                    0 & 2(h_{00} - h_{03}) & 2(h_{11} + i h_{12}) - 1 & 0 \\
                    0 & 2(h_{11} - i h_{12}) - 1 & 2(h_{00} + h_{03}) & 0 \\
                    1 & 0 & 0 & 0
                \end{pmatrix}.
            \end{equation}  
            For this observable and the state $\ket{\Phi^+_N}$, one obtains $\langle H_0 \rangle = N$ and $\Delta^2 H_0 = 1 - N^2 = 1/I_q(\Phi^+_N)$.
    
            Therefore, our method finds this or an equivalent observable $H$ giving accurate predictions and saturating qCRB.
            Thus, one should be careful while analyzing observables obtained via \eqref{eq:H_opt} from an SLD operator, as the latter can be not unique. Indeed, with the SLD $L$ being the solution of~(\ref{eq:sld_def}), the uniqueness is guaranteed when $\rho_\alpha$ is strictly positive definite~\cite{Personick71}. However, this is not the case as the states~(\ref{eq:bell_neg}) are pure. Therefore, more accurately, one should consider all the solutions of~(\ref{eq:sld_def}) and then pick only those which give, by construction~\eqref{eq:H_opt}, an observable independent of the parameter. The whole procedure gets cumbersome, and direct solving of~(\ref{eq:opt_obs}) seems to be more preferable for obtaining the global optimal observable.
    
            For the considered task, we also simulated the minimization of the Bayesian MSE \eqref{eq:bmse-comp}.
            In this case, the obtained observable saturates the qCRB \eqref{eq:cramer-rao_quantum}, but gives inaccurate predictions of the labels.
            Moreover, if one puts the optimal observable \eqref{eq:opt_ham_bell} with $k=1$ into the Bayesian qCRB \eqref{eq:bqcrb}, one gets
            \begin{equation*}
                \Delta^2_B\alpha = \Delta^2_p\alpha - I_B = \frac{2\pi^2-8\pi+4}{9\pi^2-108}\approx0.0726799,
            \end{equation*}
            in agreement with our numerical results.

        \subsubsection{Isotropic states}
        \label{sec:iso}
    
            \begin{figure*}
                \centering
                \includegraphics[width=.495\textwidth]{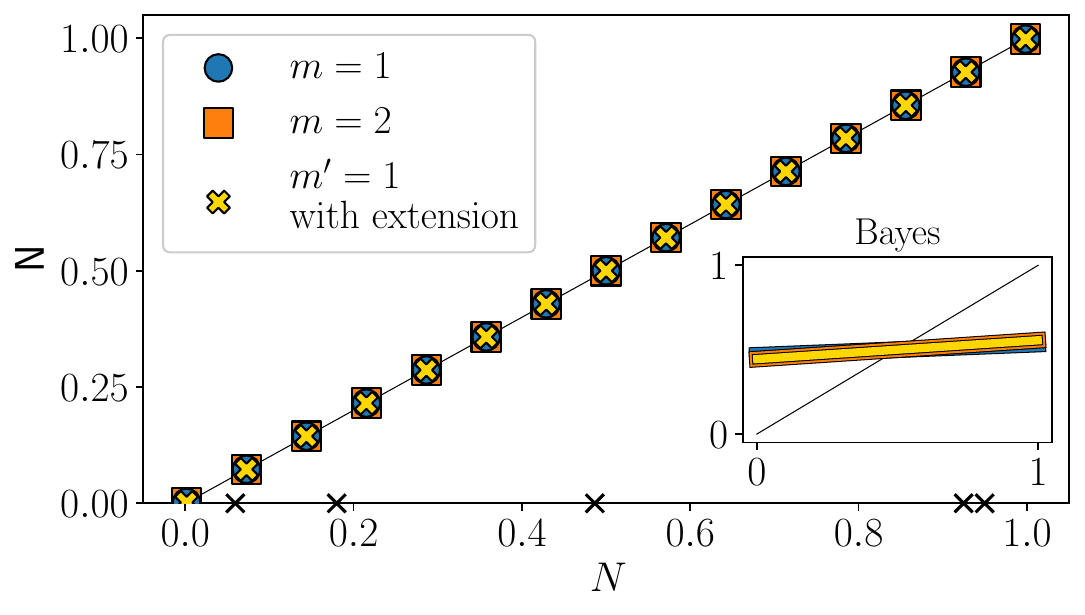}
                \includegraphics[width=.4725\textwidth]{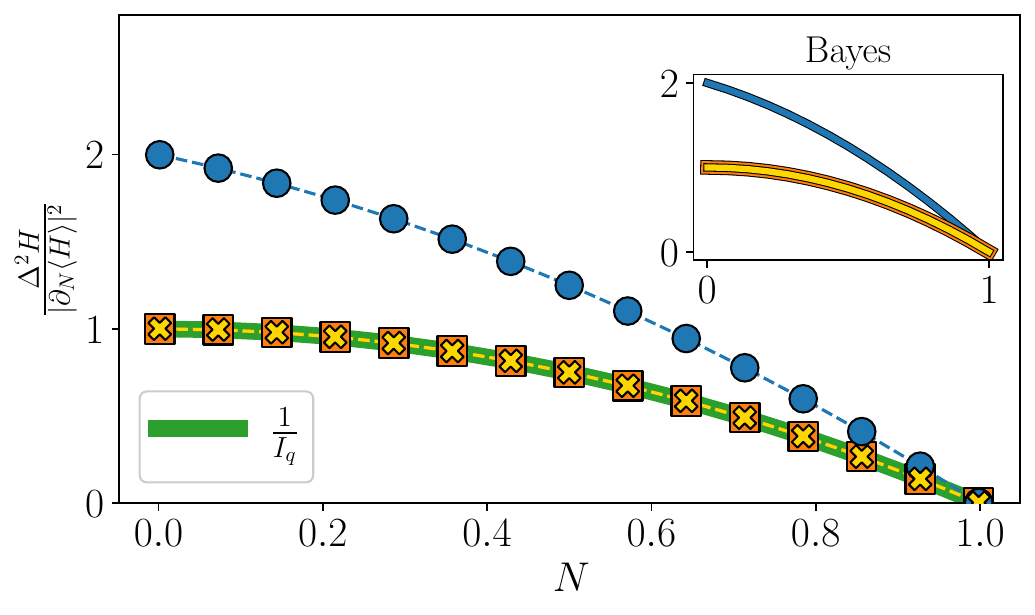}
                \caption{
                    Left: Predicted $\mathsf{N}$ vs. true $N$ negativity of the isotropic state \eqref{eq:iso} using trained observables of the form \eqref{eq:povm_ham_m} with measuring $m \in \{1, 2\}$ qubits.
                    Right: Variance \eqref{eq:var-qcrb} together with cCRB (dashed lines) and qCRB (solid green line).
                    The model was trained on a set $\mathcal{T} = \{\rho_{q_j}, q_j\}_{j=1}^5$ of isotropic states \eqref{eq:povm_ham_m} with random $q > 1/3$ with the corresponding negativities indicated as black crosses in the left panel.
                    The yellow crosses in both panels stand for the case of $m'=1$ measured qubit introduced via the Naimark's extension \eqref{eq:naimark-variational}.
                    The insets show the results obtained using the Bayesian approach \eqref{eq:bmse}.
                }
                \label{fig:iso_ent-m}
            \end{figure*}
            
            The next example is dedicated to negativities of the two-qubit isotropic state
            \begin{equation}
                \label{eq:iso}
                \rho_q = q\ketbra{\Phi} + \frac{1-q}{4}\Id, 
            \end{equation}
            where $\ket{\Phi} \equiv \ket{\Phi^+_{1/2}}$, the Bell state \eqref{eq:bell} with $p=1/2$.
            Generally, isotropic states are used as a standard noise model. In addition, they are invariant under the local unitary transformations $U\otimes\overline{U}$, where $\overline{U}$ denotes complex conjugation of $U$. Due to this property the isotropic states can be used as intermediate states in the protocols of entanglement distillation~\cite{Hordist99}.
            
            The negativity of the isotropic state reads 
            \begin{equation*}
                N(\rho_q) = 
                \begin{cases}
                    0,                & \text{if } q \leqslant \frac{1}{3} \\
                    \frac{3q - 1}{2}, & \text{if } q > \frac{1}{3} 
                \end{cases}
            \end{equation*}
            For $q > 1/3$, we can rewrite \eqref{eq:iso} as
            \begin{equation}
                \label{eq:iso_neg}
                \rho_N = \frac{2N + 1}{3}\ketbra{\Phi} + \frac{1 - N}{6}\Id, 
            \end{equation}
            for which one can find the spectral decomposition of the SLD operator \eqref{eq:sld_def} to be
            \begin{gather*}
                L = \frac{1}{N+1}\ketbra{l_1} + \frac{1}{N-1}\sum_{i=2}^4\ketbra{l_i}, \\
                \ket{l_{1,2}} = \frac{1}{\sqrt{2}}\big( \ket{00} \pm \ket{11}\big), \quad \ket{l_3} = \ket{01}, \quad \ket{l_4} = \ket{10},
            \end{gather*}
            and qFI is $I_q(\rho_N) = 1/(1 - N^2)$.
            This time, the eigenvectors $\ket{l_i}$ are independent of $N$, so the optimal observable constructed via \eqref{eq:H_opt} is
            \begin{equation}
                \label{eq:iso_eff_obs}
                H =\ketbra{00}{11}+\ketbra{11}{00}-\ketbra{01}-\ketbra{10}.
            \end{equation}
    
            On the other hand, solving the equation (\ref{eq:opt_obs}) one can obtain the following optimal observable:
            \begin{equation}
                \label{eq:opt_ham_iso}
                    H_0 = \frac{1}{(8+k)}
                    \begin{pmatrix}
                        4 & 0 & 0 & k \\
                        0 & 4-k & 0 & 0 \\
                        0 & 0 & 4-k & 0 \\
                        k & 0 & 0 & 4
                    \end{pmatrix}.
            \end{equation}   
            For the state $\rho_N$, this observable gives 
            \begin{equation*}
                \langle H_0 \rangle = \frac{4+N k}{8+k}, 
                \qquad 
                \Delta^2 H_0 = \frac{(1 - N^2) k^2}{(8 + k)^2}.
            \end{equation*}
            Therefore, as long as $k$ is finite, there will always be a bias in the prediction.
            This confirms the statement we made at the end of Section~\ref{sec:theor-var}.
            If we assign a small weight to the total variance in (\ref{eq:opt_obs}), i.e., $k\rightarrow \infty$, then this observable coincides with \eqref{eq:iso_eff_obs}.
    
            Interestingly, the  observable \eqref{eq:iso_eff_obs} has two distinct eigenvalues, the unique $x_1=1$ and the thrice-degenerate $x_2=-1$.
            Moreover, one can verify that this degeneracy for the eigenvalues holds also for \eqref{eq:opt_ham_iso} for any $k$.
            One can find such observable using the procedure \eqref{eq:ls_min_variational_mod} by measuring one qubit introduced via the Naimark's extension \eqref{eq:naimark-variational}, which we denote $m'=1$.
            Indeed, if we look for an optimal observable in the form \eqref{eq:povm_ham_m} with measuring one qubit of the state, we could find only an observable with two-dimensional projectors \eqref{eq:orth_POVM_variational_m}. Hence one gets two-fold degenerate eigenvalues.

            \begin{figure*}
                \centering
                \includegraphics[width=.495\textwidth]{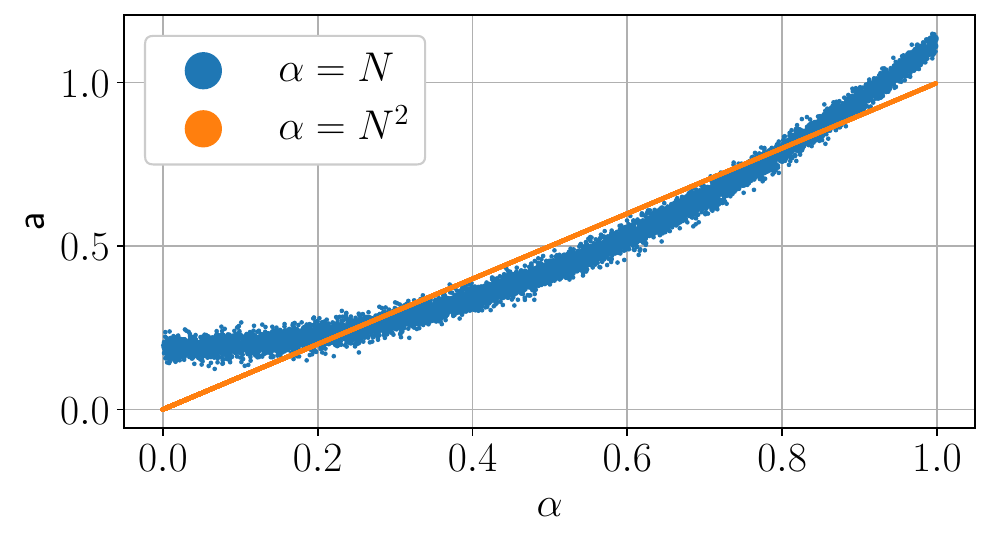}
                \includegraphics[width=.495\textwidth]{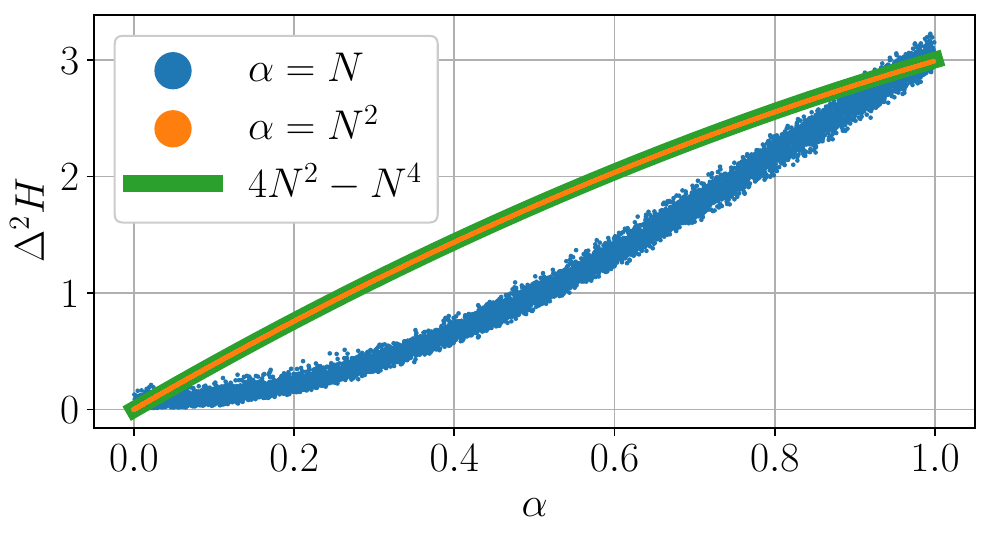}
                \caption{
                    Left: Predicted negativity $\mathsf{N}$ (blue) and the squared negativity $\mathsf{N}^2$ (orange)  of $10^4$ random pure two-qubit states vs. the corresponding true $N$ and $N^2$ for $c=2$ copies.   Right: Variance of the trained observable $H$; the solid green line indicates the variance of the observables $M_{1,2}$ in \eqref{eq:negbound}, which overlaps with the variance of $H$ trained to predict the squared negativity. 
                    The models were trained on the sets $\mathcal{T} = \big\{|\psi_j\rangle^{\otimes 2}, N_j \big\}_{j=1}^{1000}$ (blue) and $\mathcal{T} = \big\{|\psi_j\rangle^{\otimes 2}, N_j^2 \big\}_{j=1}^{1000}$ (orange) with random $|\psi_j\rangle$ and evenly distributed $N_j$.
                 }
                \label{fig:ent_rand_pure}
            \end{figure*}

            These eigenvalue degeneracies may be important in solving regression problems.
            In Fig.~\ref{fig:iso_ent-m}, we show the negativities of the state \eqref{eq:iso} predicted by our model with different numbers of measured qubits $m$.
            The model was trained on a set $\mathcal{T} = \big\{\rho_{q_j}, N_j \big\}_{j=1}^{5}$, where $q_j$ are uniformly sampled from $(\frac{1}{3}, 1]$, so that the states $\rho_{q_j}$ are always entangled.
            As we see in Fig.~\ref{fig:iso_ent-m}, by measuring $m=1$ or $2$ qubits of the isotropic state we are able to find an observable giving accurate predictions of the negativity. 
            However, with $m=1$, the variance of the optimized observable is as about twice as large as with $m=2$, which may be happening due to the eigenvalue degeneracy discussed earlier.
            Across many runs of our simulations, we obtain an observable with the two-fold degenerate eigenvalues $x_1=1$ and $x_2=-2$.
            Such an observable may have the form, e.g.,         
            $
                H = \big(\ketbra{\Phi} + \ketbra{\psi_1^\perp}\big) -2\big(\ketbra{\psi_2^\perp} + \ketbra{\psi_3^\perp}\big),
            $
            where $\ket{\psi_i^\perp}$ are any vectors orthogonal to $\ket{\Phi}$ and to each other.
            For this observable and isotropic state $\rho_N$ of the form \eqref{eq:iso_neg}, one can verify that $\langle H \rangle = N$ and $\Delta^2 H = 2 - N - N^2$, which coincides with the blue line in Fig.~\ref{fig:iso_ent-m} (right).
            In contrast, if we measure $m'=1$ \textit{additional} qubit introduced via the Naimark's extension \eqref{eq:naimark-variational}, we obtain an observable giving both accurate predictions and saturating qCRB.
    
            Additionally in the insets of Fig.~\ref{fig:iso_ent-m}, we plot the results for the observable obtained by means of the Bayesian approach \eqref{eq:bayes_min}.
            As in the previous cases, the resultant observable does not give accurate label predictions.
            At the same time, it saturates the quantum Cramer-Rao bounds \eqref{eq:cramer-rao_quantum} and \eqref{eq:bqcrb}.
            For the latter, the observable \eqref{eq:opt_ham_iso} with $k=1$ gives
            \begin{equation*}
                \Delta^2_B\alpha = \Delta^2_p\alpha - I_B = \frac{2}{27} \approx 0.074,
            \end{equation*}
            achieved in our numerical experiments for $m=2$ and $m'=1$ measured qubit introduced via Naimark's extension.
    
            An extra attention is paid to entangled isotropic states $\rho_q$ with $q > 1/3$, for which one can find an observable giving accurate predictions of the negativity.
            Unfortunately, there is no such observable if $q\leqslant1/3$, as $N=0$ on this interval.
            One can use the technique of processing several copies of the state $\rho_N$, as mentioned in Section \ref{sec:func_dep}, with the prediction for $N$ becoming $\mathsf{N} = \Tr H \rho_N^{\otimes c}$.
            In Appendix~\ref{sec:iso-c}, we consider the situation case with separable states in the training set.

        \subsubsection{Random pure states}
        \label{sec:ent-pure}
    
            The presented approach aims to assume as little knowledge about the regression problem as possible.
            In some specific cases, however, one might also try to augment the method with an extra knowledge about the system. As an example, from entanglement theory an upper bound on the \emph{square} of the negativity of an arbitrary \emph{two-qubit} mixed state~\cite{NegUpBound08} is known:
            \begin{equation}
            \label{eq:negbound}\left[N(\rho)\right]^2\leqslant\mathrm{Tr}\,\rho^{\otimes 2} M_i,\quad i = 1,\,2,
            \end{equation}
            where $M_1 = 4P_-^{(1)}\otimes I^{(2)}$ and $M_2 = 4I^{(1)}\otimes P_-^{(2)}$. Here $P^{(i)}_-$ denotes the projector onto the antisymmetric subspace of the two copies of the $i$-th subsystem. 
            Originally in ~\cite{NegUpBound08} the bound was derived for the square of another entanglement measure, the concurrence~\cite{MixedStateEnt96, WootEntForm98, ConcArSt04}. However, it is known that for the case of two-qubit mixed states the concurrence is always lower-bounded by the negativity~\cite{NegConcRel15}. There is also a similar \emph{lower} bound~\cite{MintBuch07} on the square of concurrence $C(\rho)$ of an arbitrary bipartite state:
            \begin{equation}
                \label{eq:ClowB}\left[C(\rho)\right]^2\geqslant\mathrm{Tr}\,\rho^{\otimes 2}\,V_i,\quad i = 1,\,2,
            \end{equation}
            where $V_1 = 4(P^{(1)}_- - P^{(1)}_+)\otimes P^{(2)}_-$ and $V_2 = 4P^{(1)}_-\otimes (P^{(2)}_- - P^{(2)}_+)$, with $P^{(i)}_+$ being the projector onto the symmetric subspace of the two copies of the $i$-th subsystem.
    
            \begin{figure*}
                \centering
                \includegraphics[width=.495\textwidth]{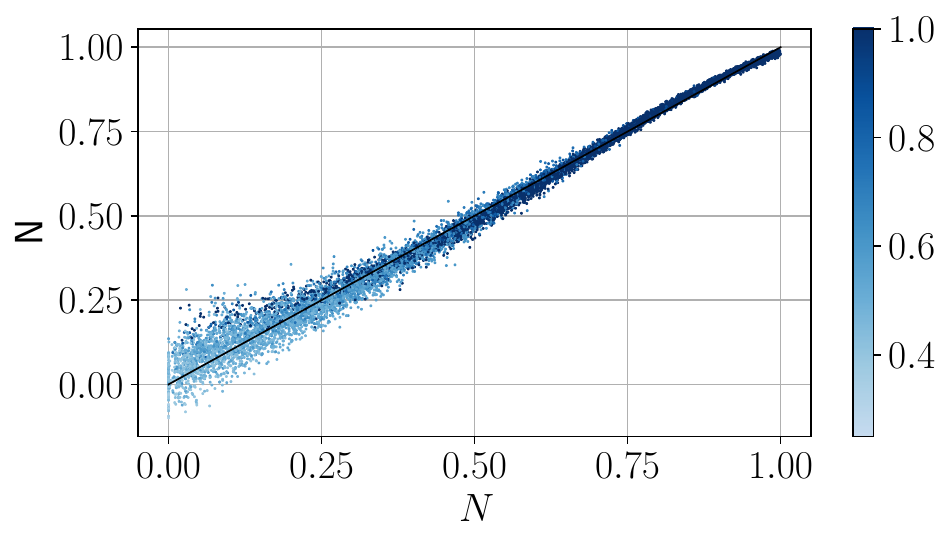}
                \includegraphics[width=.4925\textwidth]{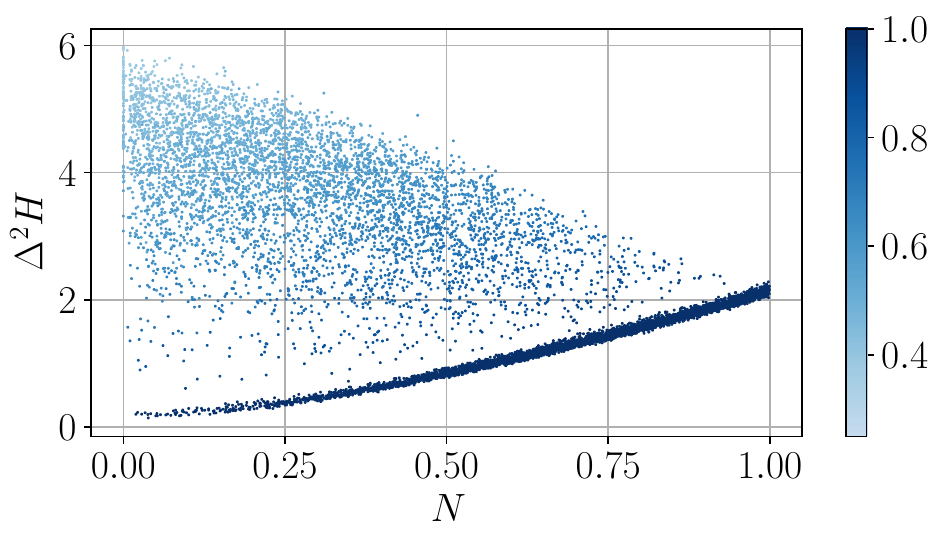}
                \caption{
                    Left: Predicted negativity $\mathsf{N}$ of $10^4$ random mixed states vs. the true negativity $N$ for $c=4$ copies. 
                    Right: Variance of the trained observable $H$.
                    The color of the points indicates the purity of the corresponding states.
                    The model was trained on a set $\mathcal{T} = \big\{\rho_j^{\otimes 4}, N_j \big\}_{j=1}^{1000}$ with random $\rho_j$ and $N_j$ evenly distributed on $[0, 1]$.
                }
                \label{fig:ent_rand_mixed}
            \end{figure*}
        
            The bounds~(\ref{eq:negbound}), (\ref{eq:ClowB}) have the form of expectation value of a Hermitian operator with respect to  \emph{two copies} of a given state fitting our approach well.
            In addition, for \emph{pure} two-qubit states the bound~gives the \emph{exact} value of squared negativity~\cite{NegUpBound08}~(and also concurrence and negativity coincide for such states~\cite{NegConcRel15}).
            Thus, we can test the performance of our method on randomly generated pure states and check how well it can variationally find the appropriate observable on two copies to obtain the exact value of the squared negativity.
            For this, we trained our regression model on a set $\mathcal{T} = \big\{|\psi_j\rangle^{\otimes 2}, N_j^2 \big\}_{j=1}^{1000}$ consisting of random pure states along with their squared negativities, which are evenly distributed on the interval $[0, 1]$.
            In our simulations, we used an ansatz of $l=2$ layers and measured all $m=4$ qubits as per \eqref{eq:povm_ham_m}.
            
            The results depicted in Fig.~\ref{fig:ent_rand_pure} show that the found observable provides very good predictions for the squared negativity and is able to reproduce numerically two-copy observables like $M_1$ and $M_2$ defined above.
            For comparison, we additionally plot the results of the performance of our model trained on the same data set, but without squaring the negativities.
            For both cases, in Fig.~\ref{fig:ent_rand_pure} we also plot the variance of the optimized observable $\Delta^2 H$.  
            For either $M_1$ or $M_2$ in \eqref{eq:negbound} and two copies of a pure two-qubit state $\ket{\psi_N}$ with negativity $N$, one can find that $\Delta^2_{\psi_N} M_{1,2} = 4 N^2 - N^4$ which is saturated by $H$ found with our method.
    
            Interestingly, one can obtain these results by measuring $m=1$ out of $n=4$ qubits, but only if it is introduced via the Naimark's extension \eqref{eq:naimark-variational}.
            Indeed, both observables $M_{1,2}$ in \eqref{eq:negbound} have only a four-fold degenerate eigenvalue $x_1 = 4$.
            In the meantime, the observables $V_{1,2}$ in \eqref{eq:ClowB} have a three-fold degenerate $x_1=-4$ and a non-degenerate $x_2 = 4$.
            If one looks for an optimal observable in the form \eqref{eq:orth_POVM_variational} by measuring $m=1$ qubit of $\rho_N^{\otimes 2}$, one could obtain only an observable with eight-fold degenerate eigenvalues (excluding the trivial case).
            This can be resolved by Naimark's extension with $m'=1$ additional qubit.

        \subsubsection{Random mixed states}
        \label{sec:ent-mixed}
    
            Prediction of the negativity of \emph{random mixed} states presents a more challenging problem. 
            To solve it, we train our model on a set $\mathcal{T}_c = \big\{\rho_j^{\otimes c}, N_j \big\}_{j=1}^{1000}$, where the states $\rho_j$ are generated such that their negativities $N_j$ are evenly distributed on $[0, 1]$.
            To parametrize the observable, we used an ansatz of $l=2$ layers.
            
            In Fig.~\ref{fig:ent_rand_mixed} we show the predictions of the negativity of randomly generated mixed states by considering $c=4$ copies.
            Additionally, we indicate the purity $P(\rho) = \Tr \rho^2$ of a given state $\rho$.
            As can be seen, for lower negativities the states are generally more mixed, which results both in higher prediction error and variance $\Delta^2 H$ of the optimized observable.
            Nonetheless, our model performs well even in a difficult case of using a training set of a relatively modest size.
            However, in this case, our method requires measuring all $m=2c$ qubits, which results in $|\boldsymbol{x}|=2^{2c}$ additional parameters to vary.
            Still, our model achieves good prediction accuracy without any knowledge about the connection between the data points $\rho_{j}$ and their labels $N_j$.          
            For comparison, in Appendix~\ref{app:random_mixed_c} we also show the results obtained for the same training set, but computed with $c=1,2,3$ copies. As one could expect, the more copies we process, the better results we get.
            Following the intuition provided by \eqref{eq:negbound}, in Appendix~\ref{app:random_mixed_c_sq} we apply our method for predicting the squared negativities which results in a significant increase of accuracy for $c=2,4$, but giving a little advantage for $c=3$ compared with $c=2$.  
    
            We make two remarks here.
            First, as mentioned above, while measuring all $m=8$ qubits of the composite state $\rho^{\otimes 4}$ indeed allows to obtain an observable giving accurate predictions of the negativity, the optimization process is computationally demanding.
            In Appendix~\ref{app:neg-rand-mixed-m}, we train our model with smaller $m$ and discuss the training complexity of this process.
            However, while our simulations show that $m=6$ gives predictions and variances one may consider acceptable, each measured qubit may contribute to the accuracy significantly.
    
            Our second remark concerns the size of the training set $T$.
            With the exception for predicting the negativity, all the regression tasks we considered in this work required few (up to ten) training states for finding good observables.
            In Appendix~\ref{app:neg-rand-mixed-T}, we train our model for predicting the negativity for different training set sizes $T$ from 100 to 1000.
            We observe that the MSE calculated on the testing set approximately scales with this size as $T^{-5/4}$.

    \subsection{Predicting transverse field of the Ising Hamiltonian}
    \label{sec:ising}

            \begin{figure*}
                \centering
                \includegraphics[width=.495\textwidth]{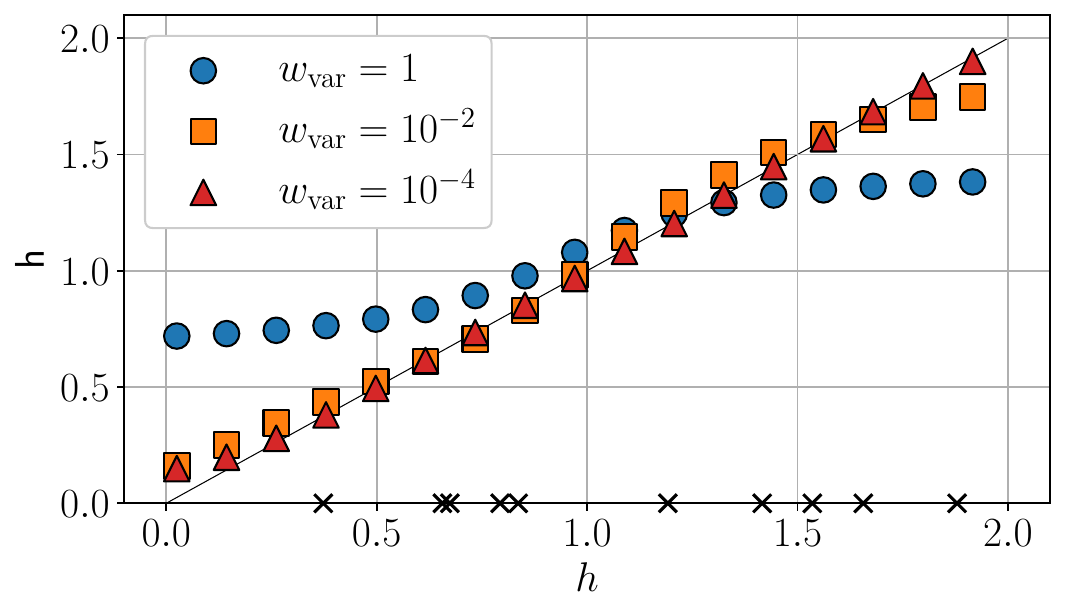}
                \includegraphics[width=.485\textwidth]{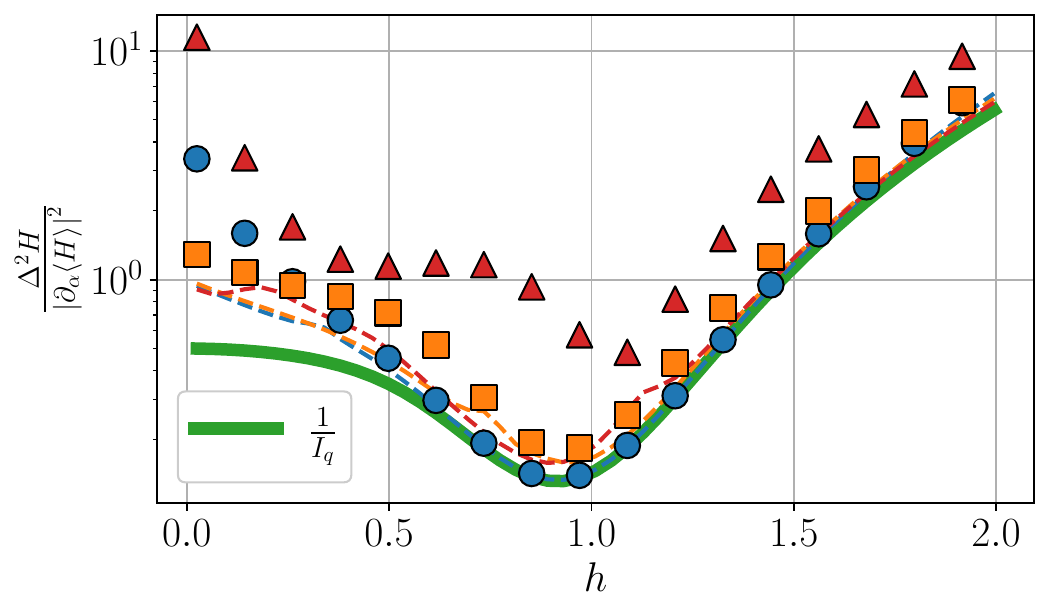}
                \caption{
                    Left: Predicted $\mathsf{h}$ vs. true $h$ transverse field of the 8-qubit Ising Hamiltonian \eqref{eq:ising} for different weights $w_\mathrm{var}$ in \eqref{eq:ls_min_variational_mod}; the black crosses on the $x$-axis indicate the training data.
                    Right: Variance \eqref{eq:var-qcrb} vs. $h$; the dashed lines indicate cCRB \eqref{eq:cramer-rao_classical}. 
                    The observable $H$ is trained on a set $\mathcal{T} = \big\{|\psi_{h_j}\rangle, h_j \big\}_{j=1}^{10}$ with $\ket{\psi_j}$ being the ground states of \eqref{eq:ising}, and the fields $h_j$ are generated randomly.
                    In the right panel, the dashed lines of the corresponding colors indicate the achieved cCRB, while the solid green line stands for qCRB \eqref{eq:cramer-rao_quantum}. 
                }
                \label{fig:ising-w_var}
            \end{figure*}

        Another interesting example of application of our approach is prediction of the transverse field $h$ of the Ising Hamiltonian
        \begin{equation}
            \label{eq:ising}
            H_h = -J\sum\limits_{i=1}^n \left( \sigma_z^i\sigma^{i+1}_z + h \sigma^i_x\right),
        \end{equation}
        where $\sigma_x^i$, $\sigma_z^i$ are Pauli operators acting on the $i$th qubit, and we apply the periodic boundary conditions $\sigma^{n+1}_z \equiv \sigma_z^1$.
        Thus, given a collection of the ground states $\ket{\psi_h}$ of $H_h$ in the form \eqref{eq:training_set}, one wishes to solve \eqref{eq:ls_min_variational_mod} finding an observable $H$ such that the expectation $\langle H \rangle_{\psi_h}$ gives a prediction of the transverse field $h$. 
        In our numerical experiments, we consider $n=8$ qubits and set $J=1$.

        In Section~\ref{sec:estimation_variance}, we mentioned that by adjusting the weights $w_\mathrm{ls}$ and $w_\mathrm{var}$ in \eqref{eq:ls_min_variational_mod}, we can trade-off between the accuracy of the prediction and its variance.
        It is already shown in Section~\ref{sec:ad_chan-res} for the case of the amplitude-damping channel, where the model is trained on a deliberately large training set of $T=500$ entries.
        Here, we make use of a much more modest set with only $T=10$ points.
        We will look for an observable in the form \eqref{eq:povm_ham_m} with $m=4$ and the ansatz of $l = 5 $ layers.

        In Fig.~\ref{fig:ising-w_var}, the results of predicting the transverse field $h$ of \eqref{eq:ising} with setting $w_\mathrm{ls} = 1$ and considering  $w_\mathrm{var} \in \{1, 10^{-2}, 10^{-4}\}$ are shown.
        As might be expected, the greater is the weight $w_\mathrm{var}$, the less accurate predictions become.
        However, with $w_\mathrm{var}=1$, one is closer to the qCRB \eqref{eq:cramer-rao_quantum}.

        Note that with $w_\mathrm{var}=10^{-4}$, one gets good results by measuring only $m=4$ qubits out of $n=8$ and utilizing an ansatz of $l=5$ layers.
        However, in Appendix~\ref{sec:ising-training} we show that if all $m=8$ qubits are measured and an ansatz of $l=2$ layers is used, the even better results are obtained, especially in terms of variance.
        Additionally in that Appendix, we discuss the training complexity of our method in solving the considered regression task.
        In short, despite the possible better prediction accuracy and variance due to greater $m$, such results are harder to achieve since each additional measured qubit exponentially increases the number of trainable parameters (see \eqref{eq:povm_ham_m}).     

        As of the minimization of the Bayesian MSE \eqref{eq:bmse}, for this case it is computationally heavy.
        However, as one would do in practice \cite{marciniak2022optimal}, one can approximate it as a sum of discrete points. 
        If we possess only the data given in the training set and assume the uniform distribution of the labels, we reduce the problem to \eqref{eq:ls_min_variational_mod} with equal weights $w_\mathrm{ls}=w_\mathrm{var}=1$.
        As can be seen in Fig.~\ref{fig:ising-w_var}, in this case the method gives too much preference to the minimization of the variance, sacrificing the accuracy.
        This is in line with the results of minimizing the Bayesian MSE for other regression problems considered in this work.

        The problem of predicting the transverse field was also solved with simulation of a finite number of measurements $\mu$.
        Namely, in Appendix~\ref{sec:shots}, the method is used with an assumption that instead of the exact expectations $\mathsf{h} = \langle \psi_h |H| \psi_h \rangle$ we get their $\mu$-shot estimates $\hat{\mathsf{h}}$  during the training.
        This of course worsens the results compared to the case of ``infinite'' $\mu$ reported in this section.
        However, with a moderate number of measurements, a good estimation accuracy is still obtained.

\section{Discussion}
\label{sec:discussion}

    In this work, using the optimization routine \eqref{eq:ls_min_variational_mod}, we solve a regression problem for labeled data points represented by quantum states $\rho_\alpha$ with $\alpha$ being the label.
    The prediction $\mathsf{a}$ of the label $\alpha$ is $\Tr H \rho_\alpha$ with $H$ being a constructed observable.

    Our approach assumes no knowledge about the link between the data points and their labels.
    We applied our method for the cases when $\rho_\alpha = \Phi_\alpha[\rho]$ where $\alpha$ parametrizes a quantum channel $\Phi_\alpha$ acting on some fixed input state $\rho$ (see Section~\ref{sec:ad_chan-res}), $\alpha$ is an entanglement measure of $\rho_\alpha$ (see Section~\ref{sec:ent}), and $\rho_\alpha = \ketbra{\psi_\alpha}$ with $\ket{\psi_\alpha}$ being the ground state of a parametrized Hamiltonian $H_\alpha$ (see Section~\ref{sec:ising}).
    Appendix~\ref{app:add_res} contains additional numerical results related to various aspects of the proposed method, such as the number of measured qubits $m$ in \eqref{eq:povm_ham_m}, number of copies $c$ of the labeled state processed as one composite state, and finite number of measurements $\mu$.
    Additionally in that Appendix, we study the problem of predicting the parameter of the depolarizing channel and the angle of a Pauli rotation.
  
    To assess the performance of our method, we employ quantum metrology techniques. Alongside the prediction accuracy, we also look at the error propagation \eqref{eq:estimation_variance} and Cramer-Rao bounds \eqref{eq:cramer-rao_quantum}, which together produce \eqref{eq:var-qcrb}.
    For some cases, the proposed method allows finding an observable $H$ such that it saturates qCRB.
    Additionally, by tuning $w_\mathrm{ls}$ and $w_\mathrm{var}$ in \eqref{eq:ls_min_variational_mod}, one can trade off between prediction accuracy and its variance, as we show in Sections~\ref{sec:ad_chan-res} and~\ref{sec:ising}.

    Interestingly, to solve the most regression problems considered in this work we needed training sets of a few states (up to $T=10$).
    The exception was the prediction of the negativity of two-qubit states (see Section~\ref{sec:ent-mixed}), which required a training set of the size $T=1000$.
    In Appendix~\ref{app:neg-rand-mixed-T}, we numerically study the dependence of the performance of our method on $T$ for the latter problem.
    We observe that MSE for the optimized observable calculated on the training set scales as $T^{-5/4}$.
    This is in line with known upper bounds on generalization error, which typically scale as $T^{-1/2}$ (see, e.g., \cite{caro2022generalization}).

    In Appendix~\ref{app:neg-rand-mixed-m} we discuss the training complexity of our method for solving the regression problem.
    Namely, we study the dependence of the number of iterations of the BFGS algorithm on the number of measured qubits $m$ when we look for an observable in the form \eqref{eq:povm_ham_m}.
    While for each $m$ it takes up to 2000 iterations for the algorithm to converge, the number of calls of the cost function in \eqref{eq:ls_min_variational_mod} grows exponentially.
    Indeed, each increment of $m$ doubles the number of varied eigenvalues in \eqref{eq:povm_ham_m}.
    One may try to look for an optimal observable on fewer qubits $m$, as we did for the problem of predicting the transverse field in Section~\ref{sec:ising}.
    However, for the problem of predicting the negativity of random states, we observed that increasing $m$ may contribute significantly to both the prediction accuracy and variance.

    One possible way of circumventing this issue is taking into account the knowledge about the problem to be solved. 
    Indeed, as is shown in Section~\ref{sec:ent-pure} and Appendix~\ref{app:random_mixed_c_sq}, learning to predict the \textit{squared} negativity gives a significant advantage.
    Moreover, for pure states, one can obtain nearly perfect accuracy by measuring only $m'=1$ qubit introduced via the Naimark's extension \eqref{eq:naimark-variational}.
    Additionally, one could construct the variational ansatz $U(\boldsymbol{\theta})$ such that it preserves certain symmetries (an obvious one in our case is the permutation symmetry on $c$ copies of the labeled state, $\rho_\alpha^{\otimes c}$) \cite{larocca2022group, meyer2023exploiting, nguyen2024theory}.
    Alternatively, to construct an ansatz, one could apply various techniques proposed in the literature \cite{liu2022quantum, heese2022representation}, including the optimization of the ansatz's structure \cite{lu2021markovian, grimsley2019adaptive}. 
    
    Throughout this work we assume that the expectations $\mathsf{a} = \Tr H \rho_\alpha$ can be computed exactly.
    In a measurement experiment, however, one would obtain a $\mu$-shot estimation $\hat{\mathsf{a}}$ of it.
    In Appendix~\ref{sec:shots}, we numerically study the performance of our method with simulating a finite number of measurements $\mu$.
    As one may expect, larger $\mu$ give better performance  both in terms of the estimation accuracy and the variance.
    For an easier regression problem with single-qubit labeled states and simple optimal observable, even a modest number of shots $\mu=2^6$ allows to achieve good results.
    In contrast, for a harder problem with eight-qubit states, a rather large number of measurements $\mu=2^{14}$ does not allow to train the model such that it gives the results close to the case of ``infinite'' $\mu$.
    This issue of the shot noise is a common challenge for variational quantum algorithms \cite{scriva2024challenges}.
    To overcome it, a variety of techniques have been proposed, such as clever initialization \cite{grimsley2023adaptive, scriva2024challenges}, adaptive tuning of the variance weight in \eqref{eq:ls_min_variational_mod} as proposed in \cite{quantumKrepRoth2024}, exploiting symmetries in the labeled states \cite{zhang2024inferring}, and adjusting the optimizer's hyperparameters (e.g., learning rate in the Adam algorithm \cite{kingma2014adam}) \cite{sung2020using, bonet2023performance}.
    We note that in this work we applied the optimization algorithms from the SciPy library \cite{2020SciPy-NMeth} ``out of the box'' with random initializations of the variational parameters.

    We studied the proposed method also analytically, assuming a large size of the training set \eqref{eq:training_set}.
    In Section~\ref{sec:theory}, first we obtained an equation for an observable which, under the above-mentioned assumption, solves the minimization problem \eqref{eq:ls_min_variational_mod}.
    For several regression problems, we solved this equation and numerically verified the solutions against the method \eqref{eq:ls_min_variational_mod}.
    Furthermore, for optimal observables we derive an upper bound on the total variance, which depends on the ratio $k = w_\mathrm{ls} / w_\mathrm{var}$ and the range of the label $\alpha$.
    Another interesting finding is that the observable obtained with the proposed method cannot be unbiased on the whole range of the label $\alpha$.

    For a large training set size $T$ and $k=1$, our method \eqref{eq:ls_min_variational_mod} reduces to the minimization of the Bayesian MSE \eqref{eq:bmse} if the prior distribution of the labels $\alpha$ is uniform (see Section~\ref{sec:bayes-comp}).
    When applicable, we numerically compare our method to the Bayesian approach.
    We find that the observable obtained via this approach gives more biased predictions, as it prefers to minimize the variance. 
    At the same time, in numerical minimization of the Bayesian MSE we obtain observables saturating the Bayesian quantum Cramer-Rao bound \eqref{eq:bqcrb}.

    In analytical considerations one might as well take into account finite training set size $T$, as in~(\ref{eq:training_set}). In this case the equation for the optimal observable~(\ref{eq:opt_obs}) is straightforwardly rewritten to
    \begin{multline}\label{eq:opt_obs_fin}
        \frac12\left(\Tilde{\rho}_f H_0+H_0\Tilde{\rho}_f\right) - \frac{k}{T}\sum_{j=1}^T \alpha_j \rho_{\alpha_j}  \\  + \frac{(k-1)}{T} \sum_{j=1}^T \Tr (H_0\rho_{\alpha_j}) \rho_{\alpha_j} 
         = 0,
    \end{multline}
    where
    \begin{equation}
    \label{eq:rhotil_fin}
        \Tilde{\rho}_f = \frac1{T}\sum_{j=1}^T \rho_{\alpha_j} 
    \end{equation}
    is a finite sum density operator defined in analogy with~(\ref{eq:rhotil}). Eq.~(\ref{eq:opt_obs_fin}) could be beneficial in analyzing the generalization error of the presented regression  method in application to specific tasks with arbitrarily distributed~ (possibly sparse)  training sets.
    
    As mentioned earlier, the application of our method results in an observable giving generally a biased prediction $\Tr H \rho_\alpha$.
    This bias can sometimes be mitigated, as we show in Appendix~\ref{app:bias_comp}.
    Alternatively, one can find $H$ such that it acts on $\rho_\alpha^{\otimes c}$, several copies of the labeled state.
    As we show in Appendices~\ref{sec:iso-c}, \ref{app:random_mixed_c} and \ref{app:unitary_chan}, the more copies we process simultaneously, the higher is the prediction accuracy.

    In particular, we have observed that the prediction of the negativity of random mixed two-qubit states works better in a multi-copy scenario.
    The prediction gets more accurate with an increase of the number of copies, as can be seen from comparing results for $1,\,2,\,3,\,4$ copies (see Appendix~\ref{app:random_mixed_c}). 
    The possible explanation could be devised from recent results on nonlinear multipartite witnesses~\cite{NonLinWit24}.
    The witnesses of this kind are also suited for a multi-copy scenario.
    In Appendix~\ref{sec:app:nonlinearwit}, we adapt the theory to demonstrate that some specific witnesses show better entanglement detection~(averaged over random states) with higher number of copies. 
    Although this is just entanglement detection without estimating any entanglement measure, the inequalities (\ref{eq:negbound}) and (\ref{eq:ClowB}) suggest\footnote{In fact, the lower bound in (\ref{eq:ClowB}) is an example of nonlinear witness.} that there might be lower and upper bounds on entanglement measures associated with such witnesses. Also, the higher the number of copies is, the tighter and more accurate these bounds are. 
    As a possible direction of further research it would be interesting to establish  connections between entanglement measures and multi-copy witnesses.  

    Among the trainable parameters in \eqref{eq:ls_min_variational_mod} there are eigenvalues of the observable $H$. 
    Thus, if the labeled states are of $n$ qubits the number of parameters to vary would be exponential.
    As we mentioned earlier, one may search for an observable $H$ such that it is enough to measure a fewer number of qubits $m<n$ in some cases (see Section~\ref{sec:m-qubits}).
    This may be advantageous also in the way that if $m=1$, then one can have the variational circuit $U$ in \eqref{eq:orth_POVM_variational_m} to be in the form of a QCNN \cite{cong2019quantum, nagano2023quantum} which is known to be immune to barren plateaus \cite{pesah2021absence}.
    However, the usefulness of such ans\"atze was recently put under question \cite{bermejo2024quantum}.

    In this regard, our work opens new directions for further studies.
    As such, one can investigate the applicability of QCNNs for regression tasks we consider.
    In Section~\ref{sec:iso} for the prediction of the negativity of isotropic states \eqref{eq:iso} of $n=2$ qubits, we discuss an observable $H$ requiring the measurement of only $m=1$ qubit and providing the same prediction accuracy as that of an efficient observable with $m=2$, but with about twice greater variance. 
    It could be interesting to understand for which regression problems there exists an observable $H_m$ of the form \eqref{eq:povm_ham_m} requiring a measurement of $m<n$ qubits and providing accurate predictions with $\langle H_m \rangle_{\rho_\alpha} = \langle H_n \rangle_{\rho_\alpha}$, \textit{and} $\Delta^2_{\rho_\alpha} H_m = \Delta^2_{\rho_\alpha} H_n$, presumably also saturating the Cramer-Rao bound.
    Similar results were recently demonstrated for states that have symmetries, for which one can find observables with the same expectation but generally a lower variance \cite{zhang2024inferring}.

    It would be also interesting to consider predictions of some hard-to-compute quantities. 
    For instance, unlike the qubit-qubit case, for mixed qutrit-qutrit states~(and also ones of higher dimensions) there is no effective procedure of computing entanglement measures~\cite{Huang14}. 
    This is also true for multipartite quantum states, in which case a relevant question is the calculation of various measures~\cite{SenGeEnt10,GuhnAccEnt20,Ant21} of genuine entanglement, i.e., entanglement present in each bipartite cut of a quantum state~\cite{Svet87,DurVidEnt00}. 
    In the context of quantum channels, important quantities for transmitting classical and quantum information include channel capacities~\cite{wilde2011classical}. 
    The question is how effective the regression method will be in predicting the properties listed above.

\section{Acknowledgements}
    The work was supported in the framework of the Roadmap for Quantum Computing (Contracts No. 868-1.3-15/15-2021 and No. R2163).

    
\newpage
    
\bibliography{bibliography}
\bibliographystyle{unsrt}

\newpage
\onecolumngrid
\appendix

\section{Hardware-efficient ansatz}
\label{app:hea}

    In this work, we represent the parametrized quantum circuit $U(\boldsymbol{\theta})$ by the so-called hardware-efficient ansatz (HEA) \cite{kandala_hardware-efficient_2017, Kardashin_2021}.
    The idea behind such ans\"atze is in alternation between single-qubit rotations and operations capable of introducing entanglement into the system.  
    In our implementation, an example of which is shown in Fig.~\ref{fig:hea}, the ansatz consists of several layers of parametrized gates.
    Each layer contains an array of $x$- and $z$- of Pauli rotations acting on individual qubits, followed by a cascade of controlled $y$-rotations which produce entanglement.
    In our ansatz, we also have the ``zeroth layer'' containing no entangling gates.
    Therefore, the total number of the parameters in an $n$-qubit $l$-layered ansatz is $3ln - l + 2n$. 

    \begin{figure*}[h]
        \begin{equation*}
            \Qcircuit @C=1em @R=1em {
                & \qquad\quad\mbox{layer 0} & & & & & \qquad\qquad\quad\mbox{layer 1} & & & & & & & \qquad\qquad\mbox{layer 2} & & & & & & \\
                & \gate{R_x} & \gate{R_z} & \qw & \qw & \ctrl{1}   & \qw        & \qw        & \gate{R_x} & \gate{R_z} & \qw & \qw & \ctrl{1}   & \qw        & \qw        & \gate{R_x} & \gate{R_z} & \qw \\
                & \gate{R_x} & \gate{R_z} & \qw & \qw & \gate{R_y} & \ctrl{1}   & \qw        & \gate{R_x} & \gate{R_z} & \qw & \qw & \gate{R_y} & \ctrl{1}   & \qw        & \gate{R_x} & \gate{R_z} & \qw \\
                & \gate{R_x} & \gate{R_z} & \qw & \qw & \qw        & \gate{R_y} & \ctrl{1}   & \gate{R_x} & \gate{R_z} & \qw & \qw & \qw        & \gate{R_y} & \ctrl{1}   & \gate{R_x} & \gate{R_z} & \qw \\
                & \gate{R_x} & \gate{R_z} & \qw & \qw & \qw        & \qw        & \gate{R_y} & \gate{R_x} & \gate{R_z} & \qw & \qw & \qw        & \qw        & \gate{R_y} & \gate{R_x} & \gate{R_z} & \qw 
                \gategroup{2}{2}{5}{3}{0.75em}{--} \gategroup{2}{5}{5}{10}{0.75em}{--} \gategroup{2}{12}{5}{17}{.75em}{--}
            }
        \end{equation*}
        \caption{Example of a two-layered hardware-efficient ansatz for four qubits. Here, $R_i(\theta) = e^{-i \theta \sigma_i}$, and the rotation angles $\theta$ are omitted in the circuit.}
        \label{fig:hea}
    \end{figure*}
    
    As one may notice, all gates in the ansatz are parametrized.
    This could be helpful for applying optimization techniques such as layerwise training \cite{campos2021training, skolik2021layerwise}.
    That is, having trained an ansatz of $l$ layers, one adds another layer with the parameters initialized to zeros, and continues the training of the resultant $(l+1)$-layered ansatz.
    Similarly, in our setting one could first train the ansatz for $c$ copies of a state $\rho_\alpha$. Then add another copy and, correspondingly, add new gates to the ansatz with the parameters set to zeros, and continue the optimization for $\rho_\alpha^{\otimes (c+1)}$. 

    In Section~\ref{sec:m-qubits}, we discussed measurements of $m$ out of $n$ qubits of the transformed state $U(\boldsymbol{\theta}) \rho_\alpha U^\dagger(\boldsymbol{\theta})$ for reduction of the number of parameters $\boldsymbol{x}$ in \eqref{eq:povm_ham_m} from $2^n$ to $2^m$.
    For such a purpose, one can represent $U(\boldsymbol{\theta})$ in the form of a so-called quantum convolutional neural network (QCNN) \cite{cong2019quantum, hur2022quantum, nagano2023quantum}.
    A QCNN can be viewed as a sequence of convolutional layers, which usually consist of two-qubit gates acting on neighboring qubits, and pooling layers, in which one traces out a subset (usually, a half) of qubits. 
    At the end of the QCNN, one measures an observable of interest on the remaining qubits.
    In Fig.~\ref{fig:hea-qcnn}, we show an example of a QCNN for a state $\rho_\alpha$ of $n=8$ qubits and an observable $H$ of the form \eqref{eq:povm_ham_m} for $m=1$ measured qubit.
    
    \begin{figure*}[h]
        \begin{equation*}
            \rho_\alpha
            \begin{cases}
                \Qcircuit @C=1em @R=.5em {
                    & \qw & \multigate{1}{U_1} & \push{\rule{2.em}{0em}} &     &                    &                         &     &                       &     & \\
                    & \qw &        \ghost{U_1} &      \multigate{1}{U_5} & \qw & \multigate{2}{U_8} & \push{\rule{2.em}{0em}} &     &                       &     & \\
                    & \qw & \multigate{1}{U_2} &             \ghost{U_5} &     &       \nghost{U_8} &                         &     &                       &     & \\
                    & \qw &        \ghost{U_2} &      \multigate{1}{U_6} & \qw &        \ghost{U_8} &      \multigate{2}{U_9} & \qw & \multigate{4}{U_{10}} &     & \\
                    & \qw & \multigate{1}{U_3} &             \ghost{U_6} &     &                    &            \nghost{U_9} &     &       \nghost{U_{10}} &     & \\
                    & \qw &        \ghost{U_3} &      \multigate{1}{U_7} & \qw & \multigate{2}{U_8} &             \ghost{U_9} &     &       \nghost{U_{10}} &     & \\
                    & \qw & \multigate{1}{U_4} &             \ghost{U_7} &     &       \nghost{U_8} &                         &     &       \nghost{U_{10}} &     & \\
                    & \qw &        \ghost{U_4} &                     \qw & \qw &        \ghost{U_8} &                     \qw & \qw &        \ghost{U_{10}} & \qw & \meterB{H}                    \gategroup{1}{3}{8}{4}{0.75em}{--} \gategroup{2}{6}{8}{7}{0.75em}{--} \gategroup{4}{9}{8}{9}{0.75em}{--} %
                }
            \end{cases}
        \end{equation*}
        \caption{
            Example of an eight-qubit QCNN with measuring an observable $H$ of the form \eqref{eq:povm_ham_m} with $m=1$ measured qubit. In this implementation considered in \cite{hur2022quantum}, adjacent convolutional and pooling layers are combined into one layer shown by dashed frames, after each of which a half of the qubits is traced out. The two-qubit unitaries $U_i$ can be constructed, e.g., from one- and two-qubit gates shown in Fig.~\ref{fig:hea}.
        }
        \label{fig:hea-qcnn}
    \end{figure*}

    Despite allowing to have fewer parameters in both $\boldsymbol{x}$ and $\boldsymbol{\theta}$, it is not always advantageous to measure $m<n$ qubits, even if the corresponding observable $H$ gives $\Tr H \rho_\alpha = \alpha$.
    In Section~\ref{sec:iso}, we consider a regression task in which, if one measures $m=1$ qubit, one gets the variance $\Delta^2 H$ about twice as large as if measuring $m=2$ qubits.

    \section{Mean squared error and Cramer-Rao bound for biased estimators}
    \label{app:estimation_bias}

        In Section~\ref{sec:func_dep}, we mentioned that for a given $\rho_\alpha$ the estimator obtained from the expectation $\Tr H \rho_\alpha$ of $\alpha$ can be biased. 
        In this Appendix, following the works \cite{shettell2022quantum, sidhu2020geometric, shettell2022cryptographic, shettell2022delegated}, we derive the expressions for the MSE \eqref{eq:error_prop-biased} and cCRB \eqref{eq:ccrb-biased} for a biased estimator.

        First, let us introduce a bias into our prediction,
        \begin{equation}
          \label{eq:expectation-biased}
            \mathsf{a}(\alpha) \equiv \Tr H \rho_\alpha = \alpha + b(\alpha),
        \end{equation}
        where we have a bias $b(\alpha)$ and an observable in the form $H = \sum_i x_i \Pi_i$ with the operators $\{\Pi_i\}_i$ forming a POVM, and $x_i \in \R$.
        Denoting $p_i = \Tr \Pi_i\rho_\alpha$, we can rewrite \eqref{eq:expectation-biased} as
        \begin{equation*}
            \mathsf{a}(\alpha) = \sum_i x_i p_i.
        \end{equation*}
        Now, suppose that we perform $\mu$ measurements for obtaining this expectation value in an experiment.
        Each measurement results in an outcome $m_k = x_i$ which is observed with probability $p_i$.
        Hence, the estimation of $\mathsf{a}$ can be written as
        \begin{equation*}
            \hat{\mathsf{a}} \equiv \frac{1}{\mu} \sum_{k=1}^\mu m_k.
        \end{equation*}
        In this expression, $m_k$ are independent and identically distributed (i.i.d.) random variables, each on average giving $\langle m_k \rangle = \sum_i x_i p_i$ for all $k$. 
        Then $\langle \hat{\mathsf{a}} \rangle = \mathsf{a}$ as in \eqref{eq:expectation-biased}.
        The MSE of the estimation $\hat{\mathsf{a}}$ is
        \begin{align*}
            \Delta^2 \hat{\mathsf{a}} 
            &\equiv \big\langle (\hat{\mathsf{a}} - \mathsf{a})^2 \big\rangle
            = \langle \hat{\mathsf{a}}^2 \rangle - \langle \hat{\mathsf{a}} \rangle^2
            = \left\langle \left( \frac{1}{\mu} \sum_{k=1}^\mu m_k \right)^2 \right\rangle - \left\langle \frac{1}{\mu} \sum_{k=1}^\mu m_k \right\rangle^2 
            = \frac{1}{\mu^2} \left( \sum_{k=1}^\mu \langle m_k^2 \rangle - \sum_{k=1}^\mu \langle m_k \rangle^2 \right) \\
            &= \frac{1}{\mu^2} \left( \sum_{k=1}^\mu \sum_i x_i^2 p_i - \sum_{k=1}^\mu \left(\sum_i x_i p_i\right)^2 \right)
            = \frac{1}{\mu^2} \Big( \mu \langle H^2 \rangle - \mu \langle H \rangle^2 \Big)= \frac{1}{\mu} \Delta^2 H. \\
        \end{align*}
        In the derivation above, we used the property of i.i.d. random variables that $\langle m_k m_l \rangle = \langle m_k \rangle \langle m_l \rangle$ for $k \neq l$.

        So, we have an expression of the error of the estimation $\hat{\mathsf{a}}$.
        However, after training and testing the model, one would observe neither $\Delta^2 \hat{\mathsf{a}}$ nor $\Delta^2 \hat{\alpha}$, but rather $\big\langle (\hat{\mathsf{a}} - \alpha)^2 \big\rangle$, i.e., the error between the estimation $\hat{\mathsf{a}}$ and the labels $\alpha$ from the training set.
        One can obtain it after the following chain of transformations:
        \begin{align*}
            \Delta^2 \hat{\mathsf{a}} 
            &\equiv \big\langle (\hat{\mathsf{a}} - \mathsf{a})^2 \big\rangle
            = \big\langle \big[(\hat{\mathsf{a}} - \alpha) - (\mathsf{a} - \alpha)\big]^2 \big\rangle
            = \big\langle (\hat{\mathsf{a}} - \alpha)^2 \big\rangle - (\mathsf{a} - \alpha)^2 = \big\langle (\hat{\mathsf{a}} - \alpha)^2 \big\rangle - b^2,
        \end{align*}
        where we have inserted the bias from \eqref{eq:expectation-biased} at the end.
        Now, since we obtained $\Delta^2 \hat{\mathsf{a}} = \Delta^2 H / \mu$, we can write
        \begin{equation}
            \label{eq:mse-biased-app}
            \big\langle (\hat{\mathsf{a}} - \alpha)^2 \big\rangle = \frac{\Delta^2 H}{\mu} + b^2,
        \end{equation}
        which is the equation \eqref{eq:error_prop-biased}. Finally, recalling the error propagation formula \eqref{eq:estimation_variance} and qCRB \eqref{eq:cramer-rao_quantum}, as well as noticing that 
        \begin{equation*}
            \frac{\Delta^2 \hat{\mathsf{a}}}{\big|\partial_\alpha \langle H \rangle\big|^2} = \frac{\Delta^2 H}{\mu \big|\partial_\alpha \langle H \rangle\big|^2} = \Delta^2 \hat{\alpha} \geqslant \frac{1}{\mu I_q},
        \end{equation*}
        we arrive to the desired inequality \eqref{eq:ccrb-biased}:
        \begin{equation}
            \label{eq:ccrb-biased-app}
            \big\langle (\hat{\mathsf{a}} - \alpha)^2 \big\rangle \geqslant \frac{\big|\partial_\alpha \langle H \rangle\big|^2}{\mu I_q} + b^2.
        \end{equation}

        Interestingly, combining \eqref{eq:mse-biased-app} and \eqref{eq:ccrb-biased-app} and noticing that $\partial_\alpha b = \partial_\alpha \langle H \rangle - 1$, one can also obtain
        \begin{equation}
            \frac{\Delta^2 H}{\big|1 + \partial_\alpha b\big|^2} \geqslant \frac{1}{I_q},
        \end{equation}
        which can be used for estimating the Fisher information.
        Indeed, having trained the observable $H$, one obtains the variances $\Delta^2 H$ via measurements according to \eqref{eq:observable_variance_probs}, and the bias $b = \Tr H \rho_\alpha + \alpha$ is observed on the labels from the training set $\mathcal{T} = \{\rho_{\alpha_j}, \alpha_j\}_{j=1}^T$.
        Therefore, if the training set size $T$ is sufficiently large, and the labels $\alpha_j$ are distributed uniformly, then one could try to approximate $\partial_\alpha b$ via finite differences.
        As an instance, for $1 < j < T$ and assuming $\alpha_j < \alpha_{j+1}$, the central difference approximation would be
        \begin{equation*}
            \partial_{\alpha} b(\alpha)\Bigr|_{\alpha=\alpha_j} = \frac{b(\alpha_{j+1}) - b(\alpha_{j-1})}{\alpha_{j+1} - \alpha_{j-1}} + O\Big((\alpha_{j+1} - \alpha_{j-1})^2\Big).
        \end{equation*}

\section{Additional results}
\label{app:add_res}

    Here we provide additional results of numerical application of our method for predicting the negativity with various conditions (e.g., using fewer copies of the labeled state), and for solving other regression problems, such as predicting the angle of a Pauli rotation, and predicting the strength of transverse field for the Ising Hamiltonian.  

    \subsection{Accounting for non-entangled states while predicting the negativity of isotropic states}
    \label{sec:iso-c}

\begin{figure*}[h]
            \centering
            \includegraphics[width=.4725\textwidth]{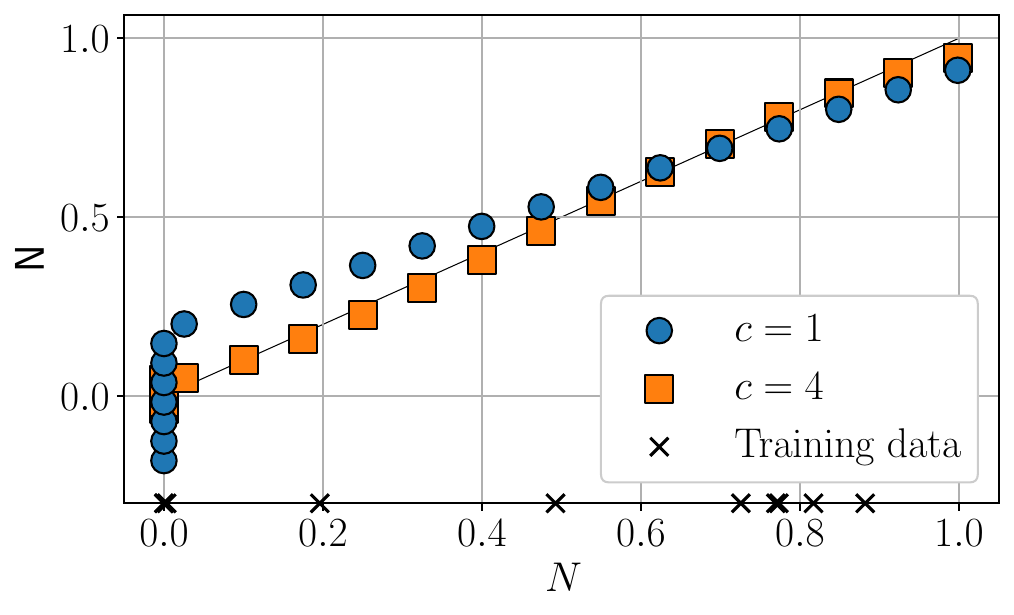}
            \includegraphics[width=.495\textwidth]{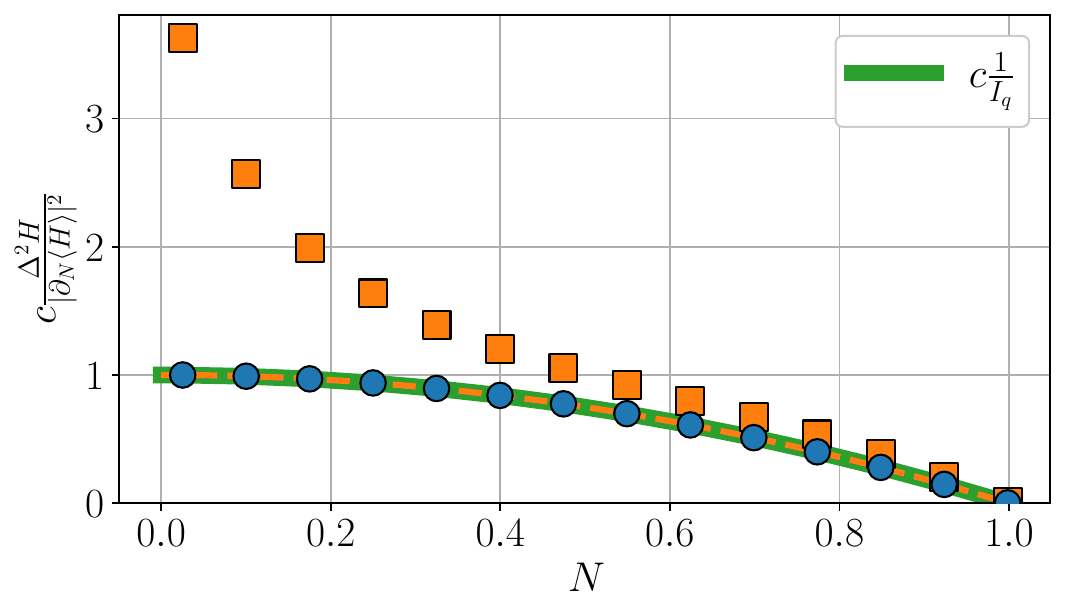}
            \caption{
                Left: Predicted $\mathsf{N}$ vs. true $N$ negativity of the isotropic state \eqref{eq:iso} estimated for different number of copies $c$ processed simultaneously. 
                Right: Variance \eqref{eq:var-qcrb} together with cCRB (dashed lines) and qCRB (solid green line), all scaled with $c$; the points with $q \leqslant 1/3$ are excluded from this plot, as they give $\partial_N \langle H \rangle = 0$.
            }
            \label{fig:iso_ent}
        \end{figure*}

        In Section~\ref{sec:iso}, we studied the prediction of the negativity $N$ of isotropic states $\rho_q$ with the restriction that there are no separable states with $q \leqslant 1/3$ in the training set.
        Importantly we would like to learn to predict not the coefficient $q$ in \eqref{eq:iso}, but the negativity $N$, which depends on $q$ non-linearly.
        In this Appendix, we train our model on the sets $\mathcal{T} = \big\{\rho_{q_j}^{\otimes c}, N_j \big\}_{j=1}^{10}$, where $q_j \in [0, 1]$ and $c\in \{1, 4\}$.
        The results are shown in Fig.~\ref{fig:iso_ent}.

        As can be seen, the model struggles to predict the negativities for $q \leqslant 1/3$, for which $N = 0$. 
        However, the more copies $c$ we process simultaneously, the more accurate our predictions for $N$ become.
        A drawback is that in this case the observable \eqref{eq:povm_ham_m} requires the measurement of all $m=2c$ qubits, for which $2^{m}$ parameters $\boldsymbol{x}$ have to be varied. 

        Additionally in Fig.~\ref{fig:iso_ent}, we show the variance of the trained observable $H$ and the Cramer-Rao bounds \eqref{eq:cramer-rao_quantum}.
        As can be seen, the quantum bound is saturated for a single copy of $\rho_N$.
        For both $c=1$ and $c=4$, our model finds the ansatz angles $\boldsymbol{\theta}$ which provide $I_c(\Pi(\boldsymbol{\theta}), \rho_N) = I_q(\rho_N)$, but the parameters $\boldsymbol{x}$ do not give the desired variance of $H(\boldsymbol{x}, \boldsymbol{\theta})$ for $c=4$.
        Note that $I_q(\rho_N^{\otimes c}) = c I_q(\rho_N)$, so the data in the plots are rescaled correspondingly.

    \subsection{Prediction of the negativity of random mixed states with fewer copies}
    \label{app:random_mixed_c}
    
        In Section~\ref{sec:ent-mixed}, we applied our method for predicting the negativity \eqref{eq:neg} of random mixed two-qubit states.
        To train the model, we processed $c=4$ copies of the labeled state $\rho_N$ simultaneously.
        Here we show how the performance of our method changes with processing a fewer number of copies. 
        To train the model, as before, we generated a training set of $T = 1000$ random states with negativities evenly distributed on $[0, 1]$.
        Additionally, having trained the model with $c$ copies, we use the optimal parameters $\boldsymbol{x}^*, \boldsymbol{\theta}^*$ as the initial ones for training the model with $c+1$ copies.
        
\begin{figure*}[h]
            \centering
            \includegraphics[width=.325\textwidth]{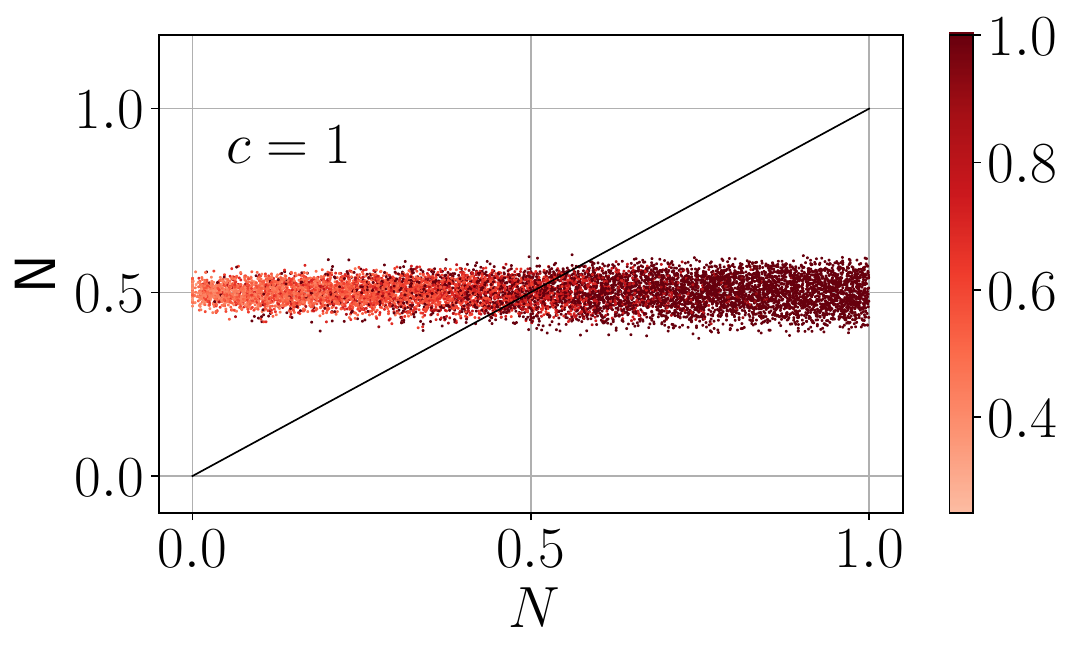}
            \includegraphics[width=.325\textwidth]{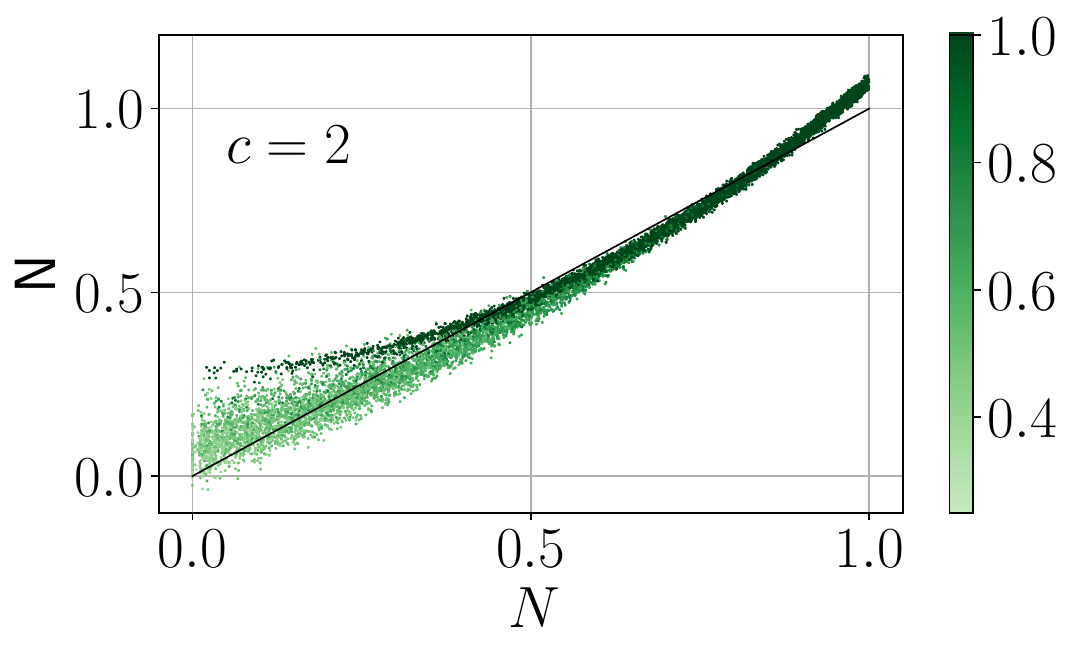}
            \includegraphics[width=.325\textwidth]{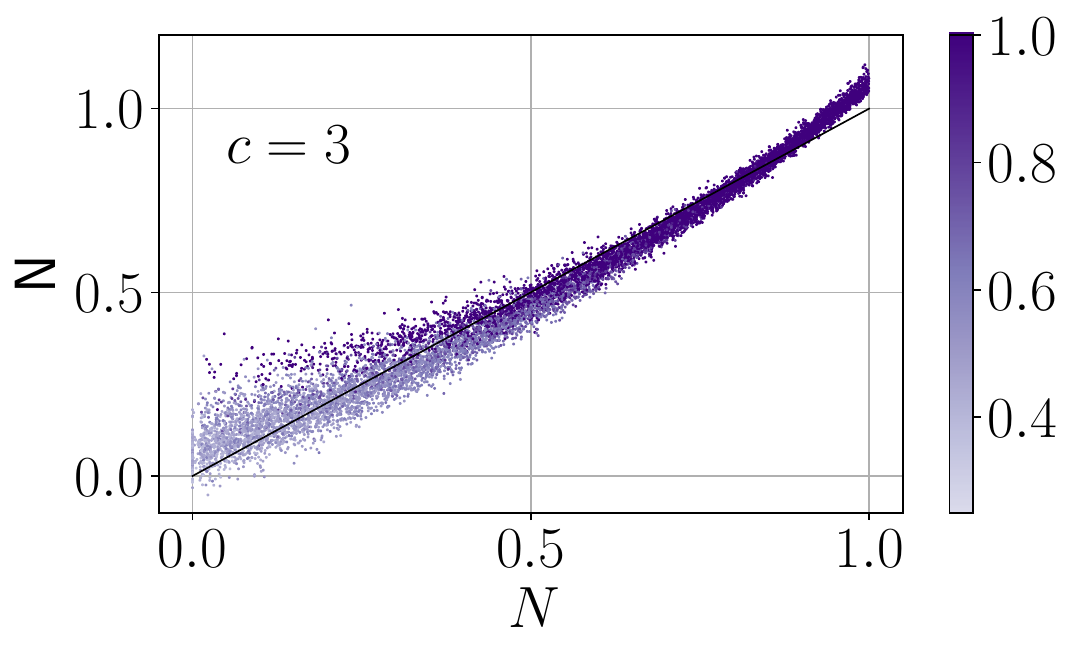}
            \caption{
                Predicted $\mathsf{N}$ vs. true $N$ negativity of $10^4$ random mixed states using an observable trained on $c=1$ (left), $c=2$ (center), and $c=3$ (right) copies of the labeled states.
                The color of points indicates the purity of the corresponding states.
            }
            \label{fig:random_mixed_c}
        \end{figure*}

        In Fig.~\ref{fig:random_mixed_c} the results of predicting the negativity of mixed states for $c \in \{1, 2, 3\}$ are shown.
        The model trained with only a single copy does not allow to predict anything, which is no surprise in light of the results presented in \cite{larocca2022group}. 
        In contrast, with two copies, the prediction accuracy is much higher which is in line with our reasoning in Section~\ref{sec:ent-pure}. 
        With three copies, the performance does not improve much in comparison to the case with two copies.
        Finally, with $c=4$ copies, as we show in Section~\ref{sec:ent-mixed}, the prediction accuracy increases significantly.

    \subsection{Prediction of the squared negativity of random mixed states}
    \label{app:random_mixed_c_sq}

\begin{figure*}[h]
            \centering
            \includegraphics[width=.325\textwidth]{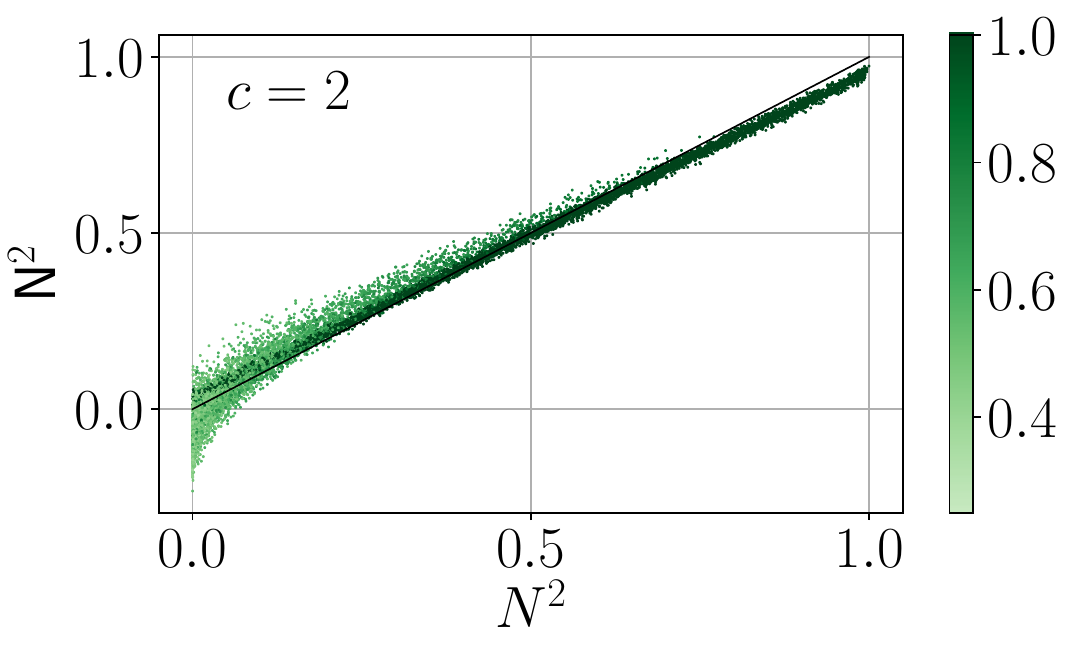}
            \includegraphics[width=.325\textwidth]{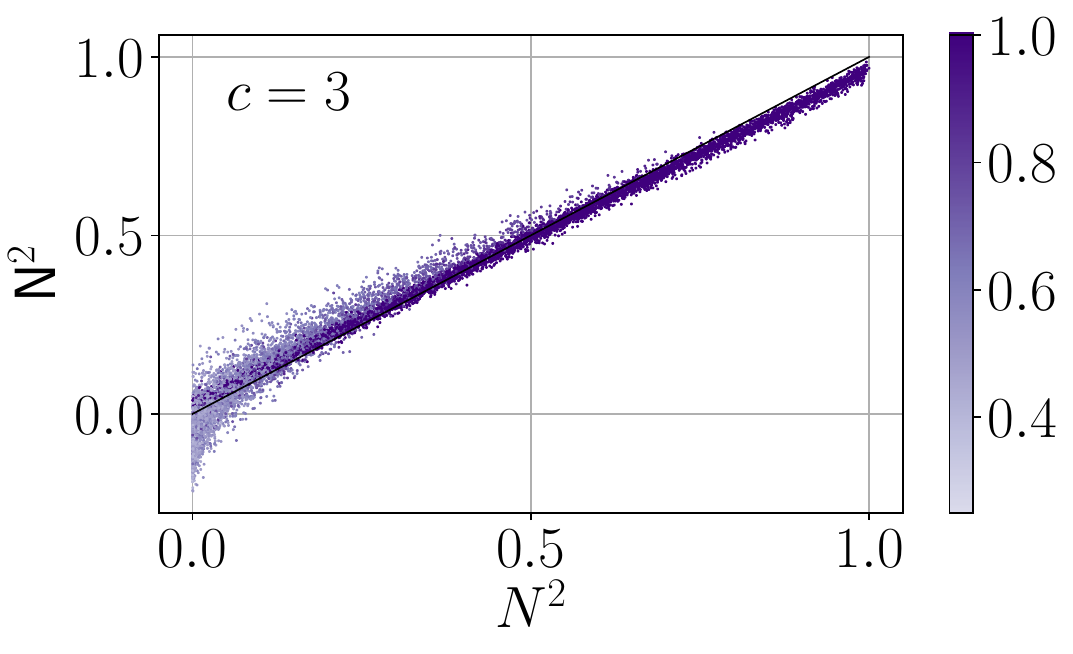}
            \includegraphics[width=.325\textwidth]{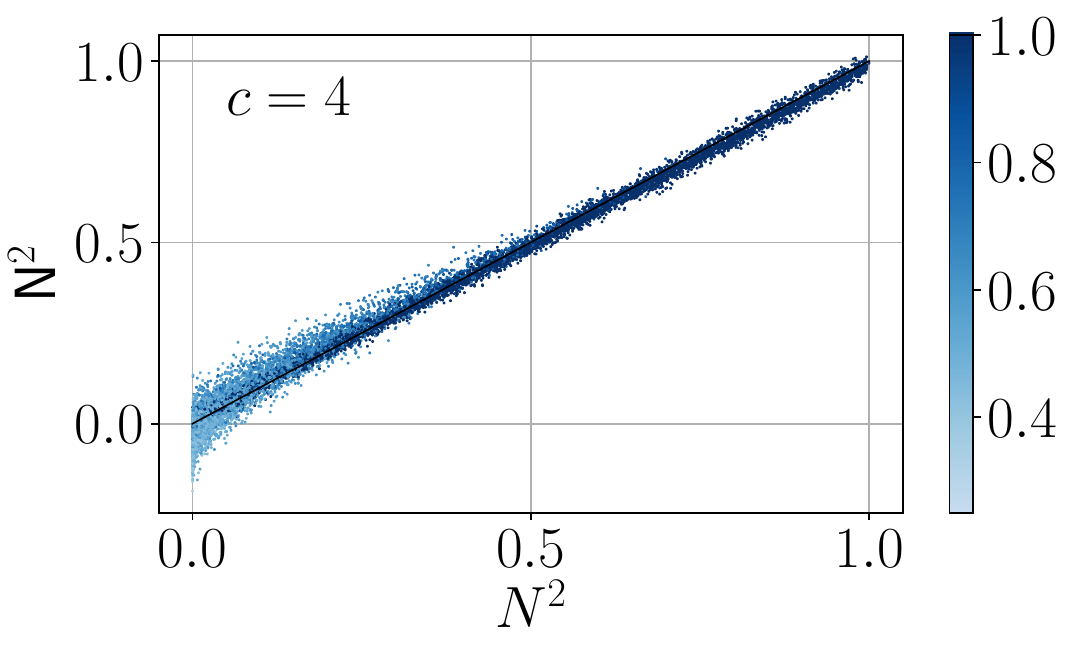}
            \caption{
                Predicted $\mathsf{N}^2$ vs. true $N^2$ negativity of $10^4$ random mixed states using an observable trained on $c=2$ (left), $c=3$ (center), and $c=4$ (right) copies of the labeled states.
                The color of points indicates the purity of the corresponding states.
            }
            \label{fig:random_mixed_c_sq}
        \end{figure*}

        In Section~\ref{sec:ent-pure} we have shown that one can predict the squared negativity $N^2$ of \textit{pure} states using our method. 
        In this Appendix, we test the performance of the method in predicting $N^2$ as well, but for random \textit{mixed} states.
        In Fig.~\ref{fig:random_mixed_c_sq}, we show the prediction accuracy for this case for $c \in \{2,3,4\}$ copies of the labeled state.
        Interestingly, while with $c=2$ copies the prediction accuracy is quite high, the performance does not increase much for $c=3$.
        In contrast, with $c=4$, the prediction of $N^2$ is very precise, especially for more pure states.
        With $c=1$ copy, we obtained the results similar to the ones presented in Fig.~\ref{fig:random_mixed_c} (left) (not shown here).
        
    \subsection{Compensation of a bias}
    \label{app:bias_comp}

        As noted in Section~\ref{sec:func_dep}, $\Tr H \rho_\alpha$ generally evaluates to a function $\mathsf{a}(\alpha)$ instead of $\alpha$ itself. 
        There is a bias $b(\alpha) = \mathsf{a}(\alpha) - \alpha$ which one can try to compensate.
        Indeed, after the training, we can observe the bias $b(\alpha_j)$ for $(\rho_{\alpha_j}, \alpha_j) \in \mathcal{T}$.
        One can plot the dependence of $b$ on the obtained predictions $ \mathsf{a}$. 
        If this dependence is monotonic, the data points can be fitted with some function (e.g., polynomial of some degree), which we denote $\tilde{b}( \mathsf{a})$.
        This function can then be used for processing the predictions $\mathsf{a}$ for unseen data.

\begin{figure*}[h]
            \centering
            \includegraphics[width=.495\textwidth]{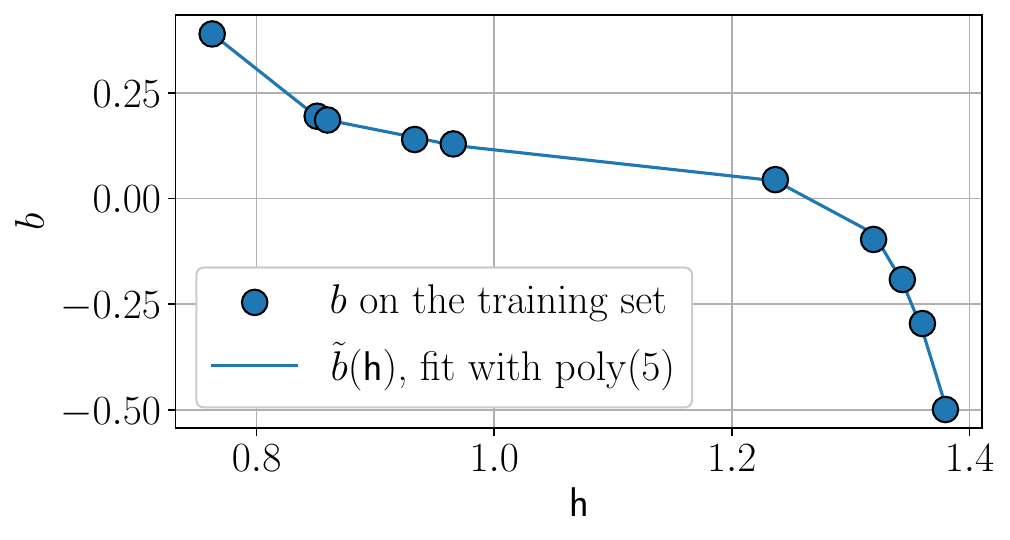}
            \includegraphics[width=.475\textwidth]{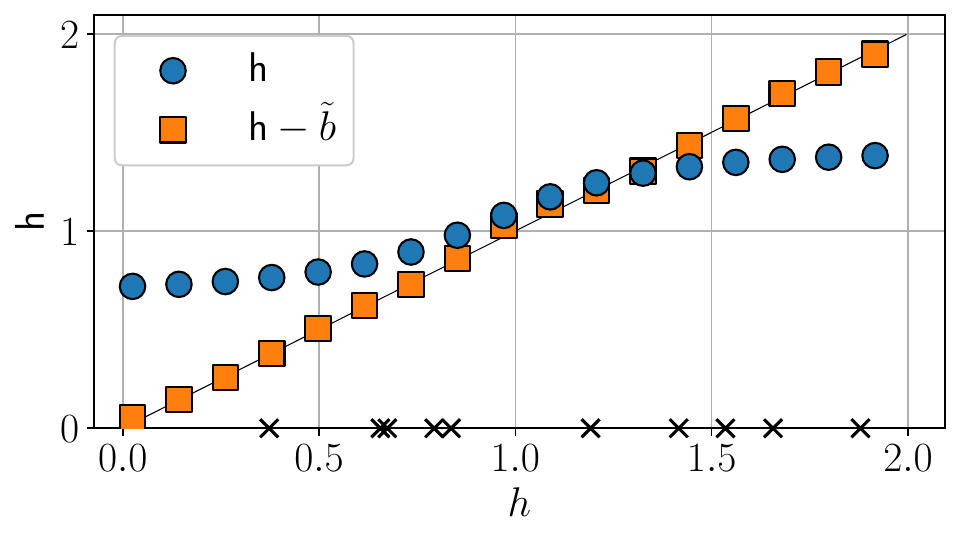}
            \caption{
                Compensating the bias of the prediction of the transverse field of the Ising Hamiltonian \eqref{eq:ising} of $n=8$ qubits.
                The observable $H$ is trained on a set $\mathcal{T} = \big\{|\psi_{j}\rangle, h_j \big\}_{j=1}^{10}$ with $\ket{\psi_j}$ being the ground states of \eqref{eq:ising}, the fields $h_j$ are generated randomly, and the weights in \eqref{eq:ls_min_variational_mod} are set to $w_\mathrm{ls} = w_\mathrm{var} = 1$.
                Left: Dependence of the bias on the predictions for the training set, fitted with a function $\tilde{b}$ being a polynomial of the degree 5.
                Right: Comparison of the ``direct'' predictions $\mathsf{h}$ with the bias-compensated ones $\mathsf{h} - \tilde{b}$ on the testing set.
            }
            \label{fig:ising-bias}
        \end{figure*}

        In Fig.~\ref{fig:ising-bias}, we show the results of testing this bias compensation method for the case considered in Section~\ref{sec:ising}, where we predicted the transverse field $h$ of the Ising Hamiltonian with the variance weight $w_\mathrm{var}=1$.
        To fit the prediction $\mathsf{h}$, we used a polynomial of the degree 5.
        As can be seen, the values $\mathsf{h} - \tilde{b}$ provide very accurate predictions.

    \subsection{Measuring fewer qubits when predicting the negativity of random states}
    \label{app:neg-rand-mixed-m}

        In Section~\ref{sec:ent-mixed}, we studied the prediction of the negativity of random mixed states. 
        We looked for an observable in the form \eqref{eq:povm_ham_m} with measuring all $m=8$ of $c=4$ copies of two-qubit states.
        In this Appendix, we numerically study the performance and the training complexity of our method in this task with smaller $m$.
        For our simulations, we set $w_\mathrm{var}=10^{-4}$, use the hardware-efficient ansatz of $l=5$ layers, and generate a training set of $T=1000$ data points. 

        To speed up our calculations, we do the following.
        We start the minimization  with $m=1$ and random initial parameters $(\boldsymbol{x}, \boldsymbol{\theta})$.
        Once the optimization is terminated, we use the optimized parameters $(\boldsymbol{x}^*_1, \boldsymbol{\theta}^*_1)$ found for $m=1$ as the initial ones for $m=2$.
        Having obtained $(\boldsymbol{x}^*_2, \boldsymbol{\theta}^*_2)$ for $m=2$, we use them to initialize the optimization for $m=3$.
        This way, we keep increasing the number of measured qubits until $m=8$.
        To quantify the performance of our method applied with different $m$, 
        we consider the MSE
        \begin{equation}
            \label{eq:testing_error}
            E(\boldsymbol{x}^*_m, \boldsymbol{\theta}^*_m) = \frac{1}{V}\sum_{j=1}^{V} \Big(N_j - \mathsf{N}_j(\boldsymbol{x}^*_m, \boldsymbol{\theta}^*_m, \rho_{N_j})\Big)^2,
        \end{equation}
        where 
        $\mathsf{N}_j(\boldsymbol{x}^*_m, \boldsymbol{\theta}^*_m, \rho_{N_j}) = \Tr H(\boldsymbol{x}^*_m, \boldsymbol{\theta}^*_m) \rho_{N_j}$ are the expectations calculated for the optimized observable $H(\boldsymbol{x}^*_m, \boldsymbol{\theta}^*_m)$ measured in a state from a testing set of the size $V = 10^4$.

\begin{figure*}[h]
            \centering
            \includegraphics[width=.485\textwidth]{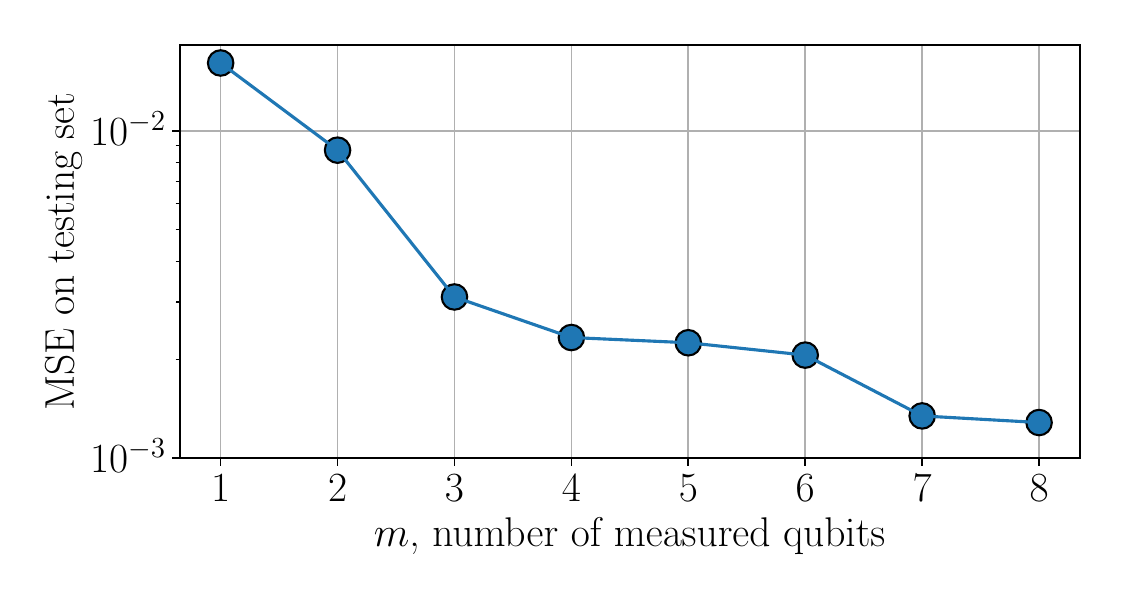}
            \includegraphics[width=.495\textwidth]{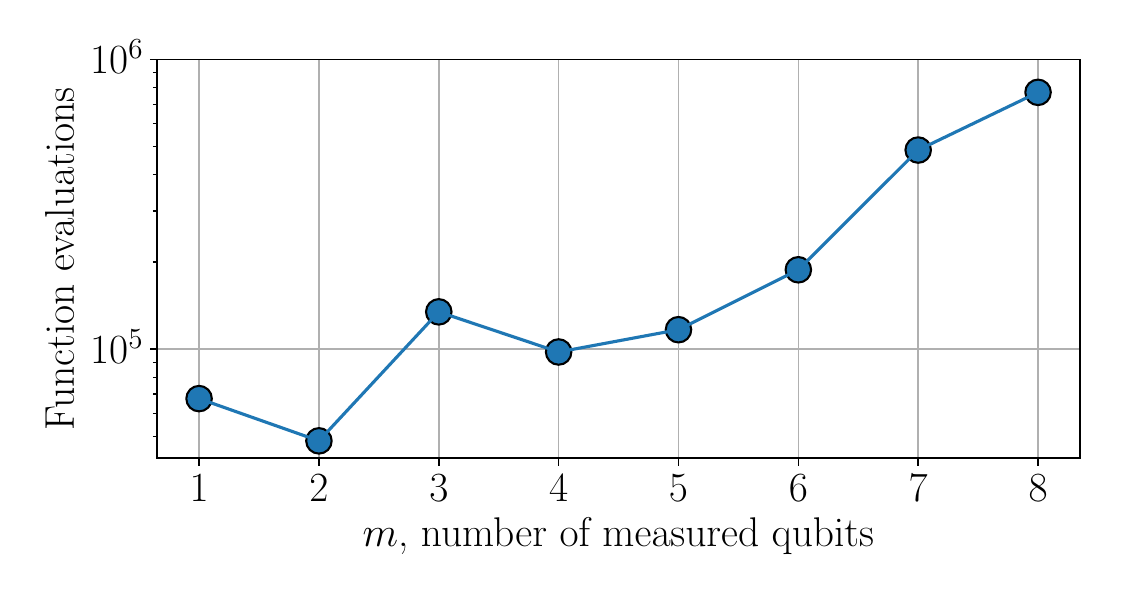}
            \caption{
                Left: Testing error \eqref{eq:testing_error}  vs. the number of measured qubits $m$ as per \eqref{eq:povm_ham_m}.
                Right: Number of evaluations of the cost function performed by the BFGS algorithm vs. the number of measured qubits $m$; the number of function calls is \textit{not} cumulative (i.e., it is obtained \textit{individually} for each $m$). 
            }
            \label{fig:ent_rand_mixed-cost-m}
        \end{figure*}

        In Fig.~\ref{fig:ent_rand_mixed-cost-m}, we plot the dependence of the error \eqref{eq:testing_error} on $m$.
        Additionally in this Figure, we show the number of times the BFGS optimizer called the cost function in \eqref{eq:ls_min_variational_mod}.
        In our numerical experiments, the optimizer required at most about 2000 iterations to terminate for all $m$.
        However, at each iteration the BFGS algorithm evaluates the cost function several times (e.g., for computing the gradient).
        With each increment of $m$, we nearly double the number of trainable parameters, as well as the number of function evaluations performed by the optimizer.

\begin{figure*}[h]
            \centering
            \includegraphics[width=.495\textwidth]{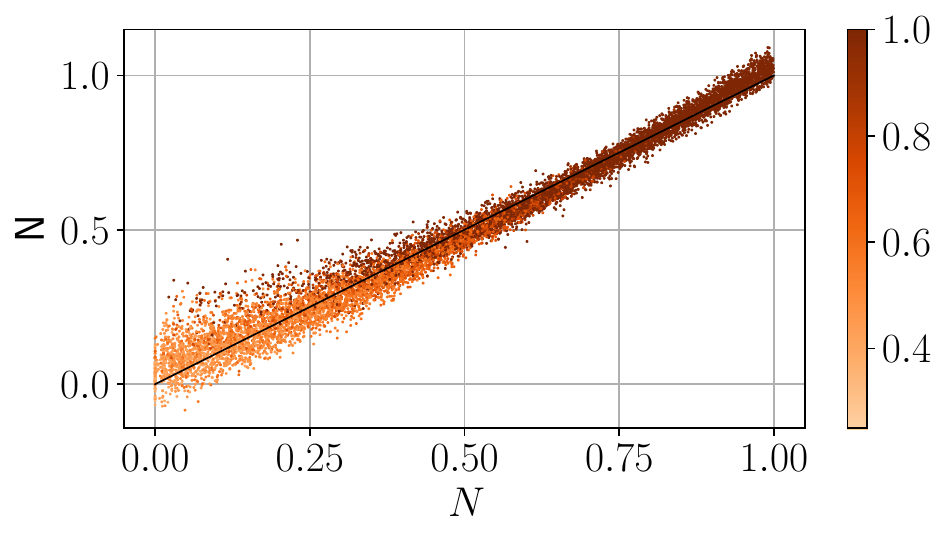}
            \includegraphics[width=.495\textwidth]{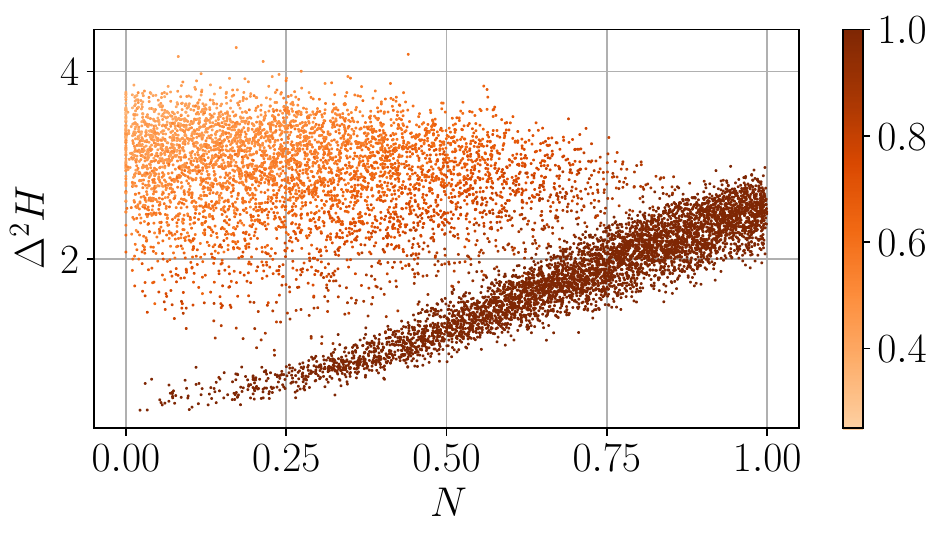}
            \caption{
                Left: Predicted negativity $\mathsf{N}$ of $10^4$ random mixed states vs. the true negativity $N$ for $c=4$ copies and the trained observable $H$ on $m=6$ qubits.
                Right: Variance of the trained observable $H$.
                The color of points indicates the purity of the corresponding states.
            }
            \label{fig:ent_rand_mixed-m=4}
        \end{figure*}

        Therefore, for large system sizes, it is prohibitive to use the proposed method with measuring all the qubits.
        In this particular case, however, each additional measured qubit may contribute to the gain of prediction accuracy, as one may notice from Fig.~\ref{fig:ent_rand_mixed-cost-m}.
        To illustrate this, in Fig.~\ref{fig:ent_rand_mixed-m=4} we show the prediction accuracy and the variance for an optimized observable on $m=6$ qubits.
        Although one may find such results satisfactory, as there is an improvement over the two-copy case (see Fig.~\ref{fig:random_mixed_c}), the optimized observable on $m=8$ qubits shows a better performance (see Fig.~\ref{fig:ent_rand_mixed}).

    \subsection{Training complexity and dependence on the training set size}
    \label{app:neg-rand-mixed-T}
        In the most of our case studies, it was enough to train the model on a data set of a quite modest size $T$.
        One exception is the prediction of the negativity of mixed states considered in Section~\ref{sec:ent-mixed}, for which we used a training set of the size $T=1000$.
        In this Appendix, we numerically study the dependence of the performance of our method on the training set size.
        Such numerical experiment may consume a considerable amount of time to perform, so we restrict the number of iterations of the BFGS algorithm to 1500.
        Additionally, to speed up our simulation, we execute it as follows.
        First, we generate a training set $\mathcal{T} = \big\{\rho_j^{\otimes 4}, N_j \big\}_{j=1}^{1000}$.
        Then, we sample $\mathcal{T}_1 \equiv \big\{\rho_j^{\otimes 4}, N_j \big\}_{j=1}^{100} \subset \mathcal{T}$ and train our model on this training set of $T=100$ data points.
        Once the optimization for $\mathcal{T}_{1}$ has terminated, we use the obtained optimized parameters $(\boldsymbol{x}^*_1, \boldsymbol{\theta}^*_1)$ as the initial ones for training the model on the set $\mathcal{T}_{2} \equiv \big\{\rho_j^{\otimes 4}, N_j \big\}_{j=1}^{200} = \mathcal{T}_{1} \cup \big\{\rho_j^{\otimes 4}, N_j \big\}_{j=101}^{200}$ and get $(\boldsymbol{x}^*_2, \boldsymbol{\theta}^*_2)$.
        We keep training the model this way until reaching $\mathcal{T}_{10} = \mathcal{T}$.

\begin{figure*}[h]
            \centering
            \includegraphics[width=.49\textwidth]{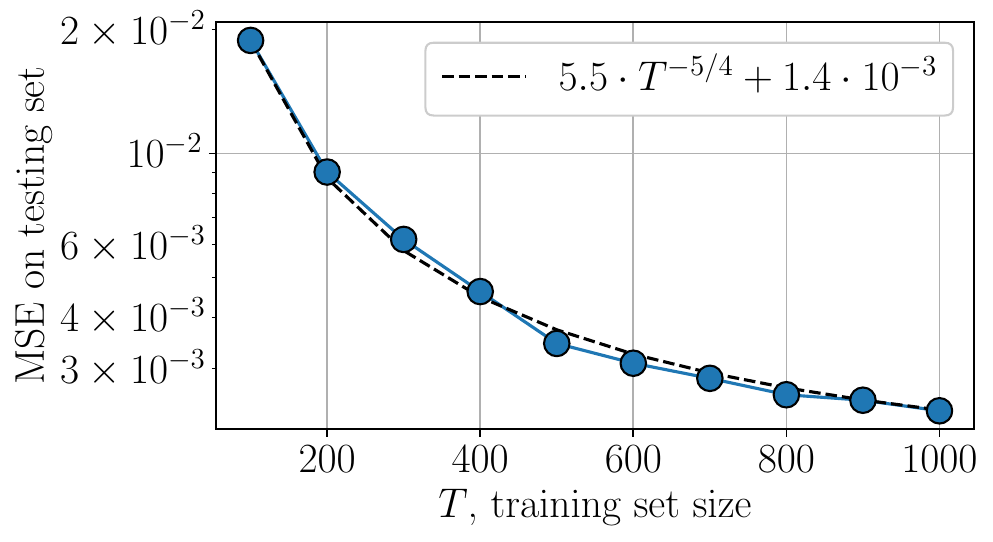}
            \includegraphics[width=.495\textwidth]{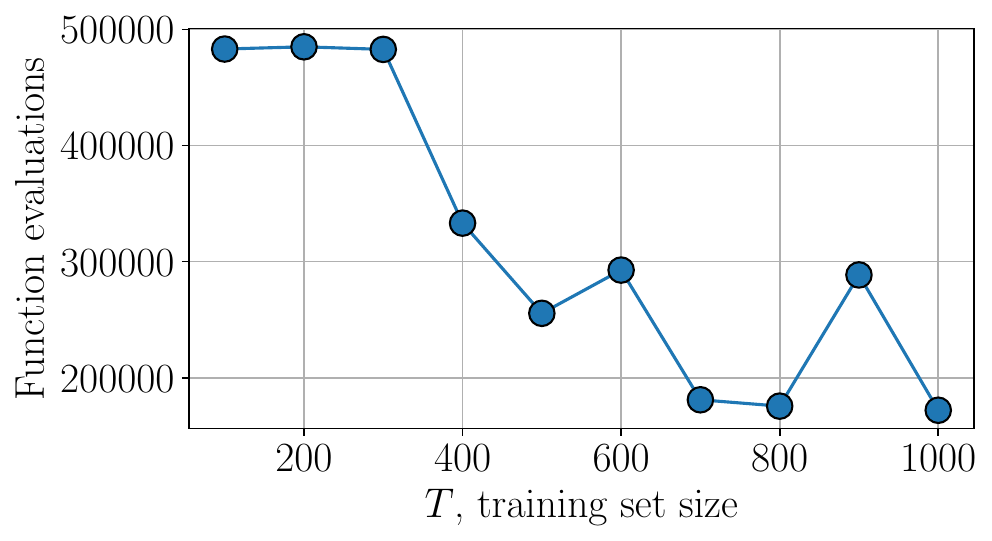}
            \caption{
                Left: Testing error \eqref{eq:testing_error} vs. the training set size $T$ in the task of predicting the negativity of random two-qubit mixed states.
                Right: Number of evaluations of the cost function in \eqref{eq:ls_min_variational_mod} performed by the BFGS algorithm vs. the training set size $T$.
                The maximum number of iterations for the BFGS algorithm was set to 1500; note that at each iteration, the algorithm may call the cost function several times.  
            }
            \label{fig:ent_rand_mixed-training}
        \end{figure*}

        To characterize the performance of our method with different $T$, we again consider the MSE \eqref{eq:testing_error} evaluated on a testing set of $V=10^4$ states for each $(\boldsymbol{x}^*_k, \boldsymbol{\theta}^*_k)$.
        As we show in Fig.~\ref{fig:ent_rand_mixed-training}, the error behaves as $T^{-5/4}$, so it expectedly decreases with the growth of the training  set size $T$.
        Additionally in Fig.~\ref{fig:ent_rand_mixed-training}, we plot the number of evaluations of the cost function in \eqref{eq:ls_min_variational_mod} the BFGS algorithm had to perform before termination.
        Note that at each iteration, the BFGS algorithm calls the cost function several times, e.g., for calculating the gradient or performing the line search.
        For $T=100, 200, 300$, the optimization stopped upon reaching the maximum number of iterations 1500 we set; 
        for the rest values of $T$, the number of cost function evaluations tends to decrease with $T$.
        Note however that the growth of $T$ leads to a bigger number of computed expectations in \eqref{eq:ls_min_variational_mod}, i.e. it may increase the simulation time.

    \subsection{Training complexity of the prediction of the transverse field of the Ising Hamiltonian}
    \label{sec:ising-training}
    
        In Section~\ref{sec:ising}, we studied the prediction of the transverse field of the eight-qubit Ising Hamiltonian \eqref{eq:ising} with different variance weights $w_\mathrm{var} \in \{1, 10^{-2}, 10^{-4}\}$.
        Here, we discuss the training complexity of this procedure.
        Recall that in the main text we performed the simulation for finding an observable in the form \eqref{eq:povm_ham_m} with $m=4$ measured qubits and the hardware-efficient ansatz of $l=5$ layers.

\begin{figure*}[h]
            \centering
            \includegraphics[width=.6\textwidth]{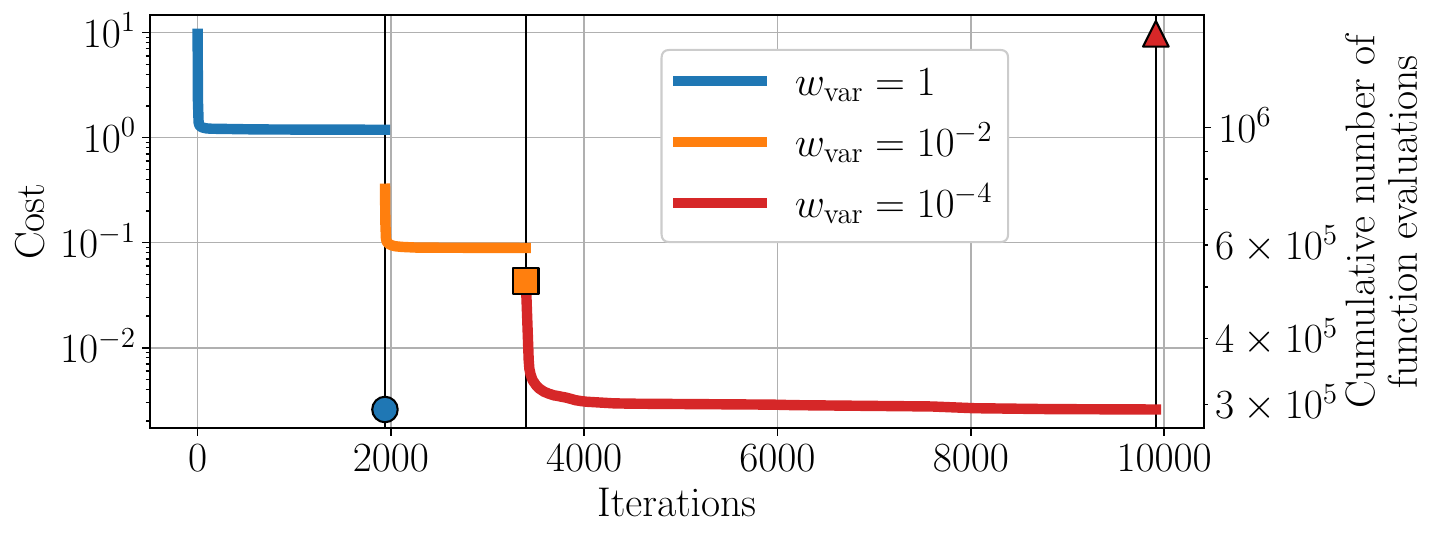}
            \caption{
                Value of the cost function in \eqref{eq:ls_min_variational_mod} vs. iteration of the BFGS algorithm when training an observable for predicting the transverse field of the Ising Hamiltonian using an ansatz of $l=5$ layers and measuring $m=4$ qubits as per \eqref{eq:povm_ham_m}.
                The markers and the right vertical axis show the \textit{cumulative} number of times the BFGS algorithm called the cost function in \eqref{eq:ls_min_variational_mod}.
                The data are gathered from the numerical simulation for Fig.~\ref{fig:ising-w_var}.
            }
            \label{fig:ising-training}
        \end{figure*}

        In Fig.~\ref{fig:ising-training}, we depict the optimization process for all the three weights $w_\mathrm{var}$.
        To speed up our calculations, we first trained the observable with $w_\mathrm{var} = 1$ starting from a random initialization of the trainable parameters, then we used the found optimal parameters $(\boldsymbol{x}^*, \boldsymbol{\theta}^*)$ as the initial ones for the optimization with $w_\mathrm{var} = 10^{-2}$, and then did the same for $w_\mathrm{var} = 10^{-4}$.
        As one sees in Fig.~\ref{fig:ising-training}, the BFGS algorithm needs several thousands of iterations to converge.
        However, in contrast to the prediction of the negativity of random states discussed in Section~\ref{sec:ent-mixed}, here we measure only $m=4$ qubits out of $n=8$.
        Despite that this significantly decreases the number of variational parameters, we had to increase the number of layers to $l=5$.
        Note that it is more advantageous to consider larger $l$ rather than larger $m$, as the number of parameters is linear in $l$ and exponential in $m$.
        In our simulations, we also observe that for all considered weights $w_\mathrm{var}$ the number of function evaluations per iteration is approximately 152, which is slightly more than the number of trainable parameters.
    
\begin{figure*}[h]
            \centering
            \includegraphics[width=.316\textwidth]{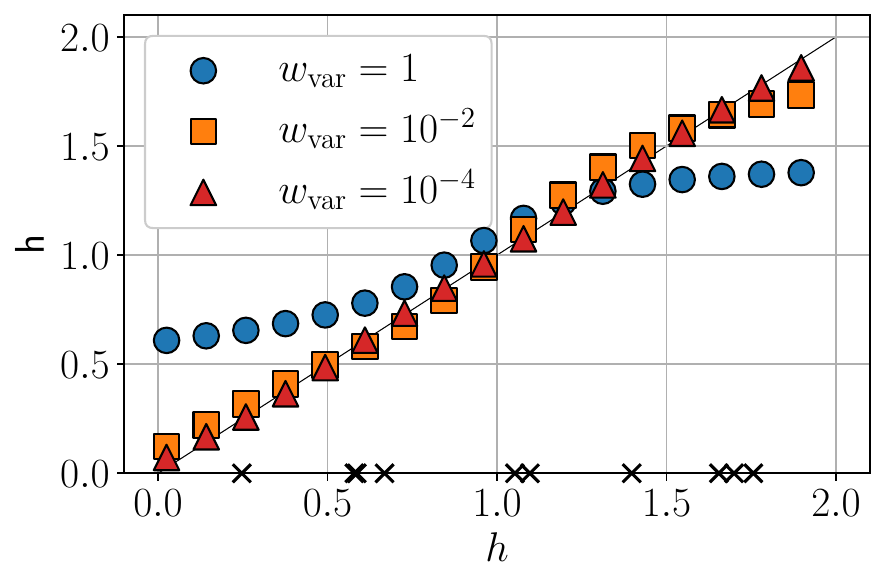}
            \includegraphics[width=.329\textwidth]{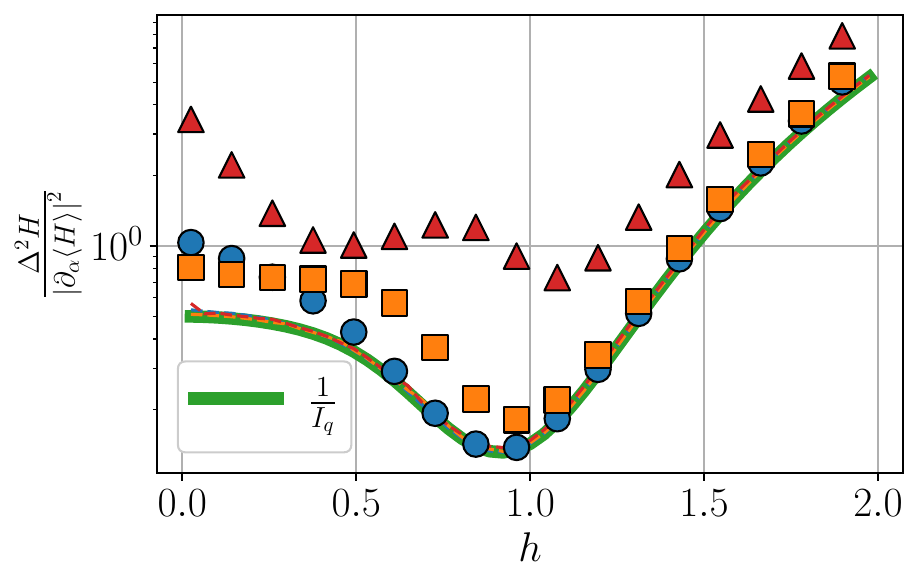}
            \includegraphics[width=.316\textwidth]{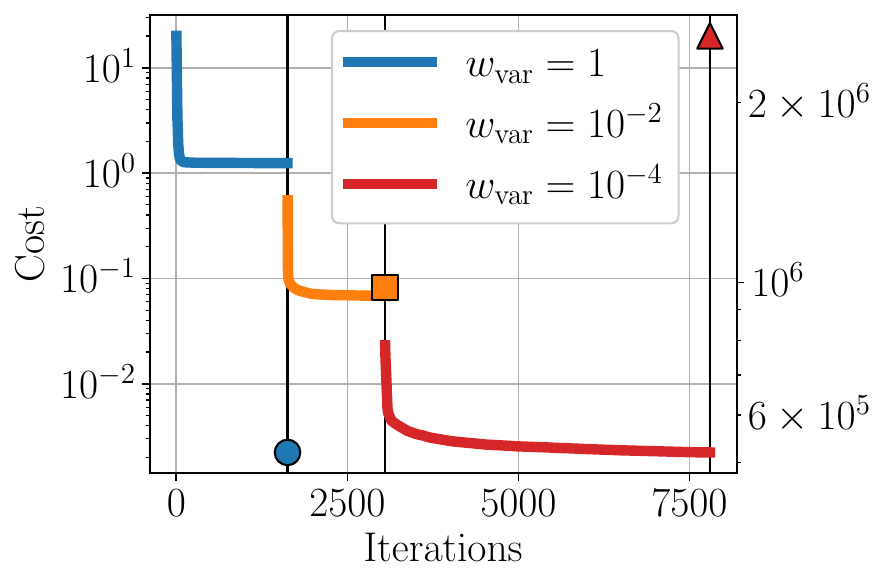}
            \caption{
                Predicting the transverse field $h$ for an 8-qubit Ising Hamiltonian \eqref{eq:ising} using an ansatz of $l=2$ layers and measuring $m=8$ qubits as per \eqref{eq:povm_ham_m}.
                Left: Predicted $\mathsf{h}$ vs. true $h$ transverse field for different weights $w_\mathrm{var}$ in \eqref{eq:ls_min_variational_mod}; the black crosses on the $x$-axis indicate the training data. 
                Center: Variance \eqref{eq:var-qcrb} vs. $h$; the dashed lines indicate cCRB \eqref{eq:cramer-rao_classical}. 
                Right: Value of the cost function in \eqref{eq:ls_min_variational_mod} vs. iteration of the BFGS algorithm during the training; the markers and the right vertical axis show \textit{cumulative} number of times the BFGS algorithm called the cost function in \eqref{eq:ls_min_variational_mod}.
            }
            \label{fig:m=8-ising-training}
        \end{figure*}

        As mentioned in Section~\ref{sec:ising}, one can obtain better results using an ansatz with fewer layers $l=2$ but measuring all $m=8$ qubits.
        The simulation results with these settings are shown in Fig.~\ref{fig:m=8-ising-training}.
        Comparing it with the case of $m=4$ and $l=5$ in Fig.~\ref{fig:ising-w_var}, one notices an improvement in terms of the variance, especially for $w_\mathrm{var}=1$.
        Moreover, we needed even less number of iterations of the BFGS algorithm to converge.
        However, one should rather look at the number of evaluations of the cost function in \eqref{eq:ls_min_variational_mod}, which is greater for the case of $m=8$.
        Indeed, the BFGS algorithm calls the function for calculating the gradient, and the number of calls therefore increases exponentially with $m$. 

        One may also notice that with $m=8$ and $l=2$ we obtain an observable such that $I_c(\Pi, \rho_\alpha) \approx I_q(\rho_\alpha)$. It shows that the ansatz is expressive enough for obtaining efficient eigenprojectors for all the weights $w_\mathrm{var}$ considered.
        This is in contrast to the case of $m=4$ with $l=5$ in Fig.~\ref{fig:ising-w_var}, where the equality between classical and quantum FI was achieved only on a smaller interval of $h$.
        Thus, the number of varied eigenvalues may play an important role in the proposed method, at least for achieving lower variance in the considered regression problem.

    \subsection{Dependence on the number of measurements $\mu$}
    \label{sec:shots}
    
        In the main text of this work, we have performed our numerical experiments with the assumption that one can exactly compute the expectations $\mathsf{a} = \Tr H \rho_\alpha$. 
        In reality, however, one can have only an estimate $\hat{\mathsf{a}}$ obtained from a $\mu$-shot measurement experiment.
        In this Appendix, we numerically study the performance of our method when $\mu$ is finite.
        Namely, during the training, whenever the minimization algorithm calls the cost function in \eqref{eq:ls_min_variational_mod}, for each data point $\rho_{\alpha_j}$ of the training set the expectation $\mathsf{a}_j = \Tr H(\boldsymbol{x},\boldsymbol{\theta}) \rho_{\alpha_j}$ is estimated as $\hat{\mathsf{a}}_j = \frac{1}{\mu}\sum_i x_i \mu_i$, where $\mu_i$ is the number of times the $i$th measurement outcome is observed.
        For the numerical experiments presented in this section, we set the variance weight in \eqref{eq:ls_min_variational_mod} to $w_\mathrm{var}=10^{-4}$.

        To study the dependence of estimation accuracy on $\mu$, here we consider two problems from the main text.
        Namely, an easier one of estimating the parameter of the amplitude damping channel (see Section~\ref{sec:ad_chan-res}), and a harder one of estimating the transverse field of the Ising Hamiltonian (see Section~\ref{sec:ising}).
        Instead of the BFGS algorithm used for obtaining the results for the main text, here we apply the COBYQA minimizer which showed a better performance for finite $\mu$ among the algorithms built into SciPy \cite{virtanen2020scipy}.

        \subsubsection{Estimating the parameter of the amplitude-damping channel}
        \label{sec:ad_shots}
        
\begin{figure*}[h]
                \centering
                \includegraphics[width=.32\textwidth]{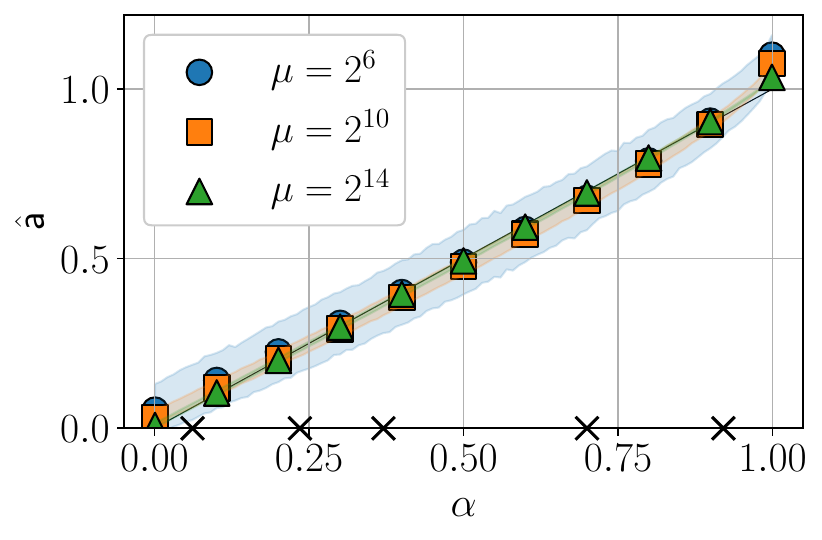}
                \includegraphics[width=.3275\textwidth]{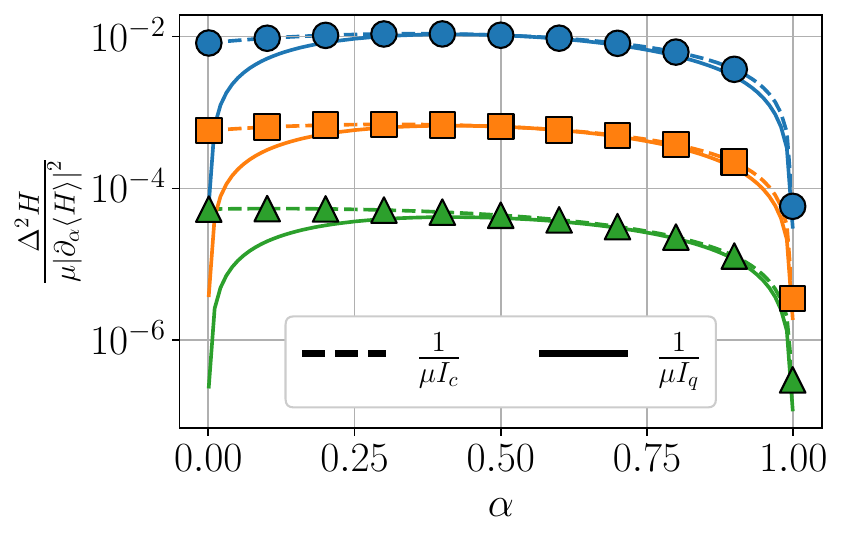}
                \includegraphics[width=.3275\textwidth]{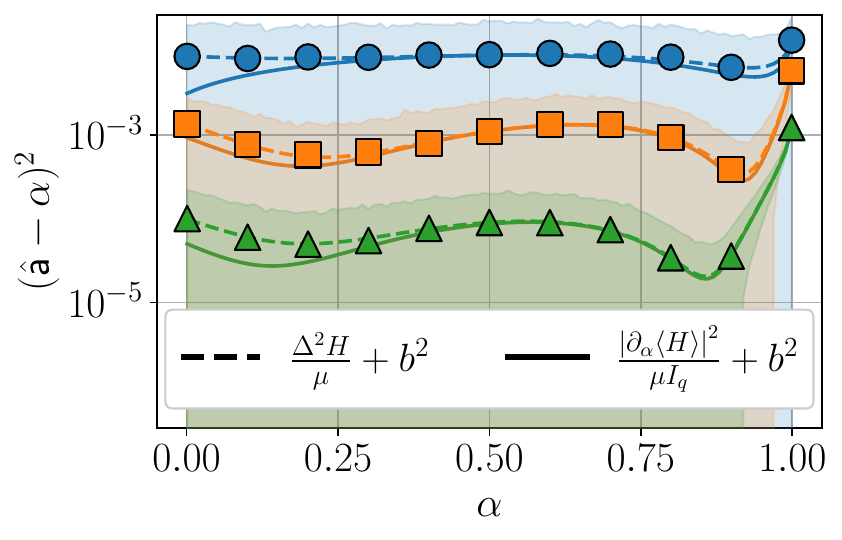}
                \caption{
                      Left: Estimate $\hat{\mathsf{a}}$ of the expectation $\mathsf{a} = \Tr H(\boldsymbol{x}^*, \boldsymbol{\theta}^*) \rho_\alpha$ vs. true $\alpha$ parameter of the AD channel \eqref{eq:ad_chan} for different numbers of measurement shots $\mu$ used during the training and testing; the black crosses on the $x$-axis indicate the training data.
                    Center: Error propagation \eqref{eq:estimation_variance} vs. $\alpha$ for the observables trained with different $\mu$; the dahsed and solid lines indicate the classical and quantum CRB, respectively.
                    Right: Average of $(\hat{\mathsf{a}} - \alpha)^2$ over 1000 measurement experiments with $\mu$ shots each; the dashed and solid lines indicate the quantities \eqref{eq:error_prop-biased} and \eqref{eq:ccrb-biased}, respectively.
                    The observables $H$ were trained on a set $\mathcal{T} = \big\{\rho_{\alpha_j}, \alpha_j \big\}_{j=1}^{5}$ with the parameters $\alpha_j$ generated randomly.
                    The shaded areas indicate the standard deviation over 1000 measurement experiments with $\mu$ shots each.
                    }
                \label{fig:ad-shots}
            \end{figure*}

            First, we show the results for a relatively simple problem of estimating the parameter of the amplitude damping channel \eqref{eq:ad_chan} given its output states $\rho_\alpha \equiv \Phi_\alpha[\rho]$  with the input $\rho=\ketbra{+}$.
            In Fig.~\ref{fig:ad-shots}, we show the estimates $\hat{\mathsf{a}}$ of $\mathsf{a} = \Tr H(\boldsymbol{x}^*, \boldsymbol{\theta}^*) \rho_\alpha$ for the number of shots $\mu \in \{2^6, 2^{10}, 2^{14}\}$.
            We note again that this number of shots was used for \textit{both} training (while obtaining the optimized observable $H(\boldsymbol{x}^*, \boldsymbol{\theta}^*)$) and evaluating (computing the estimates $\hat{\mathsf{a}}_j$ for the states of the testing set) the model.

            Expectedly, the larger is $\mu$, the more accurate is the estimation of $\mathsf{a}$.
            No surprise is also that the standard deviation of the estimation also decreases with $\mu$.
            However, our results show that the number of shots may impact significantly not only the estimation accuracy itself, but also the training process.
            That is, smaller $\mu$ may result in an observable with greater bias.
            Nonetheless, for this particular case, even the observable trained with a rather modest number of shots $\mu=2^6$ gives quite accurate estimations; with a large $\mu=2^{14}$, the results are close to those in Figs.~\ref{fig:ad_chan-bayes} and \ref{fig:ad}, where $\mu$ is ``infinite''.
            To verify the correctness of our numerical results, in Fig.~\ref{fig:ad-shots} we also plot the expressions \eqref{eq:error_prop-biased} and \eqref{eq:ccrb-biased}, and observe that they are indeed satisfied.
            Moreover, one can see that the Cramer-Rao bounds plotted in this figure resemble the ones shown in Figs.~\ref{fig:ad_chan-bayes} and \ref{fig:ad} (mind the scale of the $y$-axis).

    \subsubsection{Estimating the transverse field of the Ising Hamiltonian}
    \label{sec:ising-shots}
    
        When simulating the finite number of measurement shots $\mu$, the task of estimating the transverse field of the Ising Hamiltonian is more challenging.
        As in Section~\ref{sec:ising}, here we consider 8-qubit Ising Hamiltonian \eqref{eq:ising}.
        We look for an optimal observable in the form \eqref{eq:povm_ham_m}, where we measure $m=4$ qubits and use the variational ansatz of $l=5$ layers. 

\begin{figure*}[h]
            \centering
            \includegraphics[width=.315\textwidth]{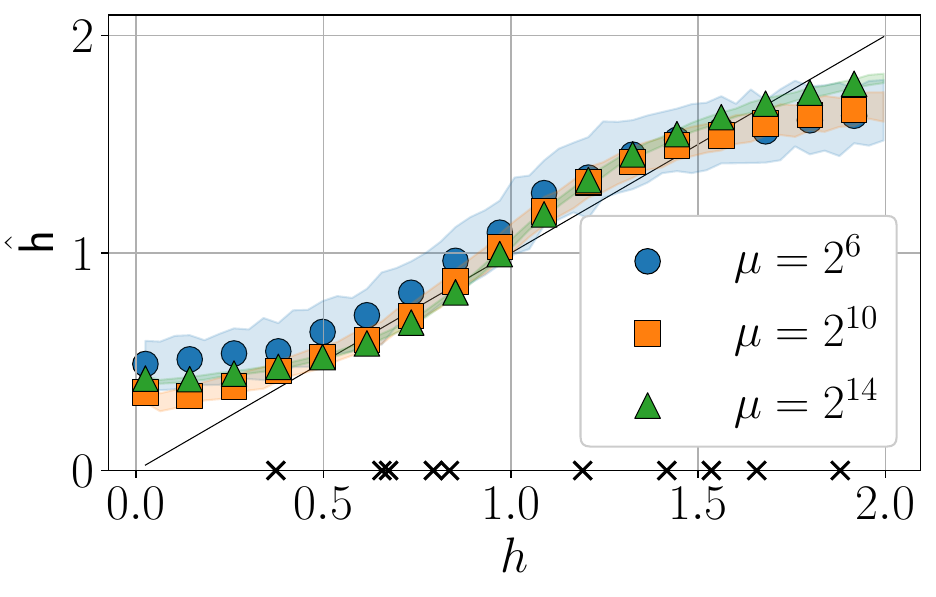}
            \includegraphics[width=.3275\textwidth]{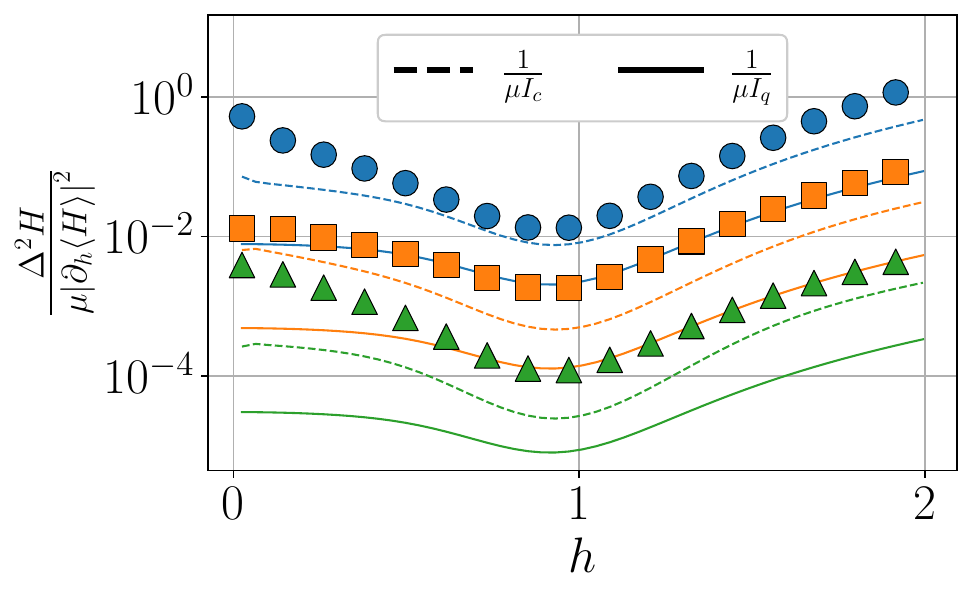}
            \includegraphics[width=.3275\textwidth]{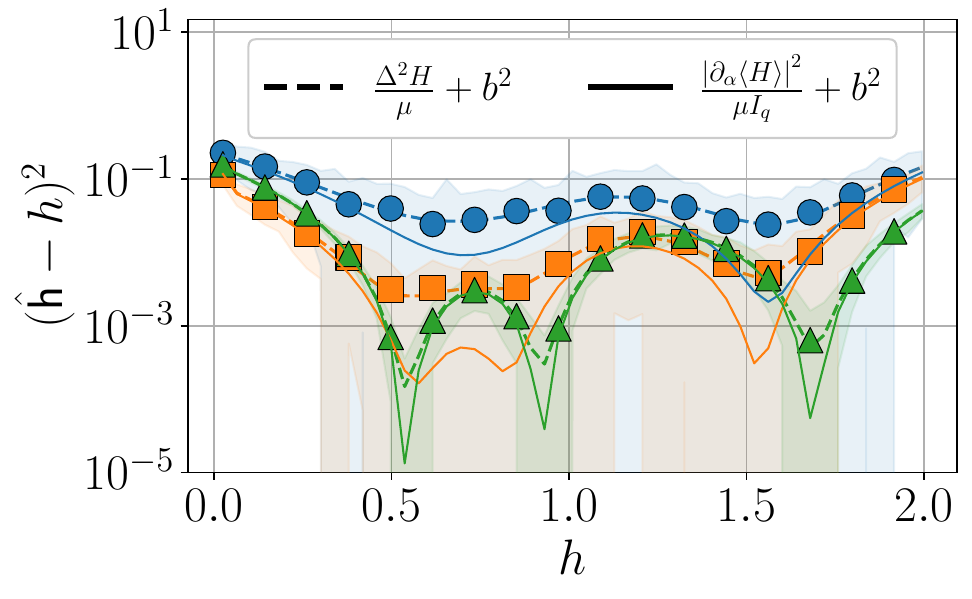}
            \caption{
                Left: Estimate $\hat{\mathsf{h}}$ of the expectation $\mathsf{h} = \langle \psi_h | H(\boldsymbol{x}^*, \boldsymbol{\theta}^*) | \psi_h \rangle$ vs. true $h$ transverse field of the 8-qubit Ising Hamiltonian \eqref{eq:ising} for different number of measurement shots $\mu$ used during the training and testing; the black crosses on the $x$-axis indicate the training data.
                Center: Error propagation \eqref{eq:estimation_variance} vs. $h$ for the observables trained with different $\mu$; the dahsed and solid lines indicate the classical and quantum CRB, respectively.
                Right: Average of $(\hat{\mathsf{h}} - h)^2$ over 100 measurement experiments with $\mu$ shots each; the dashed and solid lines indicate the quantities \eqref{eq:error_prop-biased} and \eqref{eq:ccrb-biased}, respectively.
                The observables $H$ were trained on a set $\mathcal{T} = \big\{|\psi_{h_j}\rangle, h_j \big\}_{j=1}^{10}$ with $\ket{\psi_j}$ being the ground states of \eqref{eq:ising}, and the fields $h_j$ are generated randomly.
                The shaded areas indicate the standard deviation over 100 measurement experiments with $\mu$ shots each.
            }
            \label{fig:ising-shots}
        \end{figure*}

        In Fig.~\ref{fig:ising-shots}, we show the estimates $\hat{\mathsf{h}}$ of the expectations $\mathsf{h} = \langle \psi_h | H(\boldsymbol{x}^*, \boldsymbol{\theta}^*) | \psi_h \rangle$ obtained with the number of shots $\mu \in \{2^6, 2^{10}, 2^{14}\}$.
        As for the AD channel, the estimation accuracy grows with $\mu$.
        However, in this case, with a large $\mu=2^{14}$ we do not reach the accuracy shown in Fig.~\ref{fig:ising-w_var} for ``infinite'' $\mu$.
        This may be due to the larger system size as well as because of a more complex relation between $h$ and $\ket{\psi_h}$. 

\begin{figure*}[h]
            \centering
            \includegraphics[width=.395\textwidth]{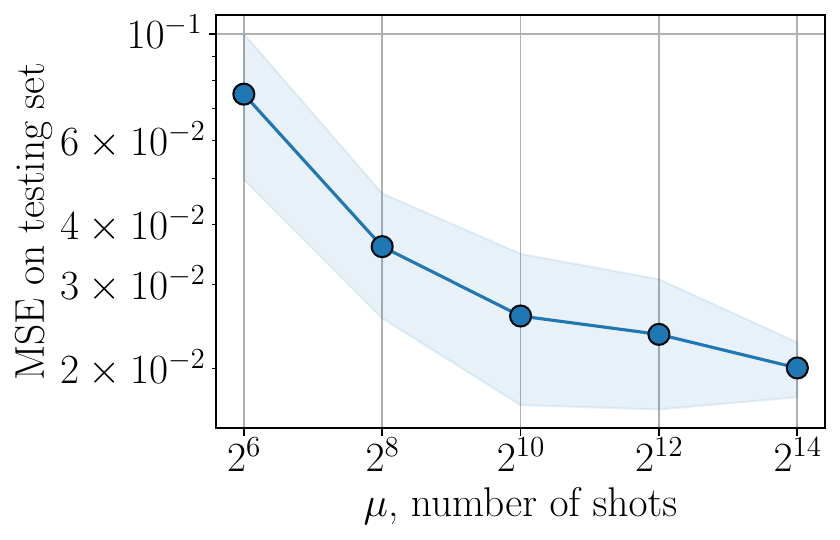}
            \hspace{40pt}
            \includegraphics[width=.395\textwidth]{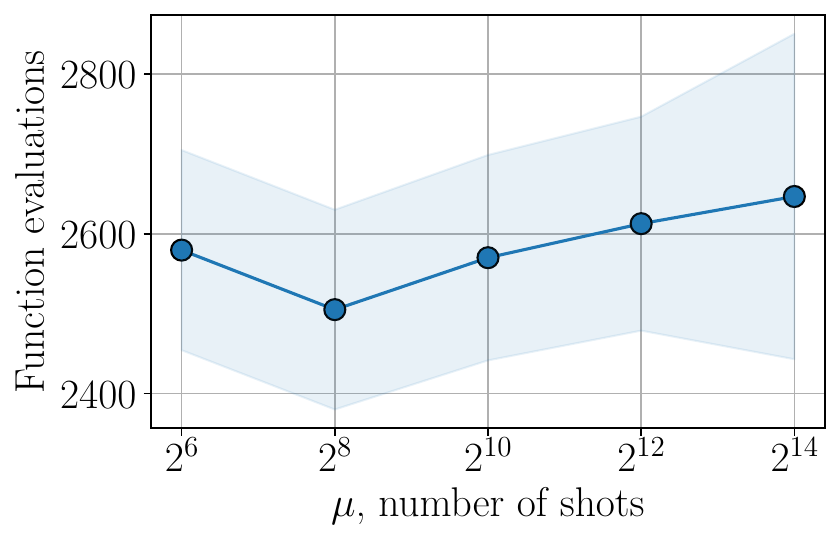}
            \caption{
                MSE evaluated on a testing set (left) and number evaluations of the cost function in \eqref{eq:ls_min_variational_mod} made by the COBYQA algorithm (right) vs. the number of shots $\mu$ used to estimate the expected values during the training.
                The shaded area shows the standard deviation obtained over 10 models trained with different initial parameters $(\boldsymbol{x}, \boldsymbol{\theta})$. 
            }
            \label{fig:ising-shots-mu}
        \end{figure*}

        To investigate the impact of the number of shots $\mu$ on training and testing, in Fig.~\ref{fig:ising-shots-mu} we plot the dependence of MSE evaluated on a testing set against $\mu$.
        Additionally in this Figure, we show the number of times the COBYQA minimizer calls the cost function in \eqref{eq:ls_min_variational_mod} during the training.
        As expected, MSE decreases with the number of shots $\mu$.
        The number of function evaluations, however, also tends to grow.
    
    \subsection{Channel parameter prediction}
    \label{sec:channels}

       Here our method is applied for learning to predict the parameter $\alpha$ of a parametrized channel $\Phi_\alpha$, i.e. the labeled states have the form $\rho_\alpha = \Phi_\alpha[\rho]$ for some \textit{fixed} input state $\rho$.

        \subsubsection{Depolarizing channel}
    
\begin{figure}[h]
                \centering
                \includegraphics[width=.495\textwidth]{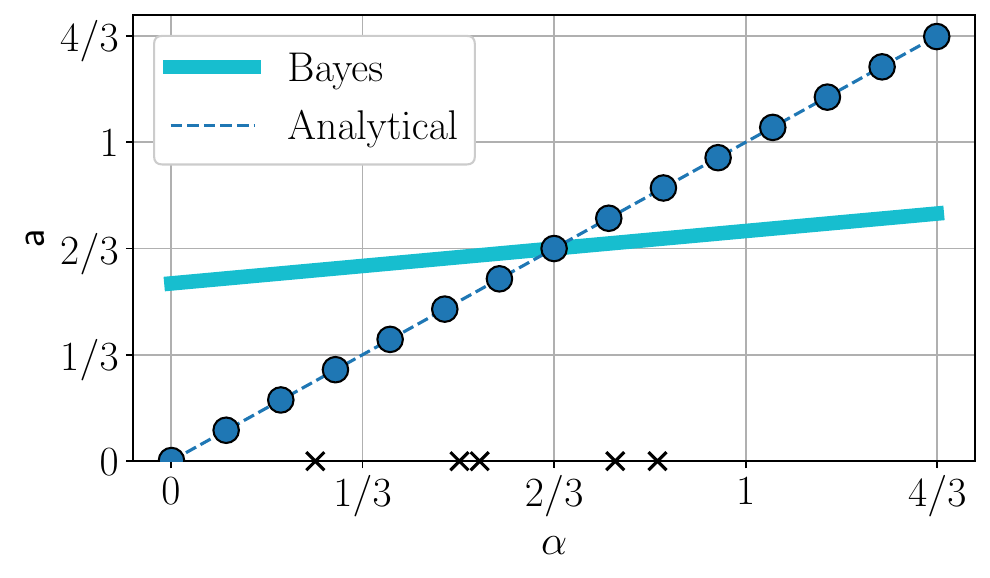}
                \includegraphics[width=.495\textwidth]{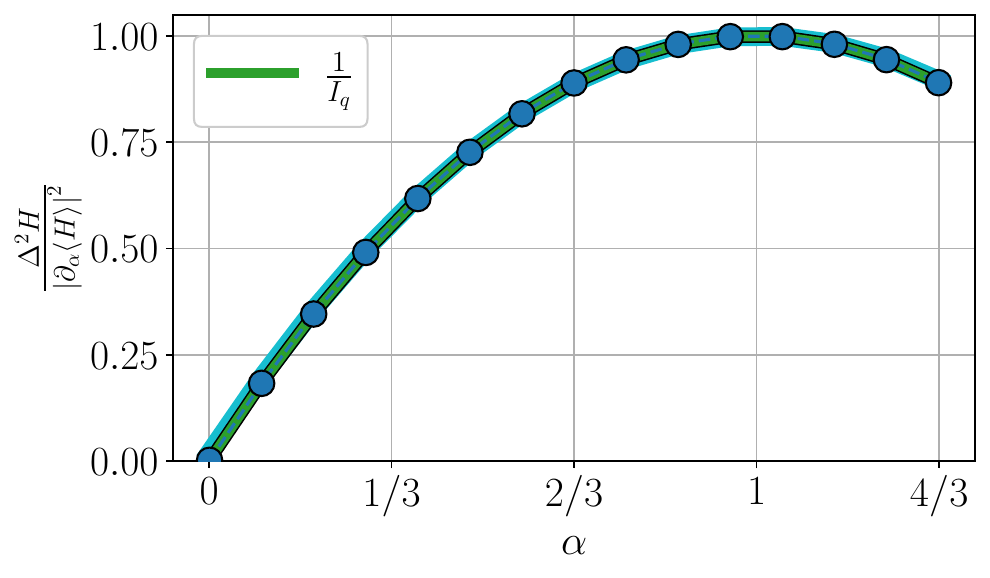}
                \caption{
                    Left: Predicted $\mathsf{a}$ depolarization strength $\alpha$ of the depolarizing channel \eqref{eq:depolarizing-app} with the input state $\rho = \ketbra{+}$. 
                    Right: Variance \eqref{eq:var-qcrb} of the optimized observable $H$.
                    The observable was trained on a set $\mathcal{T} = \big\{\rho_{\alpha_j}, \alpha_j \big\}_{j=1}^{5}$ with random labels $\alpha_j$ indicated as black crosses in the left panel.
                    In both panels, the solid blue line indicates the data for the observable obtained via the Bayesian approach \eqref{eq:bmse}.
                }
                \label{fig:depolarizing}
            \end{figure}

            Let us consider the single-qubit depolarizing channel
            \begin{equation}
                \label{eq:depolarizing-app}
                \Phi_\alpha[\rho] 
                = (1 - \alpha) \rho + \frac{\alpha}{2} \Id,
            \end{equation}
            which we have already mentioned in Section~\ref{sec:estimation_variance}.
            Representing the input state as $\rho = \frac{1}{2} \big(\Id + \sum_{j=1}^3 r_j \sigma_j\big)$ with $r_j \in \R$, one can find the (unnormalized) eigenvectors $\ket{l_k}$ of the SLD operator $L$ of the output state $\rho_\alpha = \Phi_\alpha[\rho]$ to be
            \begin{equation*}
                \ket{l_{1,2}} = \frac{r_3 \pm \sqrt{r_1^2 + r_2^2 + r_3^2}}{r_1 + i r_2} \ket{0} + \ket{1},
            \end{equation*}
            which do not depend on the label $\alpha$.
            Therefore, a POVM $\Pi$ which saturates qCRB \eqref{eq:cramer-rao_quantum} should not depend on $\alpha$ as well.
            Let us set the input state to $\rho = \ketbra{+}$.
            For this input, one can obtain the SLD operator for the output $\rho_\alpha$ as
            \begin{equation*}
                L = \frac{1}{\alpha}\ketbra{+} + \frac{1}{\alpha - 2}\ketbra{-},
            \end{equation*}
            which, by formula \eqref{eq:qfi}, gives $I_q(\rho_\alpha) = 1/(2\alpha - \alpha^2)$. 
            Note the equality of this to the reciprocal of \eqref{eq:depol_chan_obs_var} with $h_2=h_3=0$.
            In Fig.~\ref{fig:depolarizing}, we show the performance of our method in predicting $\alpha$ for the depolarized state $\rho_\alpha$.
            As can be seen, trained on a set of five states with random depolarization strengths $\alpha$, our model finds an observable giving both accurate predictions and variance saturating the qCRB.

            For the input state $\rho = \ketbra{+}$ of the depolarizing channel, we can also solve \eqref{eq:opt_obs} to obtain an optimal observable $H_0 = \frac{3k+10}{3k+15} \Id - \frac{k}{k+5}\sigma_x$.
            In the limit of large $k$ (small variance weight $w_\mathrm{var}$) and for the output state $\Phi_\alpha[\rho]$, this observable gives  $\langle H_ 0 \rangle = \alpha$ and $\Delta^2 H_0 = 2\alpha -\alpha^2$, which matches the results shown in Fig.~\ref{fig:depolarizing}.
            Additionally in this Figure, we plot prediction accuracy and the variance for the observable obtained via the minimization of the Bayesian MSE \eqref{eq:bmse}.
            For this case, putting $k=1$ into $H_0$ found earlier, one can calculate the Bayesian qCRB \eqref{eq:bqcrb} to obtain 
            \begin{equation*}
                \Delta^2_B\alpha = \Delta^2_p\alpha - I_B = \frac{10}{81} \approx 0.12345679,
            \end{equation*}
            the value achieved in our simulations of the Bayesian approach.
            
    \subsubsection{Amplitude damping channel}
    \label{app:ad-chan}

        In Section~\ref{sec:ad_chan-res}, we considered the amplitude damping (AD) channel
        \begin{equation}
            \label{eq:ad_chan-app}
            \Phi_\alpha[\rho] = \sum_{k=1}^2 V_k(\alpha) \rho V_k^\dagger(\alpha)
        \end{equation}
        with the Kraus operators
        \begin{equation*}
            V_1(\alpha) = \sqrt{\alpha} \ketbra{0}{1}, \quad V_2(\alpha) = \ketbra{0} + \sqrt{1 - \alpha} \ketbra{1}.
        \end{equation*}
        and the input state $\rho=\ketbra{+}$.
        This is a single-qubit channel which models the decay from the excited state $\ket{1}$ to the ground state $\ket{0}$ \cite{wilde2011classical}.
        Let us now consider the input state for this channel in the Bloch representation
        \begin{equation*}
            \rho=\frac12\left(\Id + \vec{r}\cdotp\vec{\sigma} \right),
        \end{equation*}
        where $\vec{r}\cdotp\vec{\sigma}  \equiv\sum_{i=1}^3 r_i \sigma_i$ with $\big|\big|\vec{r}\big|\big| \leqslant 1$.
        The output state of the channel therefore reads
        \begin{equation}
            \Phi_\alpha[\rho] = \frac12\big(\Id + \vec{r}_\alpha\cdotp\vec{\sigma} \big),
        \end{equation}
        where $\ver = \vec r(\alpha)$ is the output state Bloch vector with the components
        \begin{equation}\label{eq:appb:blochr}
            r_x(\alpha) = \sqrt{1-\alpha}\,r_x,\quad r_y(\alpha) = \sqrt{1-\alpha}\,r_y,\quad r_z(\alpha) = (1-\alpha)\,r_z + \alpha.
        \end{equation}
        
        Let us assume that $r_z = 0$, then by formula \eqref{eq:sld-bloch} we obtain the following structure of the SLD operator:
        \begin{equation}\label{eq:appb:L}
        L = c\,\Id + \vec{a}\cdotp\vec{\sigma},
        \end{equation}
        where 
        \begin{equation}
            c = \frac{r^2 - 2\alpha}{2(1-\alpha)(1 + \alpha - r^2)}, 
        \end{equation}
        with $r \equiv ||\vec{r}||$, and the components of the vector $\vec a$ are
        \begin{equation}\label{eq:appb:axyz}
            a_{x,y} = - \frac{\sqrt{1-\alpha}\,r_{x,y}}{2(1 + \alpha - r^2)},\quad a_z = \frac{2 + (\alpha - 2)r^2}{2(1-\alpha)(1 + \alpha - r^2)}.
        \end{equation}
        The eigenvector decomposition of $L$ reads
        \begin{equation}
            L = (c+a)\dyad{\vec a} + (c-a)\dyad{-\vec a},
        \end{equation}
        where $a \equiv \norm{\vec a}$ and 
        $$
            \ket{\vec a} = \cos{\frac\theta2}\ket{0} + \sin{\frac\theta2}\,e^{i\varphi}\ket{1}
        $$
        with $\theta,\,\varphi$ such that $a_z/a = \cos{\theta},\,(a_x + ia_y)/a = \sin{\theta}\,e^{i\varphi}$.
        More explicitly,
        \begin{equation}
            \label{eq:appb:Lfin}
            L = \lambda_+\dyad{\vec a} + \lambda_-\dyad{-\vec a},
        \end{equation}
        where
        \begin{equation}\label{eq:appb:Leigv}
            \lambda_\pm = \frac{r^2 - 2\alpha \pm \sqrt{(\alpha-2)^2r^4 + [(1-\alpha)^3 + 4(\alpha - 2)]r^2 +4}}{2(1-\alpha)(1 + \alpha - r^2)}.
        \end{equation}
        
        Let us consider two input states $\rho$ for the amplitude-damping channel.
        First, if $r=0$, then the input is the maximally mixed state $\rho = \frac12\Id$, for which Eqs.~(\ref{eq:appb:Lfin}) and~(\ref{eq:appb:Leigv}) yield
        \begin{equation}
            \label{eq:sld-ad-maxmixed}
            L = \frac{1}{\alpha + 1} \ketbra{0} + \frac{1}{\alpha - 1} \ketbra{1},
        \end{equation}
        with the eigenprojectors of $L$ not depending on $\alpha$, and giving $I_q(\rho_\alpha) = 1/(1 - \alpha^2)$.
        Second, if $\vec{r} = \vec e_x$, which corresponds to $\rho = \dyad{+}$, the eigenvectors of the SLD operator \eqref{eq:appb:Lfin} are characterized by the vector $\vec a$, which in this case has components
        \begin{equation}\label{eq:appb:purecaseL}
            a_x = -\frac{\sqrt{1-\alpha}}{2\alpha},\quad a_y = 0,\quad a_z = \frac1{2(1-\alpha)}.
        \end{equation}
        The eigenvalues of the SLD operator are
        \begin{equation}
             \lambda_\pm = \frac{1 - 2\alpha\pm \sqrt{(1-\alpha)^3 + \alpha^2}}{2\alpha(1-\alpha)}.
        \end{equation}
        Using \eqref{eq:qfi-basis_indep}, one also obtains $I_q(\rho_\alpha) = (1 + \alpha)/[4\alpha(1 - \alpha)]$.

\begin{figure}[h]
            \centering
            \includegraphics[width=.495\textwidth]{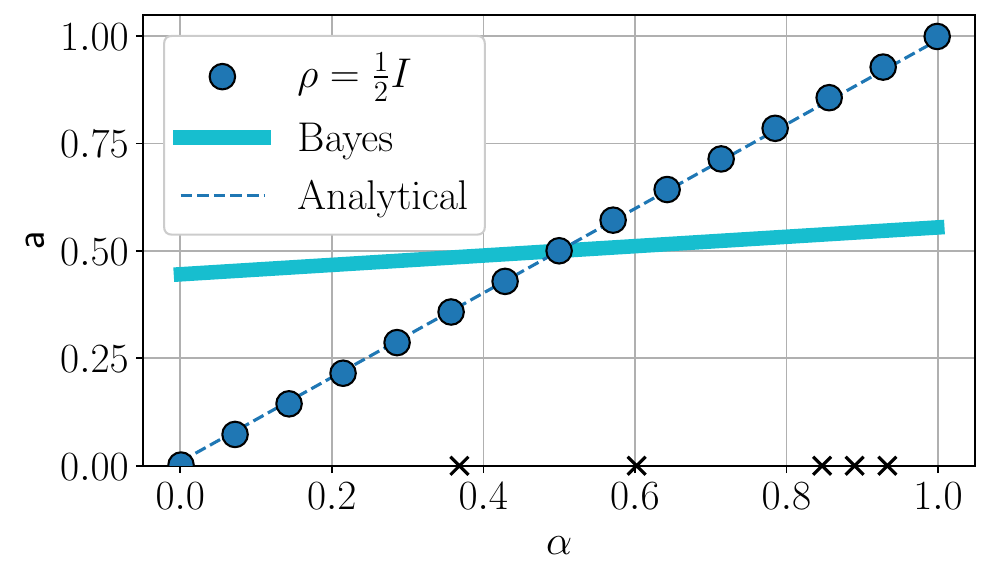}
            \includegraphics[width=.495\textwidth]{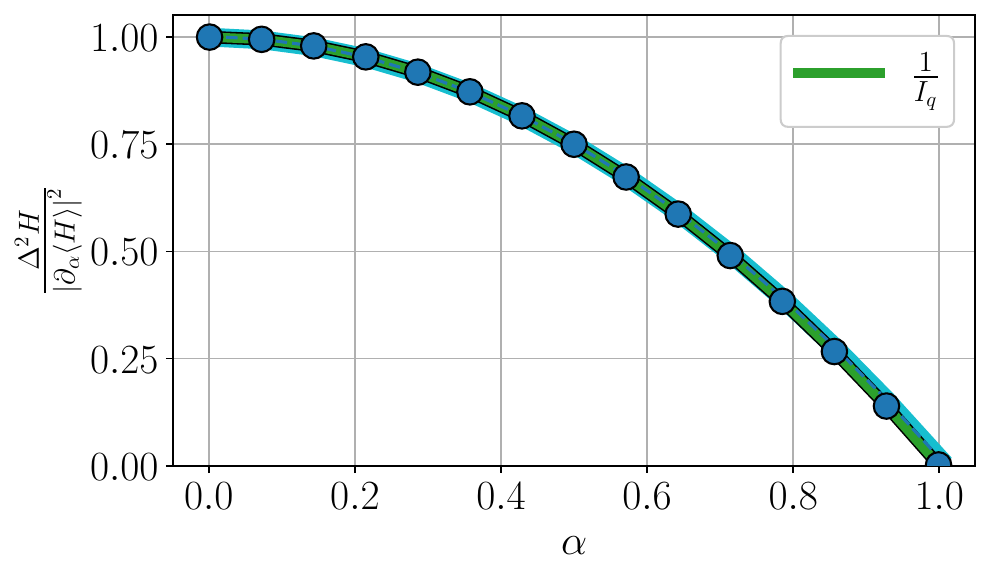}
            \includegraphics[width=.495\textwidth]{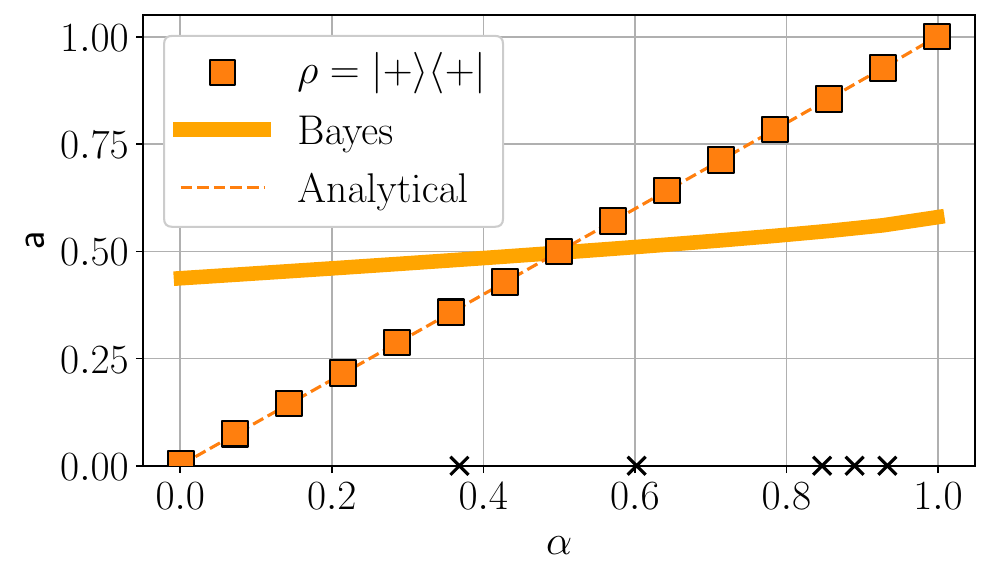}
            \includegraphics[width=.495\textwidth]{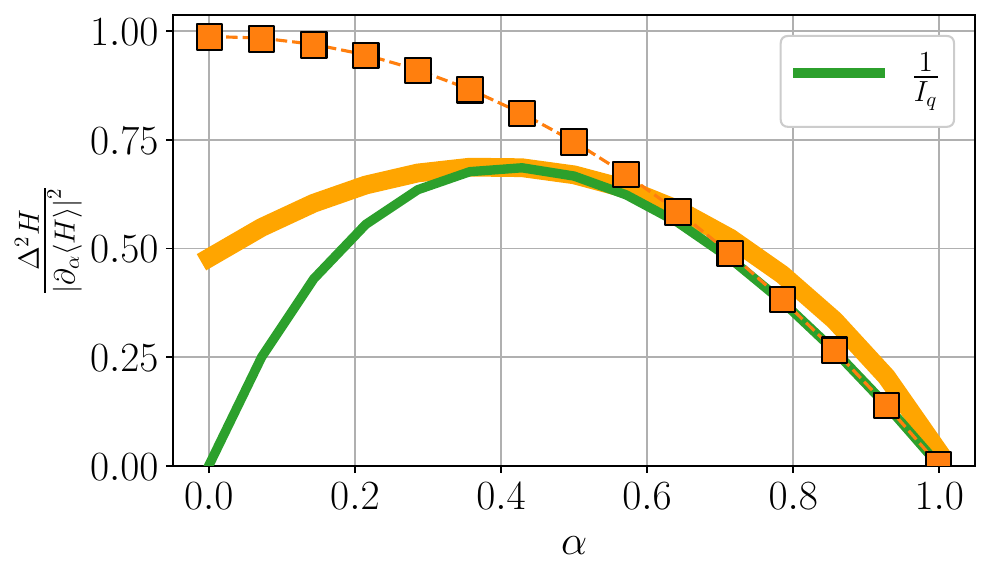}
            \caption{
                Left: Predicted $\mathsf{a}$ vs. true $\alpha$ parameters of the AD channel \eqref{eq:ad_chan-app} for the input states $\rho=\frac12 \Id$ (upper, blue) and $\rho=\ketbra{+}$  (lower, orange).
                Right: Variance \eqref{eq:var-qcrb} of the trained observable.
                The models were trained on a set $\mathcal{T}=\{\rho_{\alpha_j}, \alpha_j\}_{i=1}^5$ with random $\alpha_j$, which are indicated as black crosses in the left panels.
                In all panels, the blue and orange solid lines indicate the results of the Bayesian approach \eqref{eq:bmse}.
                The dashed lines stand for the analytical solutions obtained by solving \eqref{eq:opt_obs}.
            }
            \label{fig:ad}
        \end{figure}

        Let us look how our model predicts the amplitude-damping channel parameter for $\rho_\alpha = \Phi_\alpha[\rho]$.
        For both above-mentioned inputs $\rho = \frac12 \Id$ and $\rho = \ketbra{+}$, we find observables which give accurate predictions of $\alpha$.
        Interestingly, for either of the inputs, a simple observable $H=\sigma_z$ gives $\Tr H \rho_\alpha = \alpha$.
        Moreover, for $\rho = \frac12 \Id$, this observable saturates qCRB, which can be verified by combining \eqref{eq:sld-ad-maxmixed} and \eqref{eq:H_opt}.
        In contrast, $H=\sigma_z$ is not optimal for $\rho = \ketbra{+}$ in the sense of the variance.
        As can be seen in Fig.~\ref{fig:ad}, our model does not find an observable saturating qCRB in this case.
        From (\ref{eq:appb:Lfin}) and (\ref{eq:appb:Leigv}), one can notice that the eigenprojectors of $L$ depend on the parameter $\alpha$. If $\alpha\to 0$ the optimal POVM tends to $\{\dyad{+}, \dyad{-}\}$, while at the other end of the parameter range, $\alpha \to 1$, the optimal POVM is $\{\dyad{0}, \dyad{1}\}$.

        In Section~\ref{sec:ad_chan-res}, we obtained an optimal observable by solving (\ref{eq:opt_obs}) for the input state $\rho=\ketbra{+}$.
        Here, similarly, we find such an observable in the form $H_0=\sum_{i=0}^3h_{i}\sigma_i$ for the input $\rho=\frac{1}{2}\Id$.
        The optimal observable in this case is $H_0=\frac{h_0}{k+8}\Id+\frac{k}{8+4}\sigma_z$ yielding $\langle H_0 \rangle_{\rho_\alpha}=\frac{\alpha k+4}{k+8}$ and $\Delta^2 H_0=\frac{(1-\alpha^2) k^2}{(k+8)^2}$.
        For sufficiently large $k$, i.e. $k\rightarrow \infty$, we obtain $\langle H_0 \rangle_{\rho_\alpha}=\alpha$ and $\Delta^2 H_0=1-\alpha^2$.

        In Fig.~\ref{fig:ad}, we also show the results for the observables obtained using the Bayesian approach \eqref{eq:bmse} for both inputs $\rho=\ketbra{+}$ and $\rho=\frac{1}{2}\Id$.
        For the former, the results were shown in Section~\ref{sec:ad_chan-res}.
        For the latter, employing the optimal observable obtained above for $k=1$, one can calculate the Bayesian qCRB \eqref{eq:bqcrb} to be
        \begin{equation*}
            \Delta^2_B\alpha = \Delta^2_p\alpha - I_B = \frac{2}{27} \approx 0.074074074
        \end{equation*}
        We achieve this value of Bayesian MSE in our simulations, saturating therefore the Bayesian qCRB.

    \subsubsection{Unitary transformation}
    \label{app:unitary_chan}

        Let us consider a unitary channel $\Phi_\alpha$ which performs the $z$-rotation of a single-qubit state $\rho$ by an angle $\alpha/2$:
        \begin{equation}
            \label{eq:unitary_chan}
            \Phi_\alpha[\rho] = e^{-i \alpha \sigma_z/2} \rho e^{i \alpha \sigma_z/2}, \qquad \alpha \in [0, \pi].
        \end{equation}
        Such transformation is often considered in the area of quantum channel estimation and quantum metrology as it incarnates, e.g., the Ramsey interferometry \cite{pezze2018quantum}.
        The main obstacle for us in predicting the label for $\rho_\alpha$ with our method is that we cannot directly obtain $\alpha$ as the expected value of some observable $H$.
        Instead, as noted in Section~\ref{sec:func_dep}, our prediction is generally biased. 
        However, one can suppress this bias by training an observable acting on a number of copies of the labeled state.
        
\begin{figure}[h]
            \centering
            \includegraphics[width=.495\textwidth]{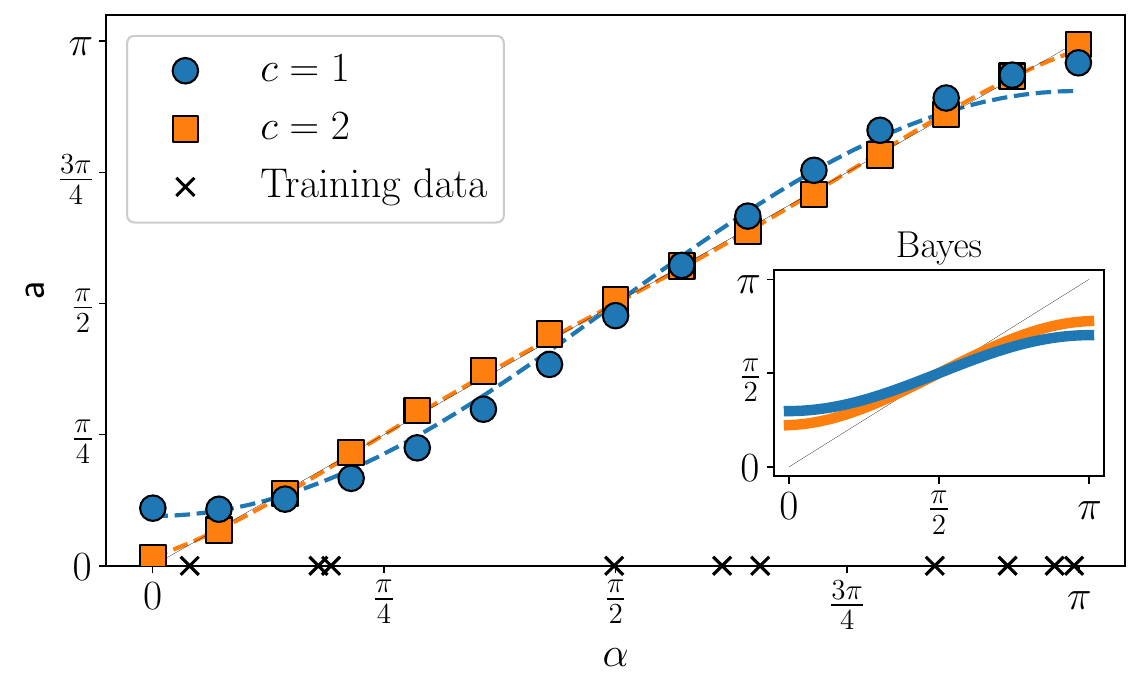}
            \includegraphics[width=.495\textwidth]{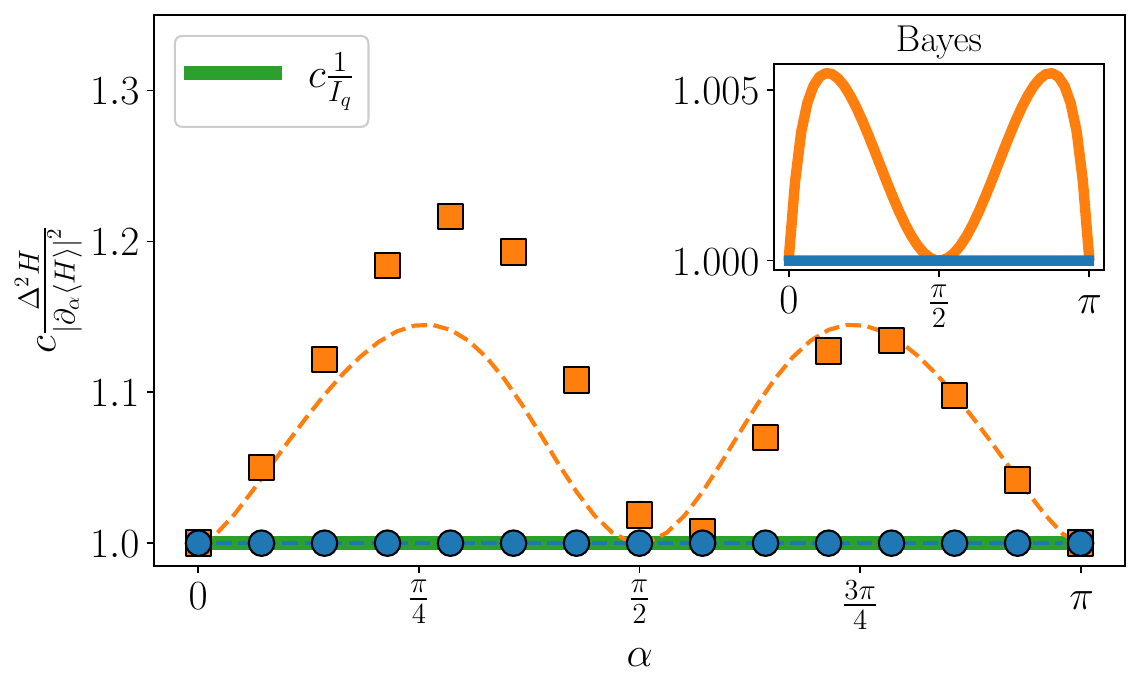}
            \caption{
                Left: Predicted $\mathsf{a}$ vs. true $\alpha$ rotation angle of the state \eqref{eq:unitary_chan}.
                Right: Variance \eqref{eq:var-qcrb} of the trained observables. 
                The observables $H$ are trained on a set $\mathcal{T}_c = \big\{\rho_j^{\otimes c}, \alpha_j \big\}_{j=1}^{10}$ with randomly generated $\alpha_j$ and $c=1, 2$ simultaneously processed copies.
                In both panels, the dashed lines of the corresponding colors indicate the data given by the optimal observables \eqref{eq:opt_ham_uni_1c} and \eqref{eq:opt_ham_uni_2c}.
                The insets show the data for the observables obtained via the minimization of the Bayesian MSE \eqref{eq:bmse}. 
            }
            \label{fig:z-rot}
        \end{figure}

        Indeed, let $\rho=\ketbra{+}$ and consider $c$ copies of this state passed through the channel \eqref{eq:unitary_chan}:
        \begin{align*}
            \rho_\alpha^{\otimes c}
            &= \left[e^{-i \alpha \sigma_z/2}\ketbra{+} e^{i \alpha \sigma_z/2} \right]^{\otimes c} \\
            &= \left[\frac{1}{2}\left(\Id + e^{i\alpha}\ketbra{1}{0} + e^{-i\alpha}\ketbra{0}{1} \right)\right]^{\otimes c} \\
            &= \frac{1}{2^c}\left[\Id^{\otimes c} +\sum_{k=1}^c\left(e^{i k \alpha}\ketbra{0}{1}^{\otimes k}+e^{-i k \alpha}\ketbra{1}{0}^{\otimes k}\right)\otimes \Id^{\otimes(c-k)} + R\right],
        \end{align*}
        where in the last line $R$ is the rest of terms to complete $\rho_\alpha^{\otimes c}$.
        From this expansion, one can \textit{guess} an observable $H_c$ such that
        \begin{equation}
            \label{eq:tracefourierham}
            \Tr \rho_\alpha^{\otimes c}H_c 
            = i\sum_{k=1}^c \frac{(-1)^k}{k} \left( e^{ik\alpha}-e^{-ik\alpha} \right),
        \end{equation}
        which is precisely the first $c$ terms of the Fourier series of the function $f(\alpha)=\alpha$.
        Indeed, if we take
        \begin{equation}
            \label{eq:hc_fourier}
            H_c = i\sum_{k=1}^{c}\frac{(-2)^k}{k}\left(\ketbra{0}{1}^{\otimes k}-\ketbra{1}{0}^{\otimes k}\right)\otimes \Id^{\otimes (c - k)},
        \end{equation}
        then we get
        \begin{align*}
            \label{eq:hc_fourier_expec}
            \Tr\rho_\alpha^{\otimes c}H_c 
            &=i\sum_{k=1}^{c}\frac{(-2)^k}{k}\Tr\left[\left(\rho_\alpha^{\otimes k}\otimes\rho_\alpha^{\otimes (c-k)}\right)\left(\big(\ketbra{0}{1}^{\otimes k}-\ketbra{1}{0}^{\otimes k}\big)\otimes \Id^{\otimes (c-k)}\right)\right] \\
            &= i\sum_{k=1}^{c}\frac{(-2)^k}{k}\Tr\left[\rho_\alpha^{\otimes k}\left(\ketbra{0}{1}^{\otimes k}-\ketbra{1}{0}^{\otimes k}\right)\right] \\
            &= i\sum_{k=1}^{c}\frac{(-2)^k}{k} \Big( \langle 1| \rho_\alpha |0 \rangle^k - \langle 0| \rho_\alpha |1 \rangle^k \Big)\\
            &= i\sum_{k=1}^{c}\frac{(-1)^k}{k} \left( e^{ik\alpha} - e^{-ik\alpha} \right). 
        \end{align*}
        Therefore, we can expect that the more copies $c$ of the state $\rho_\alpha$ we process simultaneously, the more $\Tr\rho_\alpha^{\otimes c}H_c$ gets closer to $\alpha$ (i.e. the lower is the bias). The example of rotation angle $\alpha$ prediction in \eqref{eq:unitary_chan} confirms that as shown in Fig.~\ref{fig:z-rot}.
        As can be seen, already with a single copy ($c=1$) we obtain quite an accurate estimation for $\alpha$, which significantly improves with $c=2$ copies.
        However, in contrast to the single-copy case, with two copies we do not saturate CRB \eqref{eq:cramer-rao_quantum}.

        Solving the equation (\ref{eq:opt_obs}) for the channel (\ref{eq:unitary_chan}) with the input state $\rho=\ketbra{+}$, one can obtain the following optimal observable for the case $c=1$ to be
        \begin{equation}
            \label{eq:opt_ham_uni_1c}
                H_1 =
                \frac{\pi}{2} \Id-\frac{4k}{\pi(1+k)} \sigma_x.
        \end{equation}
        Similarly for $c=2$ copies, the optimal observable takes the form
        \begin{multline}
            \label{eq:opt_ham_uni_2c}
            H_2 = 
            \left(\frac{\pi}{2} - t\right) \Id\otimes\Id
            + t (\sigma_x\otimes\sigma_x + \sigma_y\otimes\sigma_y + \sigma_z\otimes\sigma_z) \\
            -\frac{12 k \pi(k + 8)}{r (1 + 2 k)} (\Id\otimes\sigma_x + \sigma_x\otimes\Id)
            -\frac{3 k (3 \pi^2 - 32)}{r} (\sigma_x\otimes\sigma_y + \sigma_y\otimes\sigma_x), 
        \end{multline}
        where $r = -32 - 64 k + 27 \pi^2 + 9 k \pi^2$ and $t$ is arbitrary.
        The predictions obtained with these observables, as well as their variance, are shown in Fig.~\ref{fig:z-rot}.
        It may seem that these predictions do not agree with the results obtained via the method \eqref{eq:ls_min_variational_mod}, especially for the variance.
        Recall however that the observables \eqref{eq:opt_ham_uni_1c} and \eqref{eq:opt_ham_uni_2c} are obtained with the assumption of a large training set, i.e., $T\to\infty$.
        When we set $T=1000$, our simulations show almost perfect agreement with this analytical solution.
        
        It should also be noted that the Bayesian approach produces an observable with a smaller variance, but in exchange to a large bias.
        One can also verify that the both observables \eqref{eq:opt_ham_uni_1c} and \eqref{eq:opt_ham_uni_2c} with $k=1$ saturate the Bayesian qCRB \eqref{eq:bqcrb} evaluating to, respectively,
        \begin{equation*}
            \frac{\pi^4 - 48}{12 \pi^2} \approx 0.417182, \qquad
            -\frac{1}{8} + \frac{\pi^2}{12} + \frac{9}{8 - 3 \pi^2} \approx 0.28097,
        \end{equation*}
        which were achieved in our simulations.

        For a more general one-qubit unitary transformation
        \begin{equation}
            \label{eq:unitary_gen}
            \Phi_\alpha[\rho] = e^{-i _\alpha \, \vec{n}\cdotp\vec{\sigma}} \rho e^{i f_\alpha \, \vec{n}\cdotp\vec{\sigma} },
        \end{equation}
        with $\f\equiv f(\alpha)$ being an arbitrary function of parameter $\alpha$ and $\vec{n}\cdotp\vec{\sigma} \equiv\sum_i \sigma_i n_i$ the generator of rotation around the axis parallel to a given fixed unit vector $\vec n$, one can single out the properties that affect the prediction error. 
        Indeed, let the input state be given by the Bloch representation $\rho = 1/2(I + \vec{r}\cdotp\vec{\sigma})$. One can use Eqs.~(\ref{eq:qfi}) and (\ref{eq:Lformixed}) to calculate qFI of the output state:  
        \begin{equation}\label{eq:qfiuni1}
            I_q(\Phi_\alpha[\rho])=4(\fa)^2 r^2 \sin^2(\myangle{n}{r}),
        \end{equation}
        where  $\fa$ is the derivative of $f$ with respect to $\alpha$ and $\myangle{n}{r}$ is an angle between vectors  $\vec n$ and $\vec P$. 
        From this expression, it can be seen that the effects of functional dependence on the parameter and mixedness of the input state are isolated in the factors $(\fa)^2$ and $r^2$, respectively. 
        Since qFI appears in the denominator of the qCRB \eqref{eq:cramer-rao_quantum}, these two factors significantly affect the estimation error $\Delta^2 \hat{\alpha}$. 
        In particular, the case of highly mixed input states~($r \ll 1$) is expected to require higher number of measurements in comparison with pure case to achieve reasonable values of the  estimation error.

\section{Formulas for SLD operator and quantum Fisher information}
\label{app:qfi_formulas}

    For an $n$-qubit state $\rho_\alpha$, qFI can be calculated as $I_q(\rho_\alpha) = \Tr L^2 \rho_\alpha$, where the SLD operator $L$ was defined in Eq. (\ref{eq:sld_def}).   Although defined implicitly, $L$ can be expressed in the basis of $\rho_\alpha = \sum_{k=1}^{d} \lambda_k \ketbra{\lambda_k}$ as 
    \begin{equation}
        \label{eq:sld_decomp}
        L =2 \sum_{k,l=1}^{2^n} \frac{\braket{\lambda_k | \partial_\alpha \rho_\alpha | \lambda_l}}{\lambda_k + \lambda_l} \ketbra{\lambda_k}{\lambda_l},
    \end{equation}
    where the summation is taken over the indices $k$ and $l$ such that $\lambda_k + \lambda_l \neq 0$ \cite{liu2020quantum, genoni2008optimal}.     
    If $\rho_\alpha$ possesses certain properties, one can obtain $L$ more conveniently \cite{liu2016quantum}.
    For a mixed single-qubit state, one can write
    \begin{equation}
        \label{eq:Lformixed}
        L = 2 \partial_\alpha \rho_\alpha - \frac{1}{2} \rho_\alpha^{-1} \partial_\alpha \Tr \rho_\alpha^2, 
        \qquad \rho_\alpha^{-1} = \frac{2}{1 - \Tr\rho_\alpha^2} \sigma_y \rho_\alpha^T \sigma_y.
    \end{equation}
    Additionally, if $[\partial_\alpha \rho_\alpha, \rho_\alpha^{-1}] = 0$, then $L = \rho_\alpha^{-1} \partial_\alpha \rho_\alpha$.
    For a pure state $\rho_\alpha = \ketbra{\psi_\alpha}$, the SLD operator is simply 
    \begin{equation}
        \label{eq:sld_pure}
        L = 2 \partial_\alpha \rho_\alpha.
    \end{equation}
    Finally, for a single-qubit state in the Bloch representation $\rho_\alpha = \frac12\big(\Id + \vec{r}_\alpha\cdotp\vec{\sigma} \big)$, where $\vec{r}_\alpha = \vec{r}(\alpha) \in \R^3$ and $\vec{r}_\alpha\cdotp\vec{\sigma} \equiv \sum_{i=1}^3 r_i(\alpha) \sigma_i$, one can use the following expression~\cite{ChapB15}:
    \begin{equation}
        \label{eq:sld-bloch}
        L = \da\ver\cdotp\Vec{\sigma} + \frac{\ver\cdotp\da\ver}{1 - \norm{\ver}^2} \big(\ver\cdotp\Vec{\sigma} - \Id\big).
    \end{equation}

    It is also possible to obtain the qFI directly without calculating the SLD operator.
    For instance, the following expression is especially useful for numerical calculations
    \begin{equation}
        \label{eq:qfi-fidelity}
        I_q(\rho_\alpha) = 8\frac{1 - F(\rho_\alpha, \rho_{\alpha + \mathrm{d}\alpha})}{\mathrm{d}\alpha^2},
    \end{equation}
    where in the numerator we have the fidelity $F(\rho, \tau) = \Tr \sqrt{\sqrt{\rho} \tau \sqrt{\rho}}$ between quantum states $\rho$ and $\tau$ \cite{sidhu2020geometric, pirandola2018advances}.
    Additionally, if $\rho_\alpha$ is full-rank, one can also obtain the qFI as \cite{liu2020quantum}
    \begin{equation}
        \label{eq:qfi-basis_indep}
        I_q(\rho_\alpha) = \Tr(\partial_\alpha\rho_\alpha)^2 + \frac{1}{\det(\rho_\alpha)} \Tr(\rho_\alpha \partial_\alpha\rho_\alpha)^2.
    \end{equation}
    Finally, for a pure state $\rho_\alpha = \ketbra{\psi_\alpha}$, qFI is
    \begin{equation*}
        I_q(\psi_\alpha) = 4\, \left(\braket{\partial_\alpha\psi_\alpha|\partial_\alpha\psi_\alpha} - \big|\langle\partial_\alpha\psi_\alpha|\psi_\alpha\rangle\big|^2\right).
    \end{equation*}

    \section{Upper bound on the total variance} 
    \label{app:theor-var}
    
        As we mentioned in Section~\ref{sec:theor-var}, basing on Eq.~(\ref{eq:opt_obs}) one can upper-bound the total variance. This can be derived as follows.
        After multiplying both parts of Eq.~(\ref{eq:opt_obs}) by $H_0$ and taking trace, one obtains
        \begin{equation}
        \label{eq:optvar-app} 
            \Tr \Tilde{\rho}H_0^2 = \frac{1-k}{L}\int_a^b (\Tr \rho_\alpha H_0)^2 \,d\alpha 
            + \frac{k}{L}\int_a^b \alpha \Tr\rho_\alpha H_0 \,d\alpha
        \end{equation}
        Let us assume $k\leqslant1$, so that the first term in the right part of~(\ref{eq:optvar-app}) be non-negative. 
        The integral in this term can be upper-bounded with the use of inequality~(\ref{eq:trconv}) as
        \begin{equation}
        \label{eq:int_upp-app}
            \int_a^b (\Tr\rho_\alpha H_0)^2\,d\alpha\leqslant\int_a^b \Tr\rho_\alpha H_0^2\,d\alpha = L\Tr\Tilde{\rho}H_0^2,
        \end{equation}   
        where in the last step the integration was put under the trace and definition~(\ref{eq:rhotil}) of $\Tilde{\rho}$ was used. 

       With the use of Eq~(\ref{eq:int_upp-app}) one arrives at an inequality
       \begin{equation}
         \Tr\Tilde{\rho}H_0^2 \leqslant (1-k)\Tr\Tilde{\rho}H_0^2 + \frac{k}{L}\int_a^b \alpha \Tr\rho_\alpha H_0 \,d\alpha,  
       \end{equation}
       which can be further simplified to
       \begin{equation}
        \label{eq:trrhoh2-app}   
        \Tr\Tilde{\rho}H_0^2\leqslant\frac{1}{L}\int_a^b \alpha \Tr\rho_\alpha H_0 \,d\alpha,
       \end{equation}
       with $k$ having been eliminated. 
       Note that the left-hand part of \eqref{eq:trrhoh2-app} coincides with the quantum Bayes information \eqref{eq:bayes_info} when the prior is uniform.
       In its turn, the right-hand part integral can be upper-bounded by a chain of inequalities
       \begin{align}
         \int_a^b \alpha \Tr\rho_\alpha H_0 \,d\alpha  
         &\leqslant\left( \int_a^b\alpha^2\,d\alpha\right)^{1/2} \left(\int_a^b (\Tr\rho_\alpha H_0)^2\,d\alpha\right)^{1/2} \nonumber \\
         &\leqslant\left( \int_a^b\alpha^2\,d\alpha\right)^{1/2} \Big(L\Tr\Tilde{\rho}H_0^2\Big)^{1/2}  \nonumber \\
         &= \left(\frac{(b^3 - a^3)}{3}L\Tr\Tilde{\rho}H_0^2\right)^{1/2}, \label{eq:chain-app}
       \end{align}
       where the first step is due to the Cauchy-Schwarz inequality and on the second step Eq.~(\ref{eq:int_upp-app}) was used.
       Combining~(\ref{eq:trrhoh2-app}) and~(\ref{eq:chain-app}), after simple arithmetic one obtains
       \begin{equation}
           \label{eq:h2bound-app}\Tr\Tilde{\rho}H_0^2 \, \leqslant\,\frac{b^3 - a^3}{3L}.
       \end{equation}

       The total variance can be expressed as follows
       \begin{align}
            \int_a^b\Delta^2H_0\,d\alpha 
            &= \int_a^b\left(\Tr\rho_\alpha H_0^2 - (\Tr\rho_\alpha H_0)^2\right)\,d\alpha \nonumber \\
            &= L\Tr\Tilde{\rho}H_0^2 - \int_a^b(\Tr\rho_\alpha H_0)^2\,d\alpha \nonumber \\
            &= k\int_a^b\Big(\alpha \Tr\rho_\alpha H_0 -(\Tr\rho_\alpha H_0)^2\Big)\,d\alpha, \label{eq:var_expk-app}
       \end{align}
       where in the last step Eq.~(\ref{eq:optvar-app}) was used. Note that~(\ref{eq:var_expk-app}) holds for any finite $k$ as condition $k\leqslant1$ was not used here. 
        From this expression, one comes to the remark which concludes Section~\ref{sec:theor-var}.
        Namely, the optimal observable $H_0$ cannot have a bias vanishing on the whole interval $[a, b]$.
    
       Returning to the case $k\leqslant1$, the first term in the rightmost part of~(\ref{eq:var_expk-app})  can be upper bounded with the use of~(\ref{eq:chain-app}) and~(\ref{eq:h2bound-app}):
       \begin{equation}
           \label{eq:1termbd-app}
           \int_a^b\alpha \Tr\rho_\alpha H_0\,d\alpha\leqslant\,\frac{b^3 - a^3}{3}.
       \end{equation}
       As for the second term, one can exploit convexity of the function $f(x) = x^2$ and write
       \begin{equation}\label{eq:convbound-app}
           \int_a^b(\Tr\rho_\alpha H_0)^2\,d\alpha = L\int_a^b\frac{1}{L}(\Tr\rho_\alpha H_0)^2\,d\alpha
           \geqslant L\left(\int_a^b\frac{1}{L}\Tr\rho_\alpha H_0\,d\alpha\right)^2.
       \end{equation}
       The right side of Eq.~(\ref{eq:convbound-app}) can be evaluated with the use of Eq.~(\ref{eq:estsquare}), which yields
       \begin{equation}\label{eq:convbound2-app}
            \int_a^b(\Tr\rho_\alpha H_0)^2\,d\alpha\geqslant\frac{(b^2 - a^2)^2}{4L}.
       \end{equation}
    
       Finally, combining~(\ref{eq:var_expk-app}),~(\ref{eq:1termbd-app}), and~(\ref{eq:convbound2-app}), one obtains the upper bound \eqref{eq:totvar_bound} for the total variance:
       \begin{equation}
           \label{eq:totvar_bound-app}
           \int_a^b\Delta^2H_0\,d\alpha\,\leqslant\,k\left(\frac{b^3 - a^3}{3} - \frac{(b^2 - a^2)^2}{4L}\right).
       \end{equation}
       Putting $a=0$ and $b = L$, one reproduces the expression \eqref{eq:totvarprac} for the average total variance
       \begin{equation}\label{eq:totvarprac-app}
           \frac{1}{L}\int_0^L\Delta^2H_0\,d\alpha\,\leqslant\,\frac{kL^2}{12}.
       \end{equation}

\section{Bipartite entanglement}
\label{app:negativity}

Consider a bipartite state $\rho_{AB}$ defined on Hilbert space $\mathcal H_{AB} = \mathcal H_A\otimes \mathcal H_B$. The state is called \emph{separable} if it can be represented by an ensemble of product states
\begin{equation}\label{eq:sep}
    \rho_{AB} = \sum_i p_i\,\rho^i_A\otimes\rho^i_B,
\end{equation}
with  $\rho^i_A$ and $\rho^i_B$ defined on $\hil_A$ and $\hil_B$, respectively.

A bipartite state is called \emph{entangled} if it is not separable. In general, it is hard to determine whether a given  mixed state is entangled or separable even if one has full access to its density operator~\cite{Gurvitz03}.

Entanglement measures~\cite{QuantifyEnt97,  
HorEntRev09, GUHNE20091, ZiewEltRevEnt14} are used to quantify, in some sense, the amount of entanglement present in quantum states. A number of measures have been introduced, but only few of them have clear operational meaning and direct relation to protocols of quantum information processing~\cite{ConcEnt96,  MixedStateEnt96, EntTel99,  EntFisher09}. Nevertheless, even abstractly introduced measures are able to show the presence of entanglement in the sense that they vanish on separable states and take non-zero values on entangled ones. It was shown that computation of such entanglement measures for general states in high dimensions is NP-complete~\cite{Huang14}.

One of the most efficient theoretical ways to detect entanglement in a bipartite state is to analyze its partial transpose~\cite{Peres96}. The \emph{partial transpose} operation $\rho_{AB}$ with respect to party $B$ is defined by its action on elementary operators as
\begin{equation}
    \ketbra{ij}{kl}^{T_B} = \ketbra{il}{kj}
\end{equation}
and is extended on arbitrary operators by linearity. If a state is separable, i.e., is of the form~(\ref{eq:sep}), then its partial transpose is necessarily positive, i.~e. the state satisfies the \emph{positive partial transpose~(PPT) condition}.  Consequently, if $\rho^{T_B}$ has at least one negative eigenvalue, i.e. the state violates PPT condition, then its entanglement  is detected~(\emph{PPT criterion}). The entanglement quantifier, which is related to this approach,  is called \emph{negativity}~\cite{ZHSL98,VidalWernerNeg2002} and is defined as
\begin{equation}\label{eq:neg_app}
    N(\rho) = \norm{\rho^{T_B}}_1 - 1,
\end{equation}
where $\norm{A}_1 = \mathrm{tr}\sqrt{A^{\dagger}A}$ is the trace norm of $A$.

It is known that the PPT condition is necessary and sufficient for separability of qubit-qubit and qubit-qutrit quantum states~\cite{HorodNS1996}. 
For higher dimensions, there are entangled states with positive partial transpose and, correspondingly, vanishing negativity. In this respect, the negativity is a \emph{faithful} entanglement measure only for qubit-qubit and qubit-qutrit cases. While the definition of negativity can be modified to detect \emph{all} mixed entangled states also in higher dimensions~\cite{LCOK03}, we use expression~(\ref{eq:neg_app}) because of its practicality and simplicity for the qubit-qubit case, the only case we concentrate on in the present work. 

For the qubit-qubit case there is another expression~\cite{Eisert_quantum} for negativity:
\begin{equation}
    N(\rho) = 2\max\{-\lambda_{\min} (\rho^{T_B}),\,0\},
\end{equation}
where $\lambda_{\min}(\rho^{T_B})$ is the smallest eigenvalue of $\rho^{T_B}$.

    \section{\label{sec:app:nonlinearwit}Average performance of nonlinear entanglement witnesses with many copies}
        
        Here we use the results of Ref.~\cite{NonLinWit24} to get insight into the effect of multiple copies on detecting entanglement.  
        
        A linear witness is a Hermitian operator~(observable) $W$ such that $\mathrm{Tr}\, \rho_{\mathrm{sep}} W\geqslant 0$ for all separable states $\rho_{\mathrm{sep}}$ and $\mathrm{Tr}\,\rho_{\mathrm{ent}} W <0$ for \emph{some} entangled states $\rho_{\mathrm{ent}}$ that are detected by this witness. 
        In a multi-copy scenario, one can consider nonlinear witness $\mathcal W$ such that $\mathrm{Tr}\, \rho_{\mathrm{sep}}^{\otimes k} \mathcal W\geqslant 0$ for all separable states $\rho_{\mathrm{sep}}$ and $\mathrm{Tr}\, \rho_{\mathrm{ent}}^{\otimes k} \mathcal W < 0$ for some entangled states $\rho_{\mathrm{ent}}$. Such a witness is constructed as a tensor product of linear witnesses:
        \begin{equation}\label{eq:app:multiwit}
            \mathcal W = W^{(1)}\otimes\ldots\otimes W^{(n)},
        \end{equation}
        where input and output dimensions of states $\rho^{\otimes k}$ and witnesses should match. As an example, for $k$ copies of a bipartite state $\rho_{AB}$ the structure of $\mathcal W$ will be
        \begin{equation}\label{eq:app:bipkwit}
            \mathcal W = W^{(1)}_{A_1\ldots A_k}\otimes W^{(2)}_{B_1\ldots B_k}
        \end{equation}
        and the average of the observable with respect to $k$ copies of $\rho$ is
        \begin{equation}
            \langle\mathcal W\rangle_{\rho^{\otimes k}} = \mathrm{Tr}\,\rho_{A_1B_1}\otimes\ldots\otimes\rho_{A_kB_k}\mathcal W.
        \end{equation}
        In Ref.~\cite{NonLinWit24} it has also been observed that some of the linear witnesses $W^{(i)}$ in (\ref{eq:app:multiwit}) can be replaced with specific positive operators $P^{(i)}>0$, and the resulting nonlinear witness $\mathcal W$ might get even stronger.
        
        In order to consider average performance on random pure  bipartite states we follow the lines of Appendix F of Ref.~\cite{NonLinWit24} and make  special choice of operators $W^{(1)}$ and $W^{(2)}$ in eq.~(\ref{eq:app:bipkwit}). The first one will be 
        \begin{equation}\label{eq:app:antiwit}
         W^{(1)}_{A_1\ldots A_k} =   \Id - k!P^-_{A_1\ldots A_k},
        \end{equation}
        where $P^-_{A_1\ldots A_k}$ is the projector onto the antisymmetric subspace of Hilbert space of $k$ parties $A_1,\,\ldots,\,A_k$, and the second one, $W^{(2)}_{B_1\ldots B_k}$, is set just to the projector $P^-_{B_1\ldots B_k}$, which is a positive operator. Note that for $k=2$ copies the resulting nonlinear witness is proportional to the lower bound on squared concurrence in (\ref{eq:ClowB}). 
        
        The average over $k$ copies of Haar-random states $\ket{\psi}\in \mathbb C^d$, with $d=d_1\times d_2$ ($4$ for $2$ qubits), reads
        \begin{equation}\label{eq:app:expecthaar}
            \mathbb{E}\left[\langle\mathcal W\rangle\right] = \int \mathrm{Tr}\, \mathcal W \dyad{\psi}^{\otimes k} d\psi,
        \end{equation}
        and the well-known expression~\cite{harrow2013church} can be used:
        \begin{equation}\label{eq:app:haarint}
            \int \dyad{\psi}^{\otimes k} d\psi = \binom{d+k-1}{k}^{-1} P^+_{k,\,d},
        \end{equation}
        where $P^+_{k,\,d}$ is the  projector onto the symmetric subspace of Hilbert space of $k$ parties, each of dimension $d$. 
        
        Calculation of~(\ref{eq:app:expecthaar}) with the use of~(\ref{eq:app:haarint}) yields~(Ref.~\cite{NonLinWit24}, eq.~(F10)):
        \begin{equation}\label{eq:app:haarres}
            \mathbb{E}\left[\langle\mathcal W\rangle\right] = \binom{d+k-1}{k}^{-1}\left(\mathrm{Tr}\,P^-_{k,\,d_1}\right)^2 (1-k!),
        \end{equation}
        where $P^-_{k,\,d_1}$ - the projector onto the anti-symmetric subspace of Hilbert space of $k$ parties, each of dimension $d_1$ (here we assume that $d_1 = d_2$).
        One cannot use~(\ref{eq:app:haarres})  for arbitrarily large  number  of copies $k$, since  
        \begin{equation}
            \mathrm{Tr}\,P^-_{k,\,d_1} = \begin{cases}
                \binom{d_1}{k} & \text{if $k\leqslant d_1$},\\
                \,\,\,\,0              & \text{if $k > d_1$}
            \end{cases}
        \end{equation}
        is the dimension of the antisymmetric subspace of $k$ parties with local dimension $d_1$, and the projector vanishes when $k > d_1$. Therefore, for two-qubit states, i.~e., when $d_1=d_2=2$, the number of copies is limited by $2$. Nevertheless, the following argument can be used. One can take bipartite states with  local parties of sufficiently large dimensions $d_1$ and $d_2$  and then "split" each local dimension into a tensor product of smaller ones. Let us say, $d_1 = d_2 = 2^s$, then each such subsystem can be split into $s$ one-qubit systems. If the initial state $\rho_{AB}$ factorizes, under splitting operation, into tensor product of $s$ identical  two-qubit states, then using $k$ copies of $\rho_{AB}$ will result in $ks$ copies of some two-qubit state. Note also that in this case the entanglement of the initial state is fully determined by the entanglement of its factors. Of course, not every pure state factorizes under splitting operation exactly into a tensor product of states with smaller local dimensions, but averaging in~(\ref{eq:app:expecthaar}) is performed over Haar-random states, and this set includes those states that factorize. The simplest example of a factorizable entangled state is the Bell's state of dimension $d^2$
        \begin{equation}
            \frac1{\sqrt{d^2}}\sum_{l=1}^{d^2}\ket{l}_A\ket{l}_B =\frac1{\sqrt{d^2}}\sum_{i,j=1}^d\ket{ij}_A\ket{ij}_B\quad\longrightarrow\quad\left(\frac1{\sqrt{d}}\sum_{i=1}^d\ket{i}_{A'}\ket{i}_{B'}\right)\otimes\left(\frac1{\sqrt{d}}\sum_{j=1}^d\ket{j}_{A''}\ket{j}_{B''}\right),
        \end{equation}
        where the parties $A$ and $B$ of dimension $d^2$ are split into $d$-dimensional parties $A'$ and $A''$, $B'$ and $B''$, respectively. Splitting results in two identical Bell's states of smaller dimension $d$.
        
        Now we can use eq.~(\ref{eq:app:haarres}) for $k$ copies, $d_1 = d_2 = 2^s$ and, consequently, $d = d_1d_2 = 2^{2s}$. Averaging over such Haar-random pure states includes those that can be factorized into $s$ identical two-qubit states. The factorization results in $ks$ copies of such states. The reason why eq.~(\ref{eq:app:haarres}) still holds after splitting is that trace in eq.~(\ref{eq:app:expecthaar}) is stable under such an operation. When $k=2$, the averaging is performed over the set which includes $2s$ copies of identical two-qubit states and results in expression
        \begin{equation}
          \mathbb{E}\left[\langle\mathcal W\rangle\right] = \binom{2^{2s}+1}{2}^{-1}\binom{2^s}{2}^2(1-2!) = -\frac{(2^s-1)^2}{2(2^{2s}+1)}. 
        \end{equation}
        Starting from the value $-0.1$ at $s=1$ ($2$ copies), the average monotonically decreases and tends to $-0.5$ when $s\to\infty$. This decreasing of the average value of entanglement witness over random states  suggests that entanglement detection becomes more effective with the increase of number of copies of two-qubit states. We can then conjecture that, when working with many copies of  states, the  regression model variationally finds observables like those in eqs.~(\ref{eq:app:bipkwit}),~(\ref{eq:app:antiwit}) or maybe even more optimal ones connected with  some tight bounds on entanglement measures. This could  potentially  explain the increase of negativity prediction accuracy on $4$ copies of  states which is seen on  Fig.~\ref{fig:ent_rand_mixed}.

\end{document}